\documentclass[aps,prd,nofootinbib,onecolumn,superscriptaddress,preprintnumbers,balancelastpage,longbibliography,nobibnotes]{revtex4-1}

\usepackage[dvipsnames]{xcolor}
\usepackage{graphicx,color,amsmath,amssymb,flushend,bm,mathrsfs,comment}
\definecolor{linkcolor}{rgb}{0.0, 0.28, 0.67}

\usepackage[
   colorlinks=true,
    urlcolor=linkcolor,
   anchorcolor=linkcolor,
    citecolor=linkcolor,
    filecolor=linkcolor,
    linkcolor=linkcolor,
    menucolor=linkcolor,
    linktocpage=true,
    pdfproducer=medialab,
    pdfa=true
]{hyperref}

\usepackage[capitalise]{cleveref}
\usepackage{tikz}
\usepackage{ulem}
\usepackage{float}

\usepackage{amsmath}
\usepackage{physics}
\usepackage{simpler-wick}
\usepackage{amsfonts}
\usepackage{amssymb}
\usepackage[mathscr]{euscript}
\usepackage{setspace}
\usepackage{lipsum}
\usepackage{slashed}
\usepackage{cancel}
\usepackage{multirow}
\usepackage[utf8]{inputenc}
\usepackage{mathtools}

\usepackage{fontawesome5}

\usetikzlibrary{calc}
\usepackage{siunitx}
\DeclareSIUnit{\year}{yr}
\DeclareSIUnit{\parsec}{pc}
\DeclareSIUnit{\eV}{e\kern-.05em V}
\DeclareSIUnit{\Jansky}{Jy}
\DeclareSIUnit{\sr}{sr}

\newcommand{\nocontentsline}[3]{}
\newcommand{\tocless}[2]{\bgroup\let\addcontentsline=\nocontentsline#1{#2}\egroup}

\def\dbar{{\mathchar'26\mkern-12mu d}}

\newcommand{\bea}{\begin{eqnarray}\begin{aligned}}
\newcommand{\eea}{\end{aligned}\end{eqnarray}}

\newcommand{\Th}{\text{\normalfont th}}

\newcommand{\cm}{\text{cm}}

\newcommand{\Hz}{\text{Hz}}

\newcommand{\eV}{\text{eV}}

\newcommand{\HI}{\text{HI}}

\newcommand{\ma}{m_\gamma}
\newcommand{\map}{m_{A'}}

\newcommand{\res}{\text{res}}

\newcommand{\Eq}[1]{Eq.~(\ref{eq:#1})}

\newcommand{\data}{ {(\text{data})} }

\newcommand{\TTT}{{\bf T}}
\newcommand{\MMM}{{\bf M}}
\newcommand{\YYY}{{\bf Y}}

\newcommand{\ttt}{\mathcal{T}}
\newcommand{\mmm}{\mathcal{M}}
\newcommand{\yyy}{\mathcal{Y}}

\newcommand{\transfunc}{\mathscr{T}}

\newcommand{\bgf}{\bar{f}_\gamma}
\newcommand{\bgn}{\bar{n}_\gamma}
\newcommand{\bgrho}{\bar{\rho}_\gamma}
\newcommand{\bgI}{\bar{I}_\gamma}

\newcommand{\bgvecI}{\bar{\bm I}_\gamma}

\newcommand{\Gammainj}{\Gamma_{\text{inj}}}

\newcommand{\mupara}{\mu}

\newcommand{\veck}{{\bm k}}
\newcommand{\vecI}{{\bm I}_\gamma}
\newcommand{\vecQ}{\bm{\mathcal{Q}} }

\newcommand{\vecres}{\bm{\mathcal{R}}}

\newcommand{\matcov}{\bm{\mathcal{C}}}

\newcommand{\da}{\text{data}}

\newcommand{\TS}{\text{TS}}

\newcommand{\cs}{\text{CS}}
\newcommand{\dcs}{\text{DCS}}
\newcommand{\br}{\text{BR}}

\newcommand{\vis}{\mathcal{J}^*}

\newcommand{\Ps}{P_s}

\newcommand{\Tcmb}{T}
\newcommand{\Tcmbtoday}{T_0}

\newcommand{\wcmb}{\omega}
\newcommand{\wcmbtoday}{\omega_0}

\newcommand{\nucmbtoday}{\nu_0}

\newcommand{\trans}{\text{trans}}

\newcommand{\tauff}{\tau_\text{ff}}

\newcommand{\bbT}{\mathbb{T}}
\newcommand{\bbF}{\mathbb{F}}

\newcommand{\Sec}[1]{Sec.~\ref{sec:#1}}

\newcommand{\Appx}[1]{Appendix~\ref{appx:#1}}
\newcommand{\SubAppx}[1]{Appendix~\ref{appx:#1}}

\newcommand{\Source}{S}

\newcommand{\deltalike}{F}

\newcommand{\Fig}[1]{Fig.~\ref{fig:#1}}

\newcommand{\refcite}[1]{Ref.~\cite{#1}}
\newcommand{\refscite}[1]{Refs.~\cite{#1}}

\newcommand{\inj}{\text{inj}}

\newcommand{\githubicon}{\href{https://github.com/GiorgiArsenadze/CMB-Shaping-Dark-Photon-Spectral-Distortions}{\faGithub}}

\begin{document}

\preprint{CERN-TH-2024-153}

\preprint{DESY-24-139}

\title{Shaping Dark Photon Spectral Distortions}

\author{Giorgi Arsenadze}
\email{ga1348@nyu.edu}
\affiliation{Center for Cosmology and Particle Physics, Department of Physics, New York University, New York, NY 10003, USA}

\author{Andrea Caputo}
\email{andrea.caputo@cern.ch}
\affiliation{Theoretical Physics Department, CERN, 1211 Geneva 23, Switzerland}

\author{Xucheng Gan}
\email{xg767@nyu.edu}
\email{xucheng.gan@desy.de}

\affiliation{Center for Cosmology and Particle Physics, Department of Physics, New York University, New York, NY 10003, USA}

\affiliation{Deutsches Elektronen-Synchrotron DESY, Notkestr. 85, 22607 Hamburg, Germany}

\author{Hongwan Liu}
\email{hongwan@bu.edu}
\affiliation{Physics Department, Boston University, Boston, MA 02215, USA}
\affiliation{Kavli Institute for Cosmological Physics, University of Chicago, Chicago, IL 60637}
\affiliation{Theoretical Physics Department, Fermi National Accelerator Laboratory, Batavia, IL 60510}

\author{Joshua T. Ruderman}
\email{ruderman@nyu.edu}
\affiliation{Center for Cosmology and Particle Physics, Department of Physics, New York University, New York, NY 10003, USA}

\begin{abstract}
The cosmic microwave background (CMB) spectrum is an extraordinary tool for exploring physics beyond the Standard Model. The exquisite precision of its measurement makes it particularly sensitive to small effects caused by hidden sector interactions. In particular, CMB spectral distortions can unveil the existence of dark photons which are kinetically coupled to the standard photon. In this work, we use the COBE-FIRAS dataset to derive accurate and robust limits on photon-to-dark-photon oscillations for a large range of dark photon masses, from $10^{-10}$ to $10^{-4}$ eV\@. We consider in detail the redshift dependence of the bounds, computing CMB distortions due to photon injection/removal using a Green's function method. Our treatment improves on previous results, which had set limits studying energy injection/removal into baryons rather than photon injection/removal, or ignoring the redshift evolution of distortions. The difference between our treatment and previous ones is particularly noticeable in the predicted spectral shape of the distortions, a smoking gun signature for photon-to-dark-photon oscillations. The characterization of the spectral shape is crucial for future CMB missions, which could improve the present sensitivity by orders of magnitude, exploring regions of the dark photon parameter space that are otherwise difficult to access.~\githubicon

\end{abstract}

\maketitle

\tableofcontents

\section{Introduction}
\label{sec:introduction}

The dark photon $A'$ is a hypothetical particle that can open one of only a few renormalizable portals between the Standard Model (SM) sector and the dark sector. A significant research program is ongoing to detect dark photons with a mass of $m_{A'} \lesssim \SI{e-3}{\eV}$. The dark photon parameter space can be tested by terrestrial experiments, such as  Cavendish-Coulomb experiments~\cite{Bartlett:1970js, Williams:1971ms, Bartlett:1988yy, Kroff:2020zhp}, Light-Shining-Through-Walls experiments~\cite{Betz:2013dza,Berlin:2022hfx,Ortiz:2020tgs,Miyazaki:2022kxl, Berlin:2023mti, Antel:2023hkf}, helioscopes~\cite{Redondo:2008aa,Schwarz:2015lqa,Frerick:2022mjg,OShea:2023gqn}, and the direct detection of dark photons produced in the Sun~\cite{An:2014twa, An:2020bxd, Lasenby:2020goo, XENON:2021qze}. 
The dark photon parameter space can also be tested through astrophysical probes such as stellar energy loss~\cite{Redondo:2008aa, Pospelov:2008jk, Redondo:2008ec, Redondo:2013lna, An:2013yfc, Vinyoles:2015mhi,Redondo:2015iea, Hardy:2016kme,Li:2023vpv} and black hole superradiance~\cite{Cardoso:2018tly,Davoudiasl:2019nlo,Unal:2020jiy,Siemonsen:2022ivj}. Moreover, because the dark photon can impact the SM electromagnetic field through the kinetic mixing portal, the magnetosphere of Jupiter~\cite{Davis:1975mn,Marocco:2021dku,Yan:2023kdg} and Earth~\cite{Goldhaber:1971mr,Bartlett:1988yy,Fischbach:1994ir,Kloor:1994xm,Marocco:2021dku} can also be used to constrain the kinetic mixing parameter of ultralight dark photons in the \SIrange{e-16}{e-13}{\eV} mass range. For a complete review of dark photon limits, one can refer to \refcite{Caputo:2021eaa}. All dark photon searches are, however, limited by the decoupling of all interactions with the SM when the dark photon mass vanishes, $m_{A'} \to 0$, which makes detection increasingly difficult at low masses~\cite{Holdom:1985ag, An:2013yfc, An:2014twa}.

Compared with the aforementioned detection methods, distortions to the blackbody spectrum of the cosmic microwave background (CMB) serve as one of the most sensitive ways to explore the dark photon parameter space for $m_{A'} \lesssim \SI{e-4}{\eV}$~\cite{Mirizzi:2009iz,Mirizzi:2009nq,Kunze:2015noa,McDermott:2019lch,Caputo:2020bdy,Caputo:2020rnx,Garcia:2020qrp,Pirvu:2023lch,Aramburo-Garcia:2024cbz,McCarthy:2024ozh}. 
In the early universe, CMB photons were in chemical equilibrium with baryons, and their phase space density obeyed the blackbody distribution with zero chemical potential.
Within standard $\Lambda$CDM cosmology, this blackbody distribution should be mostly preserved through the process of recombination until the present day. 
The COBE-FIRAS measurement of this spectrum confirmed this prediction, finding that any deviation of the CMB phase space from a blackbody distribution has to be less than 1 part in $10^4$~\cite{Fixsen:1996nj}. Next-generation measurements of the CMB spectrum will have much better sensitivity: the proposed PIXIE~\cite{Kogut:2011xw, Kogut:2024vbi}, PRISM~\cite{PRISM:2013fvg}, Voyage 2050~\cite{Chluba:2019nxa} and SPECTER~\cite{Sabyr:2024lgg} experiments aim to measure CMB spectral distortions that are as small as $10^{-10}$--$10^{-8}$ times the CMB blackbody intensity. Any exotic process that injects energy into the CMB, after the CMB photons fall out of chemical equilibrium, potentially imprints a distortion away from the blackbody spectrum. 
Precise measurements of the CMB spectrum are therefore a powerful tool for detecting not just dark photons, but also axions~\cite{Mirizzi:2005ng, Mirizzi:2009nq,Tashiro:2013yea,Ejlli:2013uda, Mukherjee:2018oeb}, dark matter decay/annihilation/scattering~\cite{Ali-Haimoud:2021lka}, and inflation~\cite{Chluba:2019kpb}.

Significant spectral distortions caused by dark photons can arise if photons $\gamma$ and $A'$ kinetically mix, resulting in a sufficiently large probability of  $\gamma \to A'$ conversions throughout cosmic history, removing photons from the CMB blackbody spectrum. 
The conversion probability of CMB photons into dark photons is dominated by resonant conversions between the two states, which occur whenever the effective plasma mass of the SM photon $\gamma$---induced mainly by the density of free electrons---matches $m_{A'}$, the dark photon mass~\cite{Zener:1932ws, landau1932theorie, Mirizzi:2009iz, McDermott:2019lch, Caputo:2020bdy, Caputo:2020rnx}.

The impact of $\gamma \to A'$ on the CMB spectrum can be classified into two broad regimes: $m_{A'} \lesssim \SI{e-9}{\eV}$, and $m_{A'} \gtrsim \SI{e-9}{\eV}$. 
For $m_{A'} \lesssim \SI{e-9}{\eV}$, resonant conversions only occur after recombination when photons are free-streaming; the distorted spectrum produced as a result of $\gamma \to A'$ evolves only through redshifting as a function of time after the conversion has occurred. 
This regime was first studied in \refcite{Mirizzi:2009iz}, and extended to include the effect of inhomogeneities on resonance conversion by~\refcite{Caputo:2020bdy,Caputo:2020rnx} (see also~\refcite{Bondarenko:2020moh}).  
The signal from the inverse process, $A' \to \gamma$, for dark photon dark matter have been discussed in \refscite{Arias:2012az,Dubovsky:2015cca,Wadekar:2019mpc,McDermott:2019lch,Caputo:2020bdy,Caputo:2020rnx,Witte:2020rvb,An:2020jmf,An:2022hhb,An:2024wmc,An:2024kls}. Throughout the rest of the paper, we assume that there is no initial abundance of the dark photon, and consider only the $\gamma \rightarrow A'$ process.

More recently, \refscite{Pirvu:2023lch,Aramburo-Garcia:2024cbz,McCarthy:2024ozh} have examined the impact of $\gamma \to A'$ on the CMB anisotropy power spectrum. 
These results rely on the fact that $\gamma \to A'$ resonant conversion is highly sensitive to the free electron number density, and is therefore correlated with cosmic structures, leading to nontrivial spatial correlations. 
These works have demonstrated impressive limits for $m_{A'} \lesssim \SI{e-9}{\eV}$, when $\gamma \to A'$ occurs after recombination.

In this paper, we instead focus on the regime $m_{A'} \gtrsim \SI{e-9}{\eV}$, when resonant conversions happen before recombination. 
At such high redshifts, inhomogeneities are unimportant, and we can safely take the universe to be homogeneous~\cite{Caputo:2020rnx}.
Our goal is to accurately compute the spectral distortion produced by these conversions, and use the COBE-FIRAS data to set limits on $\epsilon$ as a function of $m_{A'}$\@. 
This regime was first considered in \refcite{Mirizzi:2009iz}; however, they assumed that the spectral distortion produced by resonant conversion only evolves via redshifting, \textit{i.e.}\ that photons are always free-streaming. 
In fact, any spectral distortion can potentially be redistributed as a function of time due to efficient Compton scattering between photons and electrons in the epoch before recombination.
Subsequently, \refcite{McDermott:2019lch} derived limits on $\epsilon$ by assuming that distortions from $\gamma \to A'$ conversions are identical to distortions produced in the CMB when the equivalent amount of energy is removed from baryons at the same redshift, leading to so-called pure $\mu$- and $y$-distortions. 
This, however, is also not a good approximation: the actual CMB spectral distortion from photon removal can be significantly different from those corresponding to energy removal from baryons, as pointed out in \refcite{Chluba:2015hma}. 
The spectral distortion therefore does not correspond to a pure $\mu$- or $y$-distortion, and cannot be compared directly to limits on the $|\mu|$ and $|y|$ parameters, as was done in \refcite{McDermott:2019lch}. Instead, the full distortion must be carefully computed, and compared to the full COBE-FIRAS data. 

We apply the Green's function method developed in \refcite{Chluba:2015hma} for photon injection/removal to accurately compute the spectral distortion due to $\gamma \to A'$ conversions. 
We obtain accurate results when resonant conversions occur during the $\mu$-era (when Compton scattering is highly efficient at redistributing photons, applicable to $m_{A'}$ in the approximate range of \SIrange{3e-6}{5e-5}{\eV}) and in the $y$-era (when Compton scattering is inefficient, applicable to $m_{A'}$ in the approximate range of \SIrange{e-9}{2e-8}{\eV}. 
In the intermediate $\mu$-$y$ transition era, we have obtained reliable approximations for the spectral distortion, allowing us to set a conservative limit on $\epsilon$ in the mass range \SIrange{2e-8}{3e-6}{\eV}. 
Our new limits strengthen existing limits by up to a factor of $3$, and are particularly important in setting a benchmark for upcoming Light-Shining-Through-Walls experiments such as DarkSRF, which offer a complementary search strategy for dark photons in the \SIrange{e-9}{e-5}{\eV} range~\cite{Romanenko:2023irv}. Some of our new results have been already presented elsewhere by a subset of us~\cite{Gan:2024ele, TeVPA:2024}; here we provide the complete analysis.

The rest of this paper is organized as follows. 
In \Sec{A_to_Ap}, we introduce the dark photon model and give a back-of-the-envelope estimation of its COBE-FIRAS constraints. 
In \Sec{spec_dist}, we discuss the different stages of the early universe as classified by the efficiency of Compton scattering ({\cs}), bremsstrahlung ({\br}), and double Compton scattering ({\dcs}).  We also describe some important concepts in spectral distortions. 
In \Sec{green_func}, we introduce the Green's function method and explain how the spectral distortion, in the case of $\gamma \to A'$ conversions, is calculated. In \Sec{spec_dist_AToAp}, we present our main result: the COBE-FIRAS spectral distortion limits on $\epsilon$ as a function of $\map$. 
We compare our updated COBE-FIRAS constraint using the complete treatment with previous results. We also examine the CMB spectral distortion signal from $\gamma \rightarrow A'$ in detail at different representative points in the dark photon parameter space. 
We discuss the impact of an important approximation made in deriving our limits over a small range in $\map$.
Details of our data analysis method, the complete table of the constants appearing in the CMB spectral distortion calculation, the derivation of the $\mu$ distortion for the simplified monochromatic photon injection/removal, and the detailed discussion of Green's functions in different eras can be found in
Appendices~\ref{appx:likelihood}, \ref{appx:const}, \ref{appx:mono_mu_era}, and \ref{appx:green_func}, respectively.
We use $\hbar = c = k_B = 1$ for expressions given in this paper but will use radio astronomy units in plots, where appropriate. 
The code for obtaining our results are publicly available at \url{https://github.com/GiorgiArsenadze/Shaping-Dark-Photon-Spectral-Distortions}~\githubicon.

\section{Photon-To-Dark-Photon Oscillations}
\label{sec:A_to_Ap}

In this section, we give a brief introduction to $\gamma \to A'$ conversions. We begin with the low-energy Lagrangian describing this model,
\bea
\label{eq:Lagrangian}
\mathcal{L} = -\frac{1}{4} F_{\mu \nu} F^{\mu \nu} - \frac{1}{4} F'_{\mu \nu} F'^{\mu \nu} + \frac{\epsilon}{2} F_{\mu \nu} F'^{\mu \nu} - \frac{1}{2} \map^2 A'_{\mu} A'^{\mu} + e  A_{\mu} j^{\mu}_e \,,
\eea
where $A^{(\prime)}$ is the field of the photon~(dark photon), and $F^{(\prime)}$ is the corresponding field strength. Here, $e$ is the gauge coupling of the electromagnetic $U(1)$ gauge field, $j_e$ is the electromagnetic current, and the dimensionless quantity $\epsilon$ parameterizes the size of kinetic mixing between the photon and the dark photon. Here, $\epsilon$ obeys the upper bound $\epsilon < 1$ according to the requirement of the positive-definiteness of the gauge kinetic terms~\cite{Weinberg:1996kr, Burgess:2008ri}; a small value of $\epsilon$ is technically natural, and can be realized in several ultraviolet completions of such a  model~\cite{Gherghetta:2019coi,Gan:2023wnp}. 
In our work, we investigate the minimal scenario of the kinetic mixing portal, where only the dark photon is in the dark sector, with a negligible initial abundance. 
If there are additional particles charged under an ultralight dark photon that mixes with the SM photon, these particles would appear to be millicharged and can drastically alter the expected signal. 
Such modifications include an irreducible cosmic millicharged background~\cite{Gan:2023jbs, Iles:2024zka}, a millicharged-particle-induced dark plasma mass~\cite{Berlin:2022hmt,Berlin:2023gvx}, and dark thermalization inside stars~\cite{Chang:2022gcs,Fiorillo:2024upk}.

\begin{figure}[t]
\includegraphics[width=0.8\columnwidth]{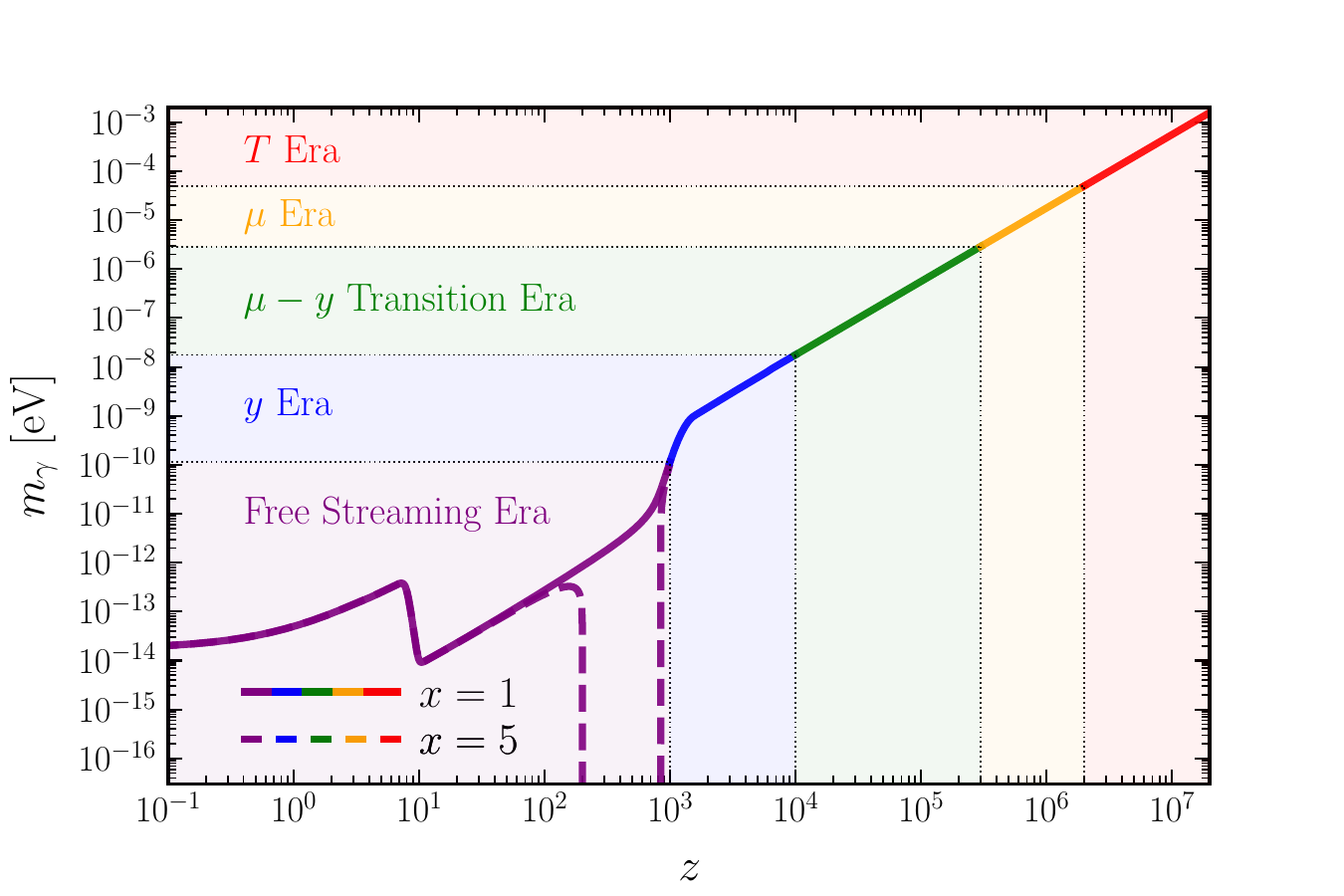}
\caption{The cosmological evolution of the photon's effective plasma mass as a function of the redshift. Based on the efficiency of the scattering channels discussed in \Sec{spec_dist}, the universe can be divided into five eras: the free streaming era~(purple), $y$-era~(blue), $\mu$-$y$ transition era~(green), $\mu$-era~(orange), and $T$ era~(red). Here, we show the evolution for two different photon energies, $x=1$ and $x=5$, which are the solid and dashed lines, respectively. The $x=5$ line dips down in the period $2 \times 10^2< z < 9 \times 10^2$, due to the negative mass squared contribution from neutral Hydrogen. In other periods, when most of the neutral Hydrogen is ionized, the difference between $x=5$ and $x=1$ is very small.}
\label{fig:z_mA}
\end{figure}

Because the photon is coupled to the electromagnetic current, the electrons induce an effective plasma mass~\cite{Born:1999ory} in the finite density environment of the early universe, given by
\bea
\label{eq:plasma_mass}
\ma^2(z) 
& = \frac{e^2 n_e(z)}{m_e} - \wcmb^2(z) \, (\vb{n}^2_\HI\left(z)-1\right)\\
& \simeq 1.4 \times 10^{-21} \eV^2 \left(\frac{n_e(z)}{\cm^{-3}}\right) - 8.4 \times 10^{-24} \eV^2 \left(\frac{\wcmb(z)}{\eV}\right)^2 \left( \frac{n_\HI(z)}{\cm^{-3}} \right),
\eea
where $\vb{n}_\HI$ is the refractive index of neutral hydrogen, $n_\HI$ is the number density of neutral hydrogen, and $n_e$ is the free electron number density. 
We obtain $n_e$ and $n_\HI$ from CLASS, which uses HyRec to model recombination~\cite{Ali-Haimoud:2010hou,Lee:2020obi}, and uses the $\tanh$ scenario to model reionization~\cite{Planck:2018vyg}.\footnote{The choice of modeling for reionization is not important for this paper, since we focus on high-redshift conversions.} We adopt cosmological parameters consistent with Planck 2018~\cite{Planck:2018vyg}. \Eq{plasma_mass} represents the modification of the dispersion relation in the presence of an optically dense medium. Here, the positive term is generated by free electrons, and the negative term comes from neutral hydrogen~\cite{Berlin:2022hmt}. The present work mainly focuses on dark photon phenomenology at high redshifts, above $z \sim 100$, where baryon fluctuations are negligible and the universe can be approximated as homogeneous. In \Fig{z_mA}, we illustrate the cosmological evolution of the photon plasma mass as a function of the redshift. Based on the efficiencies of various processes which control the evolution of spectral distortions, we can divide the universe into five distinct eras: free-streaming era~(purple), $y$-era~(blue), $\mu$-$y$ transition era~(green), $\mu$-era~(orange), and $T$-era~(red); these different eras will be described in detail in \Sec{spec_dist}. We plot the plasma frequency evolution for two different values of the photon energy $\omega$, which enters the second term in Eq.~\ref{eq:plasma_mass}. To characterize the photon energy, we introduce the redshift-independent quantity $x \equiv \omega_0/T_0$, with $\omega_0$ and $T_0$ being the present-day CMB photon energy and temperature, respectively. We show the plasma frequency evolution for $x=1$ and $x=5$. The plasma mass for $x=5$ becomes negative during $200 \lesssim z \lesssim 900$, due to the negative mass squared contribution from neutral hydrogen after hydrogen recombination. However, our focus will be on the period well before hydrogen recombination, when the contribution from neutral hydrogen is negligible and thus the plasma frequency evolution is identical for both $x=1$ and $x=5$.

The effective plasma mass, $\ma^2(z)$, evolves throughout the expansion of the universe, and it may match $\map^2$ at some moment during the cosmological history. If this happens, then a resonant $\gamma \rightarrow A'$ conversion can take place. These conversions can be modeled as  Landau-Zener transitions~\cite{landau1932theorie,Zener:1932ws}, as has been extensively explored in the context of neutrinos~\cite{Parke:1986jy,Kuo:1989qe} and dark photons~\cite{Mirizzi:2009iz,Caputo:2020bdy,Caputo:2020rnx, Berlin:2022hfx,Brahma:2023zcw}. The rate for the $\gamma \rightarrow A'$ transition reads~\cite{Caputo:2020bdy,Caputo:2020rnx,Berlin:2022hmt}
\bea
\label{eq:trans_rate}
\Gamma_{\gamma \rightarrow A'}(z) = \frac{\epsilon^2 m_{A'}^4 \Gamma_\gamma}{\left(m_\gamma^2(z) - m_{A'}^2\right)^2 + \wcmb^2(z) \Gamma_\gamma^2},
\eea
where $\Gamma_\gamma$ is the damping rate of the plasmon quanta. Under the narrow width approximation $\Gamma_\gamma \ll m_{A'}^2/\omega$, the associated conversion probability can be written as
\bea
\label{eq:trans_prob}
P_{\gamma \rightarrow A'}(\omega) = \int dt \, \Gamma_{\gamma \rightarrow A'} = \frac{\pi \epsilon^2 m^2_{\gamma}(z_\res) }{\wcmb(z_\res) (1+z_\res) H(z_\res)} \abs{\frac{d \log \ma^2}{dz}}_\res^{-1},
\eea
where ``$\res$'' labels the time when $\ma^2=\map^2$, $\wcmb(z_\res) = \wcmbtoday (1+z_\res)$, and $dt = - dz/(1+z) H(z)$, with $H(z)$ being the Hubble parameter. 

We note here that rapid scattering of $\gamma$ with free electrons in the plasma can affect the process of resonant conversion. 
In order for scattering to have a negligible effect, the resonance timescale, $\tau_r$ (the inverse of the resonant width), must be significantly smaller than the Compton scattering timescale (or, equivalently, the mean free path) $\tau_s$ for photons in the plasma. 
For the masses and kinetic mixing of interest in this work, we have checked that $\tau_r \ll \tau_s$.

In order to provide some intuition for typical values of the conversion probability, and its scaling with the dark photon mass, let us briefly  work out some simple estimates for dark photon masses $m_{A'} \gtrsim 10^{-9}\eV$\@. In this case, the $\gamma \rightarrow A'$ transition happens in the radiation-dominated~(RAD) era. At this time, hydrogen atoms are fully ionized, which gives $m_\gamma^2(z) \propto n_e(z) \propto (1+z)^3$. Using the fact that $H(z) = H_0 \,\Omega_r^{1/2} (1+z)^2$ during the RAD era, where $H_0$ is the Hubble constant, $\Omega_r = \Omega_m / (1+z_\text{eq})$, $z_\text{eq}$ is the redshift of matter-radiation equality, and $\Omega_m$ is the matter density parameter, we get
\bea
\label{eq:trans_prob_RAD}
\text{RAD:}\quad  P_{\gamma \rightarrow A'}(x) \simeq \frac{\epsilon^2 \mathcal{F}}{x} \quad \quad \text{with} \quad \mathcal{F} = \frac{\pi m_\gamma^2(z=0)}{3 \, \Omega_r^{1/2} \, H_0 \Tcmbtoday } \simeq 10^{11} \,.
\eea
To calculate the numerical value of $\mathcal{F}$, we used $m_\gamma(z=0) \simeq \SI{1.7e-14}{\eV}$. The first thing we notice in \Eq{trans_prob_RAD} is that any dependence on $\map$ has disappeared. We thus expect a bound which is roughly constant for these large masses. Moreover, we can derive a rough estimate of the bound by requiring $P_{\gamma \rightarrow A'}$ to be $\lesssim 10^{-4}$, the typical fractional uncertainty of the COBE-FIRAS measurement of the blackbody intensity. We thus find
\bea
\label{eq:eps_est}
\epsilon_\text{est} \sim 3 \times 10^{-8} \left( \frac{P_{\gamma \rightarrow A'}}{10^{-4}} \right)^{1/2} \left(\frac{10^{11}}{\mathcal{F}}\right)^{1/2},
\eea
where $\epsilon_\text{est}$ stands for the estimated COBE-FIRAS constraint in the high redshift region. A next-generation PIXIE-like experiment can, in principle, be sensitive to distortions on the order of 1 part in $10^8$, leading to a potential improvement of roughly two orders of magnitude in sensitivity to $\epsilon$. However, this assumes perfect removal of foreground distortions from \textit{e.g.}\ the epoch of reionization. 

We stress that \Eq{eps_est} merely represents a rough estimate for the constraint from CMB spectral distortions. While it gives a sense of the reach of such a probe, it is insufficient for obtaining the precise mass dependence of the bounds, and it has no information on the spectral shape of possible future signals. In the rest of this paper, we provide a thorough investigation of the CMB distortion caused by $\gamma \rightarrow A'$ oscillations using the Green's function method, for photon injection or removal processes, developed in \refcite{Chluba:2015hma}.

\section{Spectral Distortions}
\label{sec:spec_dist}

In this section, we give a brief overview of how spectral distortions are generated in the CMB by exotic energy injections or removals into photons. We intend only to introduce terminology that will be important for the reader to understand our method; for more in-depth discussions we invite the interested reader to consult the seminal works~\cite{Illarionov1975,ChanJones1975,DaneseDeZotti80,DaneseDeZotti82,Sunyaev1970,Zeldovich69} or more recent papers such as \refscite{Daly91, Hu:1992dc, Hu:1995em,Chluba:2011hw,Chluba:2013pya,Chluba:2013vsa,Chluba:2015hma}. Before we begin the discussion, we want to emphasize that much of the existing literature has focused on pure energy injection or removal processes, \textit{i.e.}\ processes that always conserve the comoving number density of photons. This can happen, for example, if the exotic process heats the baryons first, and the photons react to this change subsequently.\footnote{This can in fact be highly unrealistic; any process that heats the baryons would also likely interact with photons at the same time. See \refcite{Acharya:2018iwh} for a detailed study of the distortion produced by high-energy particles without making such an assumption.} This is certainly not the case for $\gamma \to A'$ resonant conversions, and is the key reason why our results differ from \refcite{McDermott:2019lch}; however, we still follow existing conventions for clarity. 

In the early universe, rapid scattering with electrons in the baryon plasma, together with efficient photon-number-changing processes, ensures that the photons are in thermal equilibrium with zero chemical potential. Photons therefore follow a blackbody distribution, \textit{i.e.}\ their phase space density is given by $\bar{f}_\gamma$, where\footnote{We adopt the convention of defining phase space density excluding the degeneracy factor of 2 due to the spin states of the photon.}
\bea
\label{f:black_body}
\bgf(\omega, T) \equiv \frac{1}{e^{\omega / T} - 1 }  \,.
\eea
All thermodynamic properties of the photons are determined simply by their temperature $T$.

Once $T \lesssim \SI{}{\kilo\eV}$, however, the rate of photon-number-changing processes, in the energy range relevant to the measured CMB spectrum today, drops below the Hubble expansion rate. After this point, any process that changes the number and energy densities of photons will drive the photon distribution away from the blackbody distribution, leading to a spectral distortion. Of course, if no such processes exists, then photons remain in a blackbody distribution throughout cosmological history, \textit{i.e.}\ the photon phase space remains $f_\gamma(x) = \bgf(x)$ at all redshifts, where $x \equiv \omega/T$. We therefore define the spectral distortion $\Delta f(x)$ as the distortion to the blackbody distribution that we would observe today,
\bea
\label{eq:CMB_dist}
f_\gamma(x) = \bgf(x) + \Delta f_\gamma (x) \,.
\eea 

The CMB phase space density has been measured by the FIRAS instrument aboard the COBE satellite in over 43 frequency bins ranging from $x \sim 1$ to $x \sim 10$. The FIRAS measurement confirmed that the CMB phase space density is consistent with a blackbody distribution, with a precision of 1 part in $10^4$, \textit{i.e.}\ $\Delta f_\gamma / \bgf \lesssim 10^{-4}$ in this frequency range. Any potential spectral distortion of interest, in this energy range, can therefore be taken to be small. 

\begin{figure}[t]
\includegraphics[width=0.99\columnwidth]{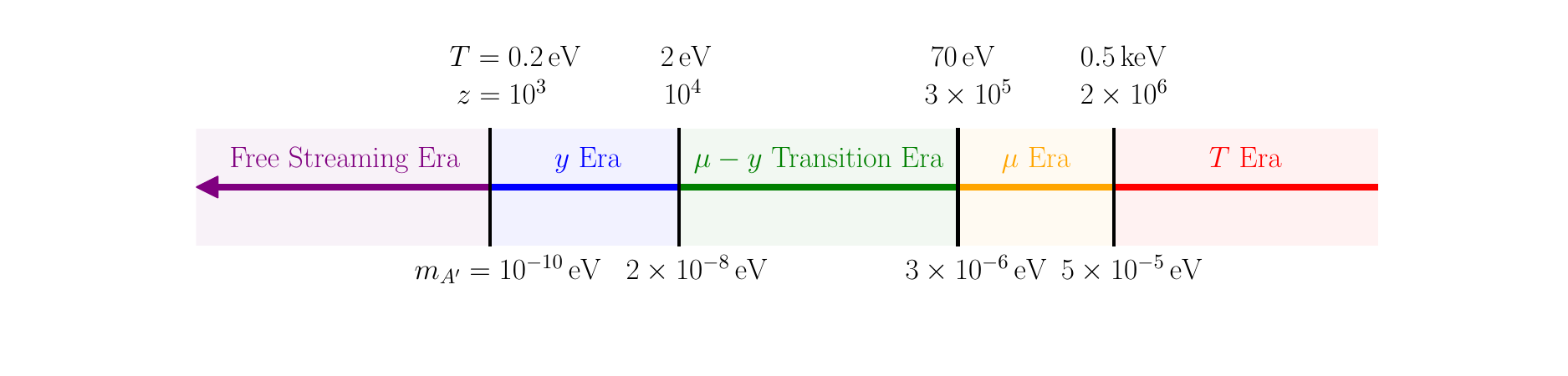}
\caption{The eras of CMB spectral distortions. The redshifts separating different eras are located above the axis. The corresponding dark photon masses for $\gamma \rightarrow A'$ resonant oscillations are shown below the axis. There are five eras depending on the efficiency of {\cs}, {\dcs}, and {\br} (in reverse chronological order): the free streaming era~(purple), $y$-era~(blue), $\mu$-$y$ transition era~(green), $\mu$-era~(orange), and $T$ era~(red). }
\label{fig:cmb_distortion_eras}
\end{figure}

The nature of the spectral distortion produced by any exotic energy injection process depends strongly on when these processes occur, and what scattering processes photons are undergoing efficiently at that time. The most relevant scattering processes are
\bea
\label{eq:channels}
\text{Compton Scattering~(\cs):}& \quad e^- + \gamma \leftrightarrow e^- + \gamma \,,\\
\text{Double Compton Scattering~(\dcs):}& \quad e^- + \gamma \leftrightarrow e^- + \gamma + \gamma \,, \\
\text{Bremsstrahlung~(\br):}& \quad e^- + X \leftrightarrow  e^- + X  + \gamma \,.
\eea
Of these processes, {\dcs} and {\br} are photon-number-changing. If they are efficient, \textit{i.e.}\ have a rate much larger than the Hubble rate, they can drive the phase space density of photons rapidly toward a blackbody distribution, \textit{i.e.}\ a Bose-Einstein distribution with zero chemical potential. On the other hand, {\cs} is number conserving; it can still however redistribute photons and change their phase space distribution. The cosmological epoch when $T \lesssim \SI{}{\kilo\eV}$ can be divided into five main eras, according to how rapid these processes are. These are the $T$ era ($T \gtrsim \SI{0.5}{\kilo\eV}$), $\mu$ era ($\SI{70}{\eV} \lesssim T \lesssim \SI{0.5}{\kilo\eV}$), $\mu$-$y$ transition era ($\SI{2}{\eV} \lesssim T \lesssim \SI{70}{\eV}$), $y$-era ($\SI{0.2}{\eV} \lesssim T \lesssim \SI{2}{\eV}$), and free-streaming era $(T \lesssim \SI{0.2}{\eV}$). The corresponding redshifts separating these eras, and the values of $m_{A'}$ which would lead to a resonant conversion at the transitions between them, are shown in \Fig{cmb_distortion_eras}.

The $T$ era corresponds to $T \gtrsim \SI{0.5}{\kilo\eV}$, when DCS and BR are highly efficient. These processes enforce thermal equilibrium and zero chemical potential in the photon distribution. Any arbitrary change to the phase space density away from a blackbody distribution by energy injection processes is quickly redistributed, resulting in a new blackbody distribution at a higher temperature. The characteristic distortion $\ttt(x)$, that is produced due to energy injection in this epoch, is simply a temperature shift to the spectrum, and is defined as
\bea
T\text{ era:} \quad \Delta f_\gamma(x) \equiv \bgf(\omega, T + t T) - \bgf(\omega, T) \simeq \bgf(x) \, t \ttt(x) \,,
\eea
where $t \ll 1$ determines the size of the distortion. Expanding to linear order in $t$ shows that
\bea
    \ttt(x) = \frac{x e^x}{e^x - 1} \,.
\eea
Note that such a distortion is fundamentally unobservable, because the blackbody temperature observed today, $T_0$, is a free parameter in $\Lambda$CDM cosmology, and can always be adjusted to absorb any unexpected energy injection processes happening during the $T$ era. The impact of exotic energy injection processes on the CMB spectrum are therefore only observable if they occur after the $T$ era. 

In the $\mu$-era ($\SI{70}{\eV} \lesssim T \lesssim \SI{0.5}{\kilo\eV}$), {\dcs} and {\br} become inefficient, while {\cs} still remains highly efficient. Immediately after any energy injection process, {\cs} rapidly drives photons to thermal equilibrium, \textit{i.e.}\ the phase space density will approach a Bose-Einstein distribution. However, the lack of number-changing-processes means that the comoving number density of photons, after the injection process, is conserved, and the equilibrium distribution reached will, in general, have a nonzero chemical potential. Injection processes can also cause a temperature shift, as in the $T$ era. The spectral distortion in this era can be written as 

\bea
\mu\text{ era:} \quad \Delta f_\gamma(x) = \bgf(\omega+\mu T, (1+t)T) - \bgf(\omega, T) \,. 
\eea
We can likewise expand in $\mu$ and $t$ to obtain the following conventional form for the distortion:
\bea
\label{eq:mu_era_dist}
\Delta f_\gamma(x) \simeq \bgf(x) \left[ \mu \mmm(x) + (t - \alpha_\mu \mu) \ttt(x) \right] \,,
\eea
where
\bea 
\label{eq:mmm_def}
\mmm(x) \equiv \ttt(x) \left( \alpha_\mu - \frac{1}{x} \right) \,,
\eea
and $\alpha_\mu \simeq 0.456$.\footnote{The convention adopted here includes the factor of $\alpha_\mu \ttt(x)$ in the definition of $\mmm(x)$. This originates from the fact that for a pure energy injection with no injection of photons, one can derive $t = \alpha_\mu \mu$, and so $\mmm(x)$ accounts for the full distortion. This relation does not hold, however, if the photon number changes during the injection, but we still follow this convention for consistency of notation.} $\mmm(x)$ is typically called the $\mu$-distortion, and $\mu$ specifies the size of this distortion. 

In the $y$-era ($\SI{0.2}{\eV} \lesssim T \lesssim \SI{2}{\eV}$), {\cs} becomes too slow for photons to remain in thermal equilibrium, \textit{i.e.}\ the photon phase space density no longer approaches a Bose-Einstein distribution. However, {\cs} is still rapid enough that photons scatter with baryons, allowing some limited redistribution of photon energies. In a pure energy injection process, where energy is dumped into heating the baryons, blackbody photons undergo {\cs} with the baryons, producing a $y$-distortion: 
\bea
y\text{ era, pure energy injection:} \quad \Delta f_\gamma(x) = \bgf(x) \, y \yyy(x) \,,
\eea
with $y \ll 1$, and where
\bea
\yyy(x) = \ttt(x) \left(x \frac{e^x+1}{e^x-1} -4\right) \,.
\eea
We stress, however, that for general energy injection processes with photons being injected or removed, the distortion can be significantly more complicated. The transition between the $\mu$ and $y$-eras ($\SI{2}{\eV} \lesssim T \lesssim \SI{70}{\eV}$) is also an epoch where the evolution of photons is complicated, and requires a numerical treatment, even in the case of pure energy injection~\cite{Chluba:2013vsa}. 

After recombination, the majority of photons cease scattering with baryons. During this free-streaming era ($T \lesssim \SI{0.2}{\eV}$), any distortion to the CMB, \textit{e.g.}\ from $\gamma \to A'$ resonant conversion, remains frozen in place, and photons are unable to redistribute themselves to any significant extent.\footnote{There are some small spectral distortions expected even in standard cosmology, such as the $y$-distortion that will be imprinted in the CMB due to scattering with free electrons after reionization, which should leave a $10^{-6}$ level distortion, well below the FIRAS uncertainty.}

\section{Green's Function Method}
\label{sec:green_func}

In this section, we give an introduction to computing the CMB spectral distortion utilizing the Green's function method, outlined in~\refcite{Chluba:2015hma}. First, we define the CMB intensity, 
\bea
\label{eq:CMB_intensity}
I_\gamma(\wcmbtoday;\Tcmbtoday) \equiv \frac{dP_\gamma}{dA \, d\Omega\, d\nucmbtoday} = \frac{\wcmbtoday^3}{2 \pi^2} f_\gamma(\wcmbtoday,\Tcmbtoday).
\eea
Here, we follow the conventions of radio astronomy: $P_\gamma$ is the power of the CMB photon received by the antenna, $\Omega$ is the solid angle along the line-of-sight, $A$ is the antenna's projected area with respect to the line-of-sight, and $\nucmbtoday = \wcmbtoday/2\pi$ is the frequency of the received CMB photons. When there is no distortion, the CMB intensity is $\bgI(\omega_0;T_0) = (\omega_0^3 / 2\pi^2) \bgf(\omega_0,T_0)$. The distortion of the phase space distribution of the CMB photon leads to the distortion of the CMB intensity away from $\bgI$. The commonly used unit of CMB intensity is $\SI{}{\Jansky\per\sr} = \SI{e-26}{\watt\per\meter\squared\per\Hz\per\sr}$, in agreement with the units expected from \Eq{CMB_intensity}.

To compute the spectral distortion due to $\gamma \to A'$, we adopt the Green's function approach described in \refcite{Chluba:2015hma}. 
If photons are injected or removed at some frequency $x'$ at some redshift $z'$, this ultimately produces a distortion in intensity with some characteristic shape $G(x; x', z')$ that we observe today at frequency $x$ (we use $^\prime$ to denote quantities related to the injection, while variables without $^\prime$ refer to quantities today).
If the number of photons that are injected or removed is small, the total distortion can be treated linearly, and is simply an integral of $G(x; x', z')$ over all values of $x'$ and $z'$, weighted by how much was injected at $x'$ and $z'$~\cite{Chluba:2015hma}. 
$G(x; x', z')$, when appropriately normalized, is precisely the Green's function mapping energy injection at $x'$ and $z'$ to a characteristic distortion. 

We now make this intuitive explanation precise. 
Let $\Source(x',z')$ be the ratio of the number density of photons injected with frequencies between $x'$ and $x' + dx'$ in the redshift interval between $z'$ and $z' + dz'$, to the number density of all photons at $z'$. 
Then the Green's function $G(x; x', z';T_0)$ is defined as 
\bea
\label{eq:Delta_I_Greens}
    \Delta I_\gamma(x; \Tcmbtoday) = \int dz' \int dx' \,\, G(x; x', z'; \Tcmbtoday) \, \Source(x', z') \,,
\eea
which has the usual structure of a solution using the Green's function method, with $\Source$ acting as a source term. 
$\Delta I_\gamma$ is the distortion to the CMB intensity, as observed today. 
With this definition, $G$ has units of intensity as well. 
$\Delta I_\gamma$ and $G$ depend on $T_0$, which we allow to float in our data analysis: see \Appx{likelihood} for more details on this.
In the context of $\gamma \to A'$ conversions, 
\bea
\label{eq:S_ATpAp}
    \Source(x', z') = -\frac{1}{\bgn} \frac{d \bgn}{dx'} P_{\gamma \to A'}(x') \delta(z' - z'_\text{res}) \,,
\eea
where $P_{\gamma \to A'}$ is defined in \Eq{trans_prob}, and $z'_\text{res}$ is the redshift at which the resonant conversion happens. 
The negative sign is consistent with the fact that photons are removed by $\gamma \to A'$ conversions. 
Note that $(d\bgn/dx')/\bgn = (1/2\zeta(3)) \, x'^2/(e^{x'}-1)$ is redshift invariant; we can therefore integrate over redshift to find
\bea
\label{eq:I_dist_AToAp_calc}
    \Delta I_{\gamma}(x;\Tcmbtoday) = -\int dx' \frac{1}{\bgn} \frac{d \bgn}{dx'} P_{\gamma \to A'}(x') G(x; x', z'_\text{res};\Tcmbtoday)  \,.
\eea

We can also show that the Green's functions obey the following normalization condition:
\bea
\label{eq:norm_condition_photon}
2 \Tcmbtoday \int dx \,\, G(x;x',z';\Tcmbtoday) = x' \alpha_\rho \bgrho(\Tcmbtoday)\,,
\eea
where $\bgrho(\Tcmbtoday)$ is the energy density of the CMB today, and $\alpha_\rho \simeq 0.3702$, defined in \Appx{const}. This is in agreement with \refcite{Chluba:2015hma}; we derive this normalization condition for completeness in \Appx{green_func_norm}, which essentially comes from requiring energy conservation during the photon injection/removal process. 

\refcite{Chluba:2015hma} provides analytic expressions for $G$ in the $\mu$ and $y$-eras. In the $\mu$-era, the Green's function $G_\mu$ for computing the $\mu$-distortion is
\bea
\label{eq:Gmu}
G_\mu(x;x',z';\Tcmbtoday) = \alpha_\rho x' \cdot \frac{3}{\kappa_c} \, \vis(z')  \left[ 1 - \Ps (x',z') \frac{x_0}{x'} \right] \, \MMM(x;\Tcmbtoday) + \frac{\lambda(x',z')}{4} \, \TTT(x;\Tcmbtoday),
\eea
where $\MMM = \bgI \cdot \mmm$, $\TTT = \bgI \cdot \ttt$, and $x_0 \simeq 3.6$. We provide the derivation of \Eq{Gmu} in \Appx{mono_mu_era}. $\Ps (x', z')$ is the probability that a photon injected with frequency $x'$ at redshift $z'$ survives before the distribution relaxes into a quasi-stationary phase---injected photons can be absorbed by {\dcs} or {\br} before {\cs} redistributes them, contributing instead to the heating of baryons~\cite{Chluba:2015hma}.
In the limit of $\Ps \to 0$, this reduces to a pure energy injection/removal process; $\Ps > 0$ includes the possibility of increasing or decreasing the comoving number density of photons. 
$\vis$ is the visibility function accounting for the washout effect of {\dcs} and {\br} on the $\mu$-type distortion.
$\lambda$ is a coefficient in front $\TTT$, which is set by the normalization condition for the Green's function (\Eq{norm_condition_photon}). 
The  formula for $\lambda$ can be found in \Eq{lambda_func_Appx}. 
The definitions and the numerical values of the other constants, such as $\alpha_\rho$ and $\kappa_c$, are worked out in \Appx{const}\@. In the weak-coupling region where $\epsilon \lesssim 10^{-6}$, as we argued in Sec.~\ref{sec:spec_dist}, $\TTT$ is fundamentally unobservable because we treat $T_0$ as a nuisance parameter when analyzing the data of CMB intensity. 
Therefore, we simply drop the second term containing $\TTT$ in \Eq{Gmu} from the Green's function, with the only observable part of the distortion coming from the first term containing $\MMM$.\footnote{We could have subtracted any multiple of $\TTT$ from the expression by the same argument; this is merely a way to simplify the expression for $G_\mu$.}
When $T \gg \SI{0.5}{\kilo\eV}$, $\vis \simeq 0$ because {\dcs} and {\br} are efficient enough to set the chemical potential of the CMB photons to zero. 
In this situation, there is no observable CMB spectral distortion. 

An analytic expression for the Green's function during the $y$-era, $G_y$, is also derived in \refcite{Chluba:2015hma}; the full expression for $G_y$, and its other important aspects, are discussed in \Appx{green_func_y}. 
Here, to give some physical intuition, we only show the approximate form of $G_y$ in the limit where {\cs} is very inefficient. 
The following dimensionless quantity describes the efficiency of {\cs}:
\bea
\label{eq:y_gamma}
y_\gamma(z) = \int_0^z dz' \, \frac{T(z')}{m_e} \frac{\sigma_T n_e(z')}{H(z')(1+z')} \,,
\eea
where $\sigma_T$ is the Thomson scattering cross-section.
$y_\gamma$ is a measure of the efficiency of energy transfer to photons due to Compton scattering.\footnote{Note that since we are in the regime where $T \ll m_e$, Compton scattering, $e^- + \gamma \to e^- + \gamma$, simply reduces to Thomson scattering. The fraction of energy transferred to photons during each Thomson scattering event is $\sim T / m_e$, which appears as the first term in the integral.} 
In the late universe, especially during the $y$and free-streaming eras, $y_\gamma \ll 1$, \textit{i.e.}\ {\cs} is inefficient at significantly changing the energy of a photon. 
In this situation, $G_y$ can be approximated as
\bea
\label{eq:Gy_small_y}
G_y(x;x',z';\Tcmbtoday) \simeq \alpha_\rho x' \cdot \left( 1 - \Ps(x',z') \right) \, \frac{\YYY(x;\Tcmbtoday)}{4} + \alpha_\rho x' \cdot \frac{\bgrho(\Tcmbtoday)}{2 \Tcmbtoday} \, \Ps(x',z') \, \delta(x-x'),
\eea
where $\YYY = \bgI \cdot \yyy$. In \Eq{Gy_small_y}, the first term leads to a $y$-distortion due to heating from the absorption of photons, while the second term leads to a free-streaming distortion, where injected photons simply redshift. 
In the full expression for $G_y$, shown in \Eq{Gy_Appx}, the $\delta(x-x')$ in the second term of \Eq{Gy_small_y} is not an exact delta function but is broadened, by {\cs}, into a Gaussian with respect to $\log x$ with width $\sim \sqrt{y_\gamma} x'$.
This effect is more pronounced for large $x'$, or at earlier epochs where the {\cs} is more efficient. 
The reader should refer to \Appx{green_func_y} for more detailed discussions.

\begin{figure}[t]
\centering
\begin{tikzpicture}
\node at (-5, -0.02){\includegraphics[width=0.495\columnwidth]{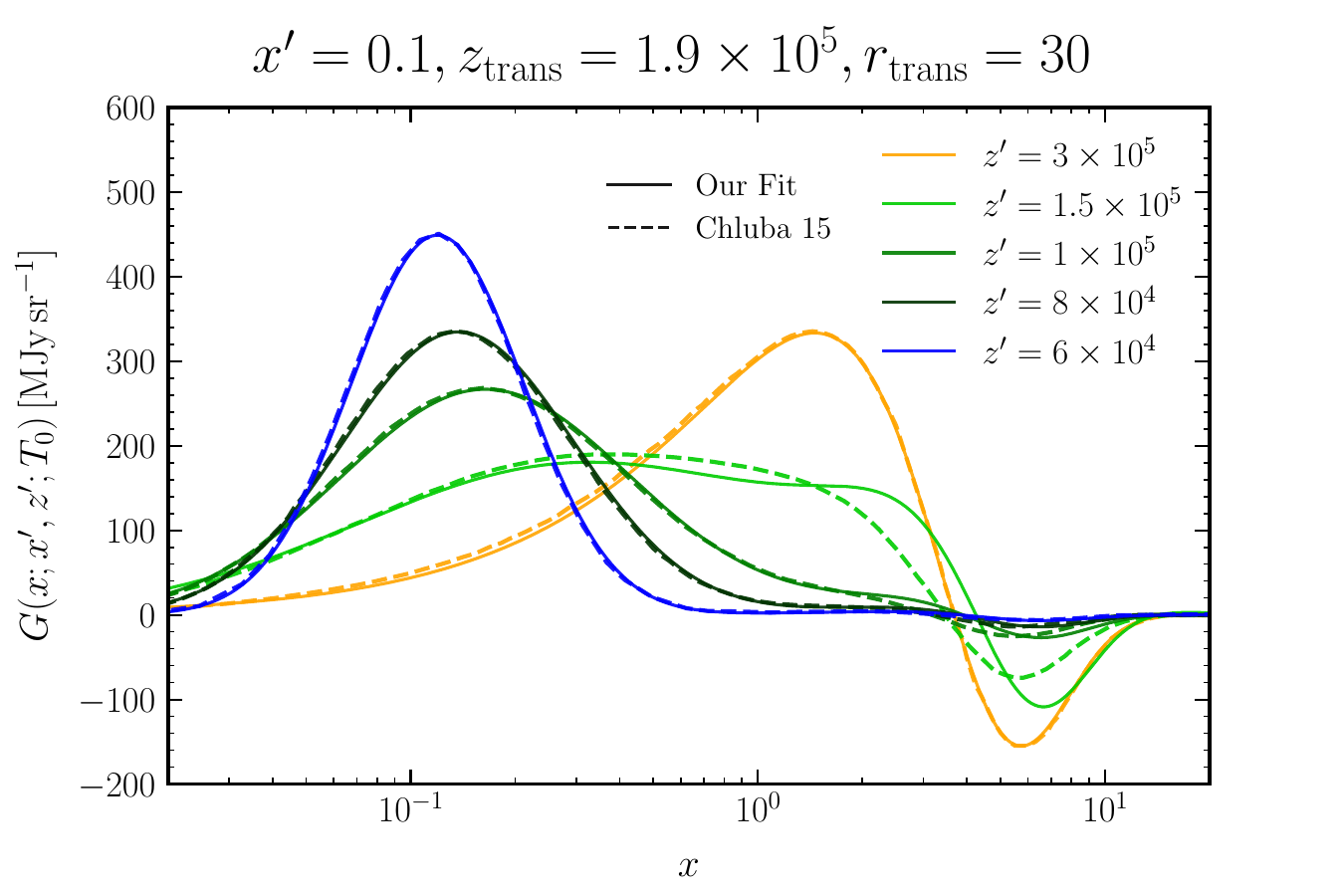}};
\node at (-9.07,-6){\includegraphics[width=0.495\columnwidth]{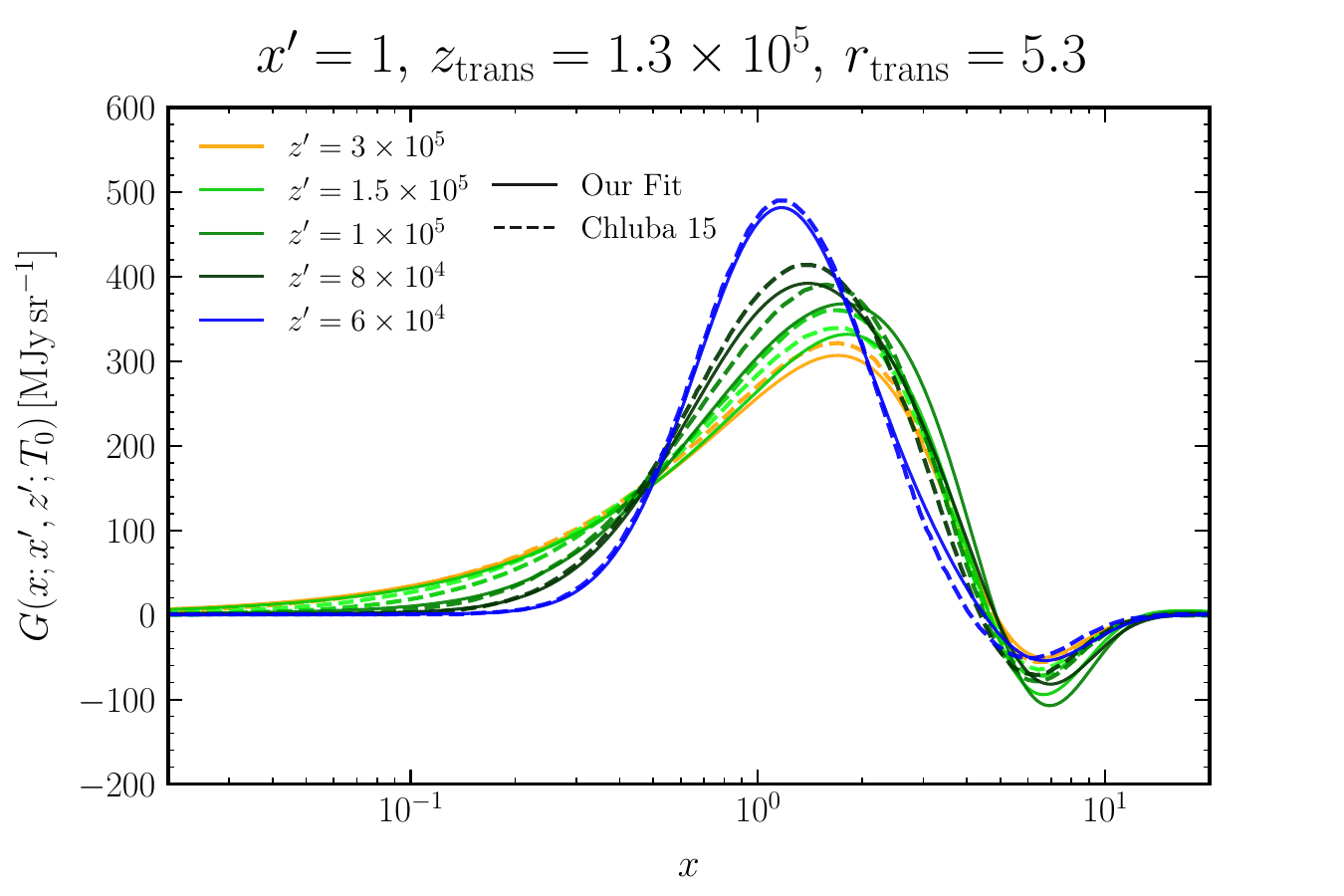}};
\node at (-0.1,-6){\includegraphics[width=0.495\columnwidth]{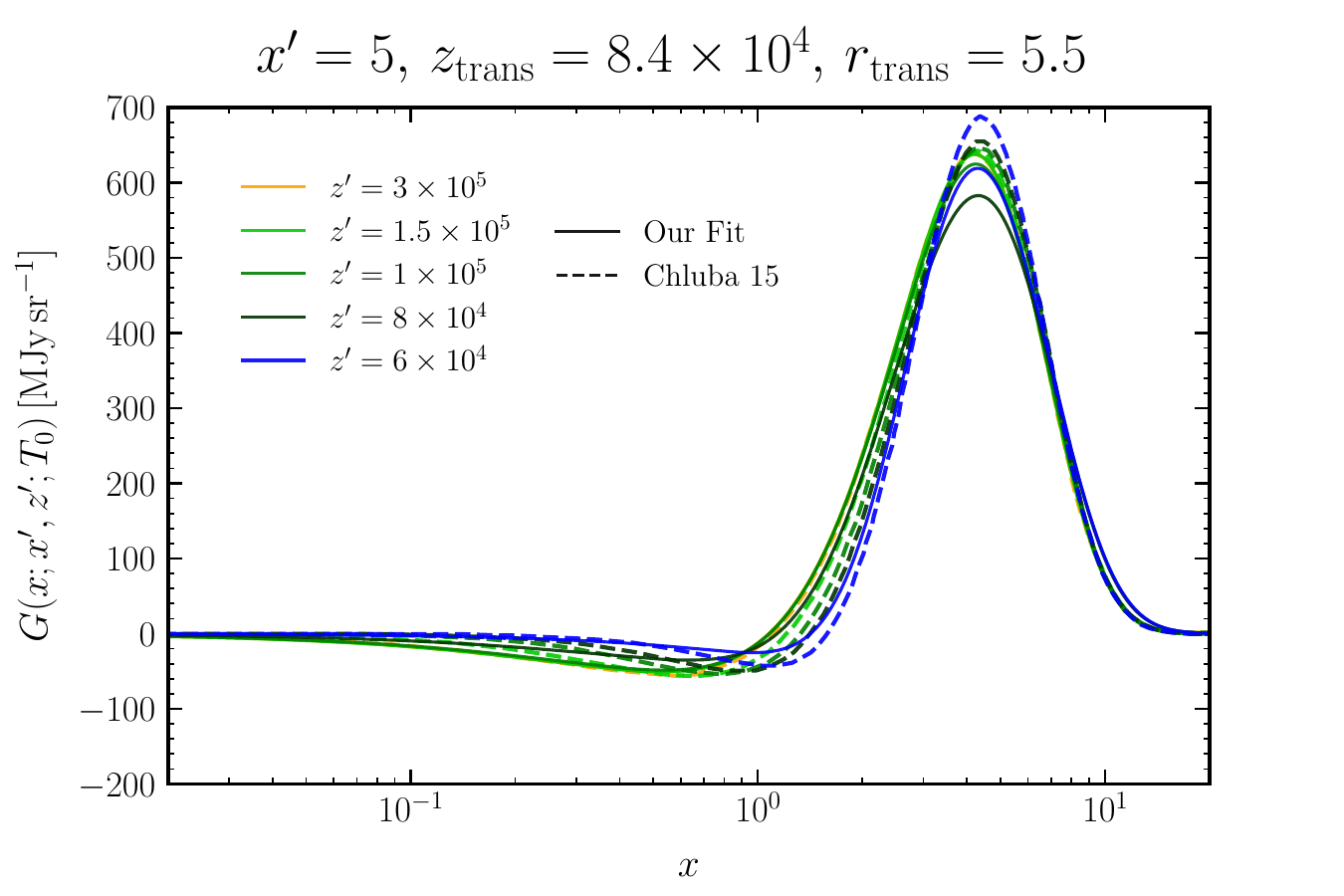}};
\end{tikzpicture}
\caption{The fitted Green's functions (solid lines) for the $\mu$-$y$ transition era listed in \Eq{Gtrans} with different frequencies $x'$ and different redshifts $z'$, compared with the numerically computed Green's functions obtained in Ref.~\cite{Chluba:2015hma} (dashed lines).
Here, for each $x'$, we find the best fit values of $z_\trans$ and $r_\trans$ compared with Ref.~\cite{Chluba:2015hma}. 
When $z' \leq 6 \times 10^4$, we have $G \simeq G_y$, which is labeled by solid blue lines. 
When $z' \geq 3 \times 10^5$, we have $G \simeq G_\mu$, which is labeled by  solid orange lines. 
The shapes of the Green's functions in the intermediate epoch, where $6 \times 10^4 < z < 3 \times 10^5 $, are represented by the green lines with different degrees of darkness.}
\label{fig:greensfunc_plot}
\end{figure}

In the $\mu$-$y$ transition era, there is no analytic form for the Green's function available. 
In order to perform a fully accurate determination of the spectral distortion, a full numerical treatment using a code package like CosmoTherm~\cite{Chluba:2011hw} is needed to track how the CMB spectrum evolves during this era. 
Since there is no publicly available package for computing spectral distortions, we have to instead rely on approximations. 
We know that the Green's function should tend toward the Green's functions of the $\mu$- or $y$-eras at higher or lower $z'$, and therefore we parametrize
\bea
\label{eq:Gtrans}
G_\trans(x;x',z';\Tcmbtoday) = \transfunc_\mu(x', z') \cdot G_\mu(x;x',z';\Tcmbtoday) + \transfunc_y(x',z') \cdot G_y(x;x',z';\Tcmbtoday) \,,
\eea
 which smoothly connects Green's functions in the $\mu$ and $y$-eras. As we approach the $\mu$-era, $\transfunc_\mu \simeq 1$ and $\transfunc_y \simeq 0$, while $\transfunc_\mu \simeq 0$ and $\transfunc_y \simeq 1$, as we approach the $y$-era.

For the fiducial treatment that we adopt in this paper, the form of $\transfunc_\mu(x', z')$ that we adopt is the same as that of a similar function used to determine the Green's functions for spectral distortions from pure energy injection in the same era~\cite{Chluba:2013vsa}:
\bea
\label{eq:Tmu}
\transfunc_\mu(x',z') = 1 - \exp\left[ - \left(\frac{1+z'}{1+z_\trans(x')}\right)^{r_\trans(x')} \right]\,,
\eea
where $z_\trans$ represents the redshift at which the Green's function transits from $G_\mu$ to $G_y$, and $r_\trans$ represents how rapidly such a transition happens. 
To maintain the proper normalization of the Green's function, as required by \Eq{norm_condition_photon}, we must always have
\bea
\transfunc_y(x',z') = 1 - \transfunc_\mu(x',z')\,.
\eea

In \Eq{Tmu}, we take $z_\trans$ and $r_\trans$ to be $x'$-dependent parameters, fitting for them at six discrete values: $x' = 10^{-3}, 10^{-2}, 10^{-1}, 1, 5$, and 15. For each value of $x'$, we find the values of $z_\trans$ and $r_\trans$ that minimize the function
\bea
\label{eq:Deviation_Appx}
\mathcal{D}(z_\trans,r_\trans;x') = \int_{-\infty}^{+\infty} d \log x \, \sum_{i}\abs{G(x;x',z'_i;z_\trans,r_\trans) - \widetilde{G}(x;x',z'_i)}^2\,,
\eea
where $G$ is as defined in \Eq{Gtrans} with $\transfunc_\mu$ given in \Eq{Tmu}, and $\widetilde{G}$ are the numerically computed Green's functions over all $z_i'$, reported in Ref.~\cite{Chluba:2015hma}. 
The sum corresponds to adding up the contributions to $\mathcal{D}$ for each reported Green's function at the redshifts $z'_i$. The best fit values of $(z_\trans,r_\trans)$ that we obtain for each $x'$ are
\bea
\label{eq:Gshape_ztrans_rtrans_Appx}
&
\left\{ 
\begin{aligned}
& x' = 10^{-3} &:\quad& z_\trans = 3.1 \times 10^5 &,\,\, r_\trans = 3.3\\
& x' = 10^{-2} &:\quad& z_\trans = 2.3 \times 10^5 &,\,\, r_\trans = 7.1  \\
& x' = 10^{-1} &:\quad& z_\trans = 1.9 \times 10^5 &, \,\,r_\trans = 30 \\
& x' = 1 &:\quad& z_\trans = 1.3 \times 10^5 &,\,\, r_\trans = 5.3 \\
& x' = 5 &:\quad& z_\trans = 8.4 \times 10^4 &,\,\, r_\trans = 5.5 \\
& x' = 15 &:\quad& z_\trans = 10^5 &,\,\, r_\trans = 2.5
\end{aligned}
\right. \, .
\eea
For $x'=0.1$, we find that the fit prefers a sudden transition between $\mu$-type and $y$-type Green's functions, which occurs at large $r$. 
Finding no significant change to the fit once $r_\trans \geq 30$, we simply choose $r_\trans = 30$. 
To get $z_\trans$ and $r_\trans$ for intermediate values of $x'$, we linearly interpolate over $\log z_\trans$ and $\log r_\trans$ as a function of $\log x'$.

The resulting fits for $G_\text{trans}(x; x', z'; T_0)$ for $x' = 0.1, 1, $ and 5, and the comparison with Ref.~\cite{Chluba:2015hma}, are shown in \Fig{greensfunc_plot}. Choices of $z'$ that are closer to the $y$-era are shown in blue, while those that are closer to the $\mu$-era are shown in orange.
Comparing these results with the numerically computed Green's functions in the $\mu$-$y$ transition shown in \refcite{Chluba:2015hma}, we find good quantitative agreement for these particular choices of $z_\trans$ and $r_\trans$, with the relative difference between our approximate Green's functions and those shown in \refcite{Chluba:2015hma} being less than $30\%$ in the region where $0.1 \leq x' \leq 5$, across all values of $z'$ and $x$ considered. For $x'< 0.1$ or $x' > 5$, energy injection, from $\gamma \rightarrow A'$ resonant conversion of photons in these ranges, does not contribute significantly to $\Delta I_\gamma$.
This is because at low $x'$, the fraction of energy removed from the CMB by $\gamma \to A'$ per $\log x'$ interval scales as $x'^2$, while at large $x'$, the CMB becomes exponentially suppressed, and yet again only a small fraction of energy is removed. 
Given that the error of the fitting in these marginal regions is less than $50\%$, the full numerical error from integrating over these regions is at most at the $5\%$ level.

So far, we have discussed the procedure we use to set our fiducial limits in the $\mu$-$y$ transition era. 
To further quantify the uncertainty associated with our approximate treatment of the transition era, we also adopt Green's functions in the $\mu$-$y$ transition era with different values of constant $z_\trans$ and $r_\trans$, \textit{i.e.}\ with no $x'$ dependence. To be more specific, we calculate the CMB spectral distortion over the following constant values of $z_\trans$ and $r_\trans$: $z_\trans = \left\{ 5.8 \times 10^4, 10^5, 1.4 \times 10^5 \right\}$ and $r_\trans = \left\{ 1.88, 3, 5\right\}$. 
These values were chosen to satisfy the following criteria: \textit{1)} $\Delta I_\gamma^\text{trans} \simeq \Delta I_\gamma^\mu$ when $z_\text{res} = 3 \times 10^5$, so that the Green's function transitions smoothly to the $\mu$-era Green's function as $z$ approaches the $\mu$-era, and \textit{2)} $y_\gamma \lesssim 1$ at $z = z_\text{trans}$, since $y_\gamma \gg 1$ characterizes the $\mu$-era when {\cs} is highly efficient.  
These other versions of the Green's functions will be used to assess how dependent the limits are on the approximation made in the $\mu$-$y$ transition range of $m_{A'}$ masses.

\section{Spectral Distortion from $\gamma \rightarrow A'$ and COBE-FIRAS Constraints}
\label{sec:spec_dist_AToAp}

The Green's function method described in the previous section allows for the computation of spectral distortions associated with any photon injection/removal process. 
We can now compute the $\Delta I_\gamma(x;\Tcmbtoday)$ associated with photon-dark photon oscillations using \Eq{I_dist_AToAp_calc}, given $\epsilon$ and $\map$. 
We then compare the predicted CMB spectrum with the spectral distortion from $\gamma \to A'$, with the intensity of the CMB measured by the FIRAS instrument aboard COBE~\cite{Fixsen:1996nj}. 
To set a limit on $\epsilon$ for each value of $\map$, we perform a profile likelihood test by constructing a test statistic from the profile likelihood ratio with model parameter $\epsilon$ and nuisance parameter $\Tcmbtoday$.  
The Gaussian likelihood that we use is constructed from the full covariance matrix provided by the COBE-FIRAS experiment~\cite{Fixsen:1996nj}. 
Further details of the statistical analysis are given in \Appx{likelihood}.

\begin{figure}[t]
\centering
\includegraphics[width=0.8\columnwidth]{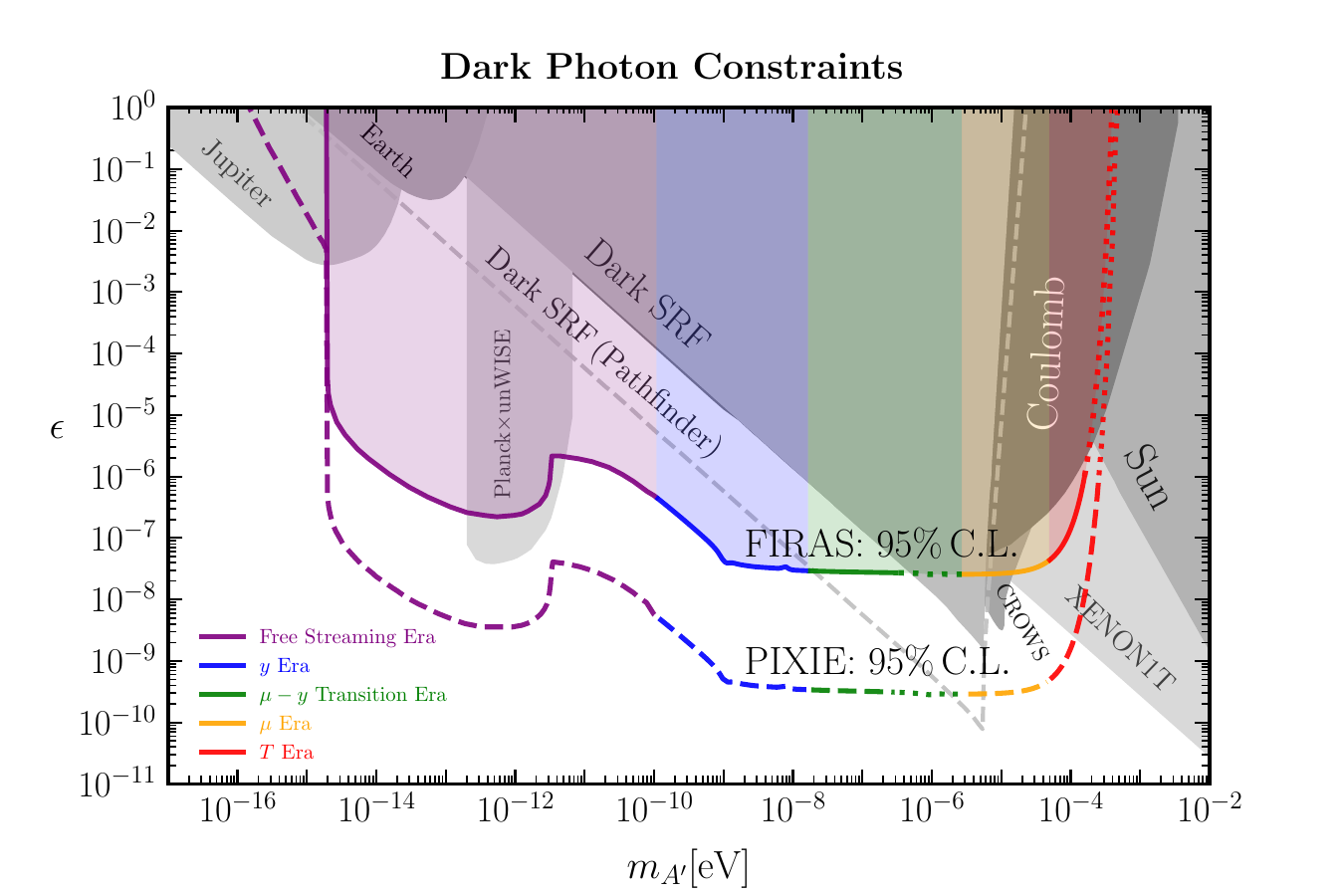}
\vspace{0.3cm}
\caption{The dark photon constraints and projections from CMB spectral distortions. The solid colored line is the CMB constraint from the COBE-FIRAS dataset at the $95\%$ confidence level. The dashed colored line is the future projection from the proposed PIXIE satellite, with a sensitivity $10^4$ better than FIRAS~\cite{Kogut:2011xw, Kogut:2024vbi}, and assuming perfect foreground removal. Different colors represent different stages of the CMB spectral distortion. The free-streaming era, $y$-era, $\mu$-$y$ transition era, $\mu$-era, and $y$-era are denoted by purple, blue, green, orange, and red, respectively. Results from the free-streaming era (purple) are identical to \refcite{Caputo:2020bdy}. The green dotted lines show where the uncertainties associated with the Green's function approximation in the $\mu$-$y$ transition era are important. We also show with a red dotted line the region of parameter space where large distortions are expected due to a high probability of conversion, where our method does not apply. Although such distortions occur deep in the $T$-era, the large $P_{\gamma \rightarrow A'}$ results in visible distortions that cannot be removed through the variation of $T_0$~\cite{Acharya:2021zhq}. Other constraints from DarkSRF~\cite{Romanenko:2023irv}, CROWS~\cite{Betz:2013dza}, Coulomb~\cite{Bartlett:1970js, Williams:1971ms, Bartlett:1988yy, Kroff:2020zhp}, XENON1T~\cite{An:2020bxd,XENON:2021qze}, Sun~\cite{An:2013yfc,Redondo:2013lna,Vinyoles:2015mhi,Redondo:2015iea,Li:2023vpv}, Jupiter~\cite{Davis:1975mn,Marocco:2021dku,Yan:2023kdg}, Earth~\cite{Goldhaber:1971mr,Bartlett:1988yy,Fischbach:1994ir,Kloor:1994xm,Marocco:2021dku} are labeled and shown in gray. We show the future projection of DarkSRF~(Pathfinder) with the gray dashed line~\cite{Berlin:2022hfx}. Conservative limits from the Planck CMB power spectrum data derived in \refcite{Aramburo-Garcia:2024cbz} found limits that are slightly weaker than those derived from COBE-FIRAS~\cite{Caputo:2020bdy,Bondarenko:2020moh}, and are therefore not shown, while a separate study found strong limits at $m_{A'} \sim \SI{e-12}{\eV}$ using a cross correlation of Planck and unWISE data~\cite{McCarthy:2024ozh} (shown in gray, labeled Planck$\times$unWISE).}
\label{fig:fullplot}
\end{figure}

We begin by showing the main result of our analysis in \Fig{fullplot}: the 95\% confidence limit on $\epsilon$ as a function of $\map$.
The solid colored line is our fiducial constraint, at the 95 $\%$ confidence level, on the dark photon mixing parameter $\epsilon$ for a broad range of dark photon masses in the ultralight region. 
Different colors represent the $\gamma \rightarrow A'$-induced spectral distortions happening in different epochs: the free-streaming era, $y$-era, $\mu$-$y$ transition era, $\mu$-era, and $T$-era are denoted by purple, blue, green, orange, and red, respectively. 
The results for the free-streaming era, shown in purple, are identical to those found in~\refcite{Caputo:2020bdy}; the limits for the other regimes constitute new limits worked out in this paper. 

The dashed colored line is the projected reach for the next-generation PIXIE satellite~\cite{Kogut:2011xw, Kogut:2024vbi}. To obtain these projected limits, we recompute the reach for the high mass region where, $m_{A'} \gtrsim 10^{-10}\,\eV$, utilizing the same statistical analysis used for our main results, and the PIXIE sensitivities provided in~\refcite{Kogut:2024vbi}. 
For the low mass region, where $m_{A'}\lesssim 10^{-10}\eV$, we show the estimated PIXIE projection from Refs.~\cite{Caputo:2020bdy,Caputo:2020rnx}. Again, we assume that we are able to accurately remove foreground contributions to the CMB spectral distortion, such as from reionization, leaving a detailed analysis of foreground removal for future work.

The green line denotes the $\mu$-$y$ transition region, in which we use the approximation scheme described in \Eq{Gtrans}. When the line is solid, we expect our approximation scheme to be highly robust, while the dotted line indicates the result obtained by our fiducial treatment. Further discussion on this region appears at the end of this section. The red line denotes the $T$-era, where {\dcs} and {\br} are efficient. Therefore, the DP-induced distortions are partially washed out. The solid line denotes the region where the DP-induced CMB distortion is perturbative. In this region, the Green's function method is valid. The dotted line denotes the region where $\epsilon \gtrsim 10^{-6}$. In this region, the photon-to-dark-photon transition probability becomes $P_{\gamma \rightarrow A'} \gtrsim \mathcal{O}(0.1)$, where the DP-induced distortion becomes nonperturbative. Furthermore, the large distortion induced by $\gamma \rightarrow A'$ in this regime takes a significant period of time to thermalize, leading to visible, present-day distortions even though the conversion is happening deep in the $T$-era~\cite{Acharya:2021zhq}. However, existing laboratory constraints exclude this regime, so we forego providing a precise treatment of the spectral distortion in this region of parameter space.

In gray, we show constraints from other probes, such as the DarkSRF~\cite{Romanenko:2023irv}, CROWS~\cite{Betz:2013dza}, modifications to Coulomb's law~\cite{Bartlett:1970js, Williams:1971ms, Bartlett:1988yy, Kroff:2020zhp}, XENON1T~\cite{An:2020bxd,XENON:2021qze}, the Sun~\cite{An:2013yfc,Redondo:2013lna,Vinyoles:2015mhi,Redondo:2015iea,Li:2023vpv}, Jupiter's magnetic field~\cite{Davis:1975mn,Marocco:2021dku,Yan:2023kdg}, and the Earth's magnetic field~\cite{Goldhaber:1971mr,Bartlett:1988yy,Fischbach:1994ir,Kloor:1994xm,Marocco:2021dku}. 
In addition, the CMB power spectrum has also recently been shown to provide strong limits on the dark photon. \refcite{Aramburo-Garcia:2024cbz} found conservative limits based on Planck data that are slightly weaker than limits from FIRAS data~\cite{Caputo:2020bdy,Bondarenko:2020moh}, while the cross-correlation between Planck and unWISE~\cite{McCarthy:2024ozh} gives the leading limits for $m_{A'} \sim \SI{e-12}{\eV}$. We also show the future pathfinder projection of the Dark SRF with the gray dashed line~\cite{Berlin:2022hfx}.
Among all the current experimental methods, the CMB spectral distortion stands out as one of the most sensitive methods in the $\SI{e-15}{\eV} - \SI{e-6}{\eV}$ mass range, even with the COBE-FIRAS dataset acquired three decades ago. 
Future projections of DarkSRF show strong sensitivity to dark photons with a mass in the range of $\SIrange{e-8}{e-5}{\eV}$. 
Future light-shining-through-wall experiments~\cite{Ortiz:2020tgs,Antel:2023hkf, Miyazaki:2022kxl, Berlin:2023mti} and a next-generation helioscope~\cite{OShea:2023gqn} are also projected to probe new parameter space for $m_{A'} \gtrsim 10^{-5}\eV$. 
That being said, a next-generation CMB spectrum experiment will be one of the most effective methods for testing this region of parameter space.

As we enter the $T$-era, in the red region of the plot, the distortions due to $\gamma \to A'$ become primarily an unobservable shift in the temperature, leaving only a small detectable $\mu$-type distortion; this explains the exponential loss of sensitivity in this region. Note that based on our estimate in Eq.~\eqref{eq:eps_est}, once $\epsilon \gtrsim 3 \times 10^{-7}$, the probability of conversion $P_{\gamma \to A'}$ for $x \sim 1$ starts to exceed 1\%, and the Green's function assumption of small distortions breaks down. Although a more detailed treatment of large distortions, as described in Ref.~\cite{Chluba:2020oip}, can be used to get accurate limits in the $T$-era with $\epsilon \gtrsim 3 \times 10^{-7}$, existing constraints from \textit{e.g.}\ XENON1T~\cite{An:2020bxd,XENON:2021qze} ($\epsilon \sim 2 \times 10^{-9}$ for $m_{A'} \sim \SI{e-4}{\eV}$) far supersede any potential limits from FIRAS or a next-generation CMB spectrum measurement, rendering a careful treatment unnecessary.

\begin{figure}[t]
\centering
\includegraphics[width=0.75\columnwidth]{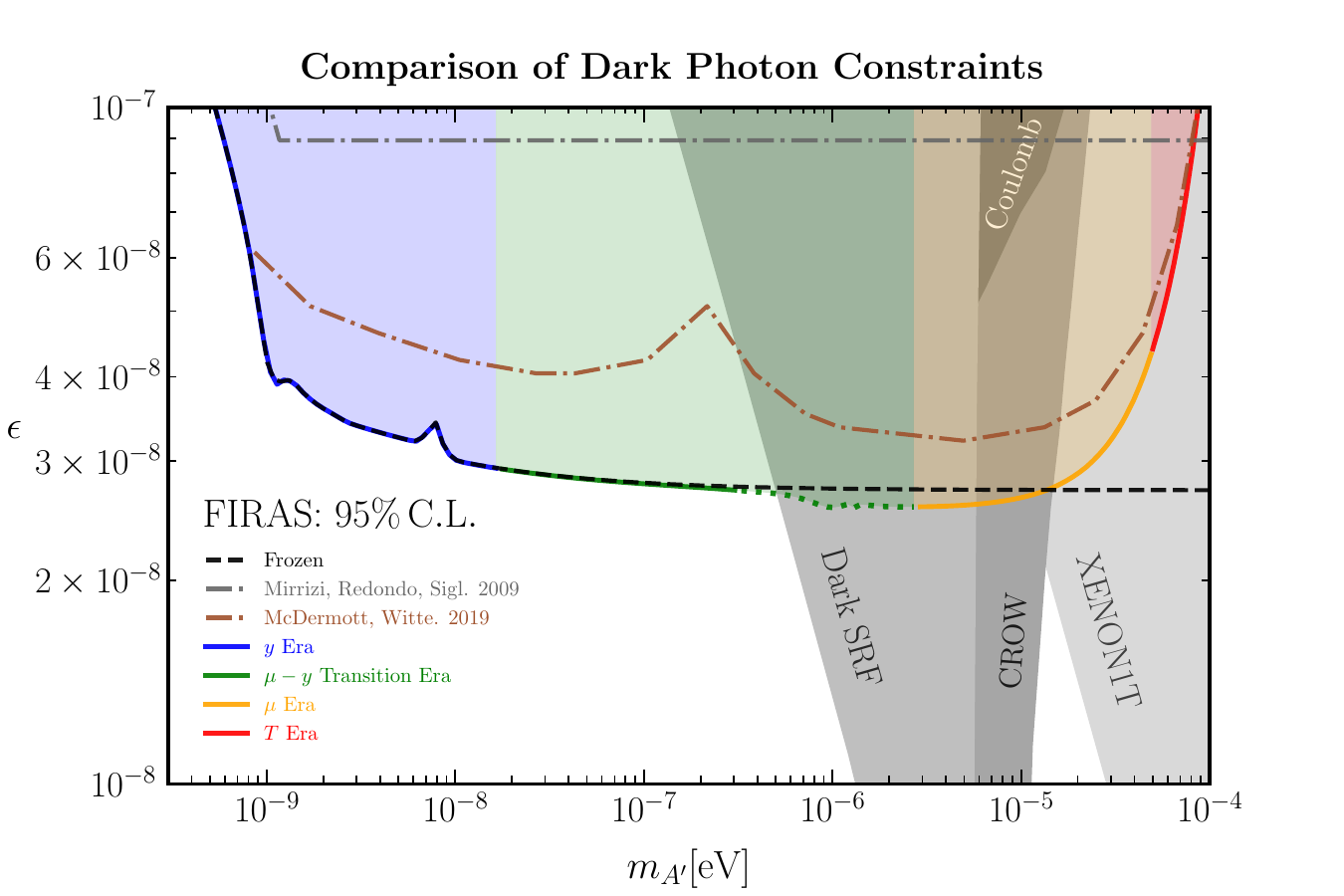}
\vspace{0.3cm}
\caption{Comparison of our fiducial constraint (rainbow colored line) with previous results. The dash-dotted gray line and the dash-dotted brown line are the constraints obtained using the free streaming calculation of \refcite{Mirizzi:2009iz}, and assuming that the effect of $\gamma \to A'$ is a pure energy removal process~\cite{McDermott:2019lch}, respectively. 
We repeated the free streaming calculation with the free electron fraction taken from CLASS~\cite{Lesgourgues:2011re,Blas:2011rf} based on Planck 2018 data, and show the result as the black dashed line. 
}
\label{fig:zoominplot}
\end{figure}

\Fig{zoominplot} shows a zoomed-in version of \Fig{fullplot}, as well as a comparison of our new COBE-FIRAS limits with previous results. 
The notch seen in our constraints at $m_{A'} \sim 10^{-8}\eV$ is due to HeIII to HeII recombination happening at $z \sim 6000$. At this point, $\abs{d m_{\gamma}^2/dz}$ is larger than those at nearby redshifts. Therefore, $P_{\gamma \rightarrow A'}$ is suppressed according to \Eq{trans_prob}, which slightly alleviates the COBE-FIRAS constraint at this value of $m_{A'}$.
 The dot-dashed gray line shows the result from \refcite{Mirizzi:2009iz}. 
These results assumed that photons were free-streaming even before recombination.
They also did not take the evolution of the free electron fraction $x_e$ into account accurately, resulting in an $m_{A'}$-independent limit, due to the approximate result for $P_{\gamma \to A'}$ that we derived in \Eq{trans_prob_RAD}. 
Updating their result with $x_e$ calculated from CLASS~\cite{Lesgourgues:2011re,Blas:2011rf} using Planck 2018 cosmology~\cite{Planck:2018vyg}, but still assuming free-streaming, leads to the black dashed curve. 
We see that free-streaming is an excellent approximation for $\map \leq \SI{e-7}{\eV}$, but starts to deviate from the actual results above this mass.
In particular, we need the full machinery of the Green's function method to calculate how the limit relaxes at high masses. 

The dash-dotted brown line shows the result from \refcite{McDermott:2019lch}, which assumes that the effect of $\gamma \to A'$ is equivalent to a pure energy removal process, producing $\mu$- and $y$-type distortions in their respective eras. 
They then directly recast the COBE-FIRAS constraints on $\abs{\mu}$ and $\abs{y}$ parameters to the dark photon parameter space.
This assumption is also inaccurate: the actual distortion differs significantly from a pure $\mu$- or $y$-type distortion, as we discuss in detail below. 
Furthermore, \refcite{McDermott:2019lch} did not attempt to treat the $\mu$-$y$ transition period consistently, leading to an artificial notch at around $\map \sim \SI{2e-7}{\eV}$. 

\begin{figure}[t]
\centering
\begin{tikzpicture}
\node at (-9,0){\includegraphics[width=0.495\columnwidth]{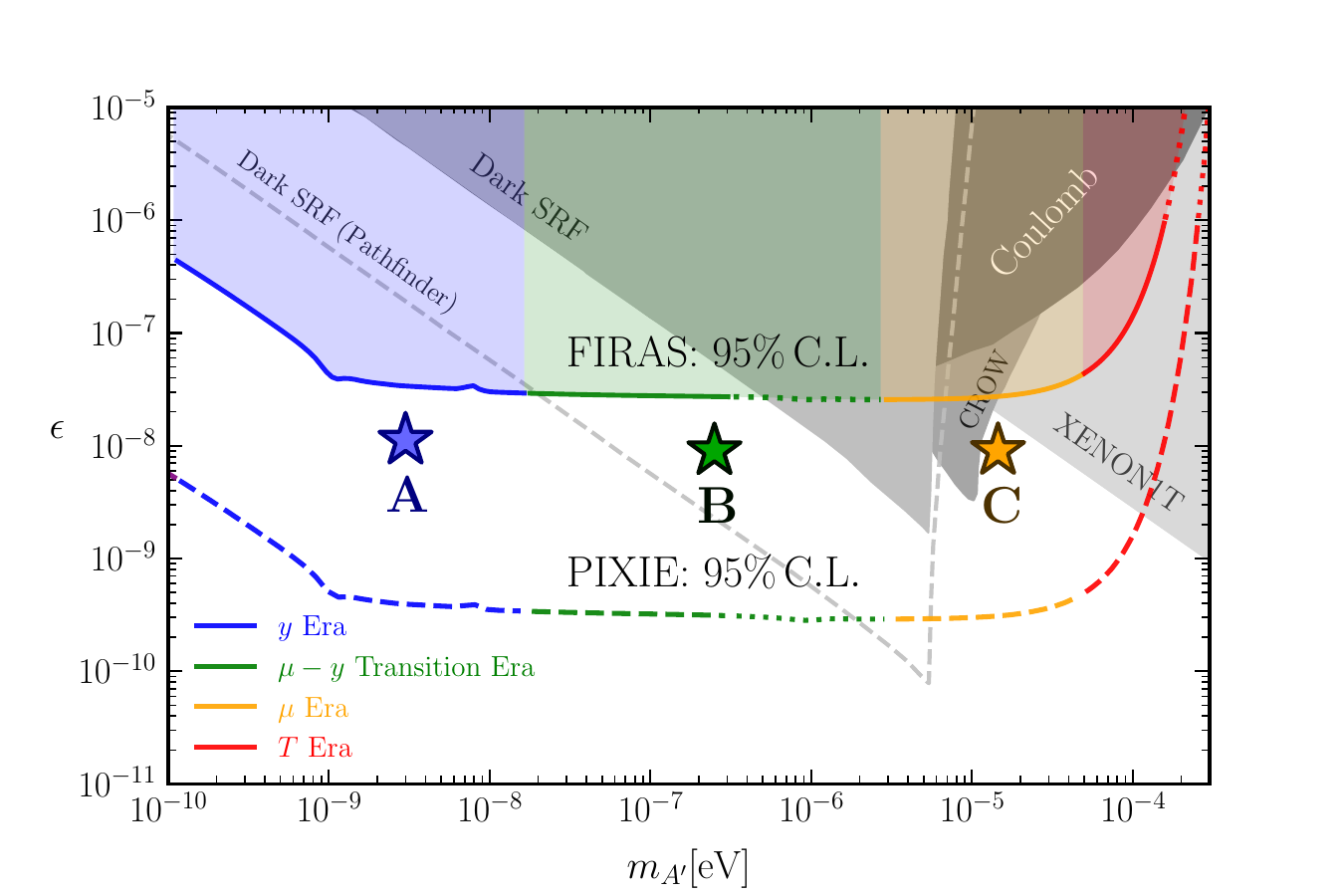}};
\node at (-0.1, -0.02){\includegraphics[width=0.495\columnwidth]{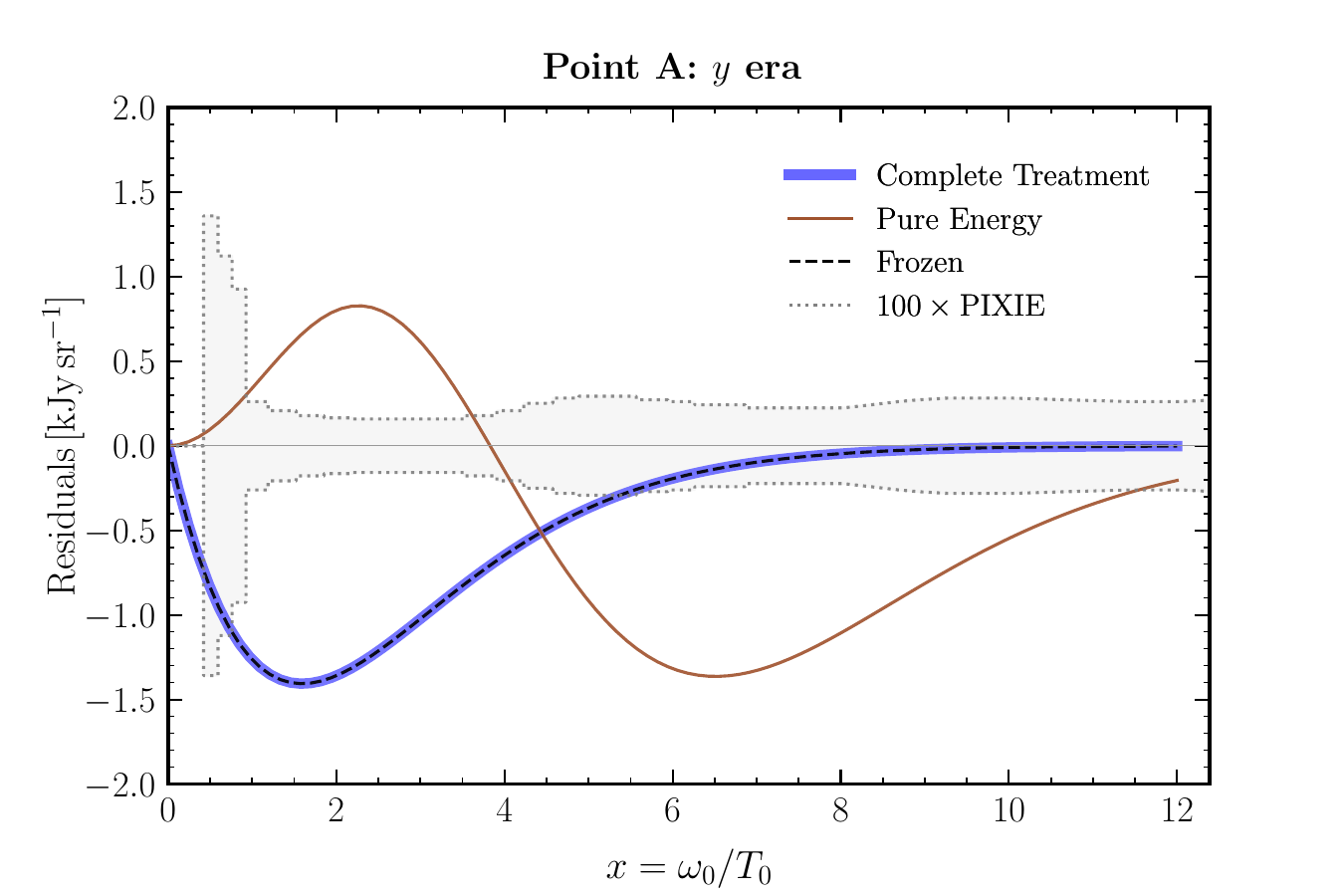}};
\node at (-9.07,-6.2){\includegraphics[width=0.495\columnwidth]{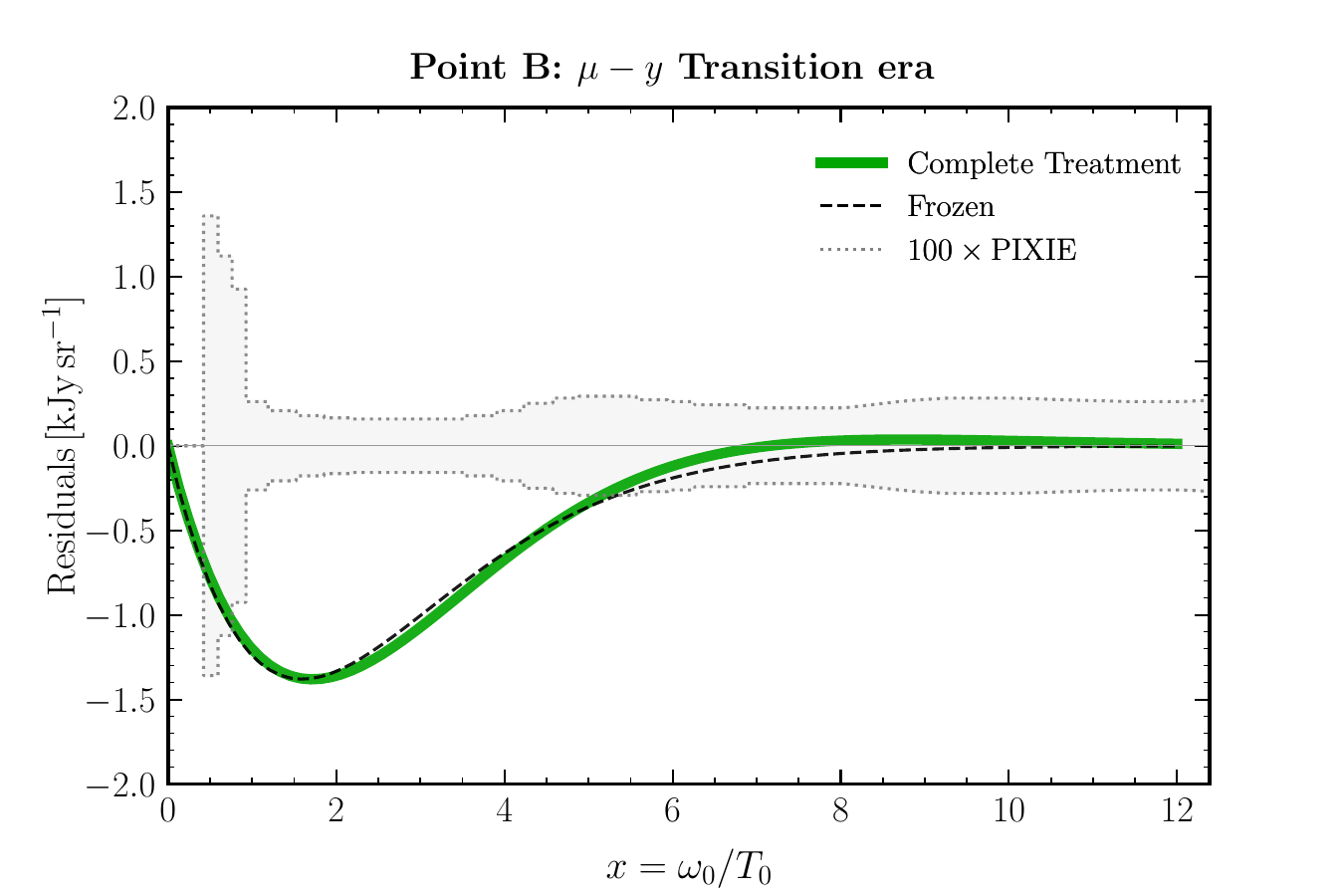}};
\node at (-0.1,-6.2){\includegraphics[width=0.495\columnwidth]{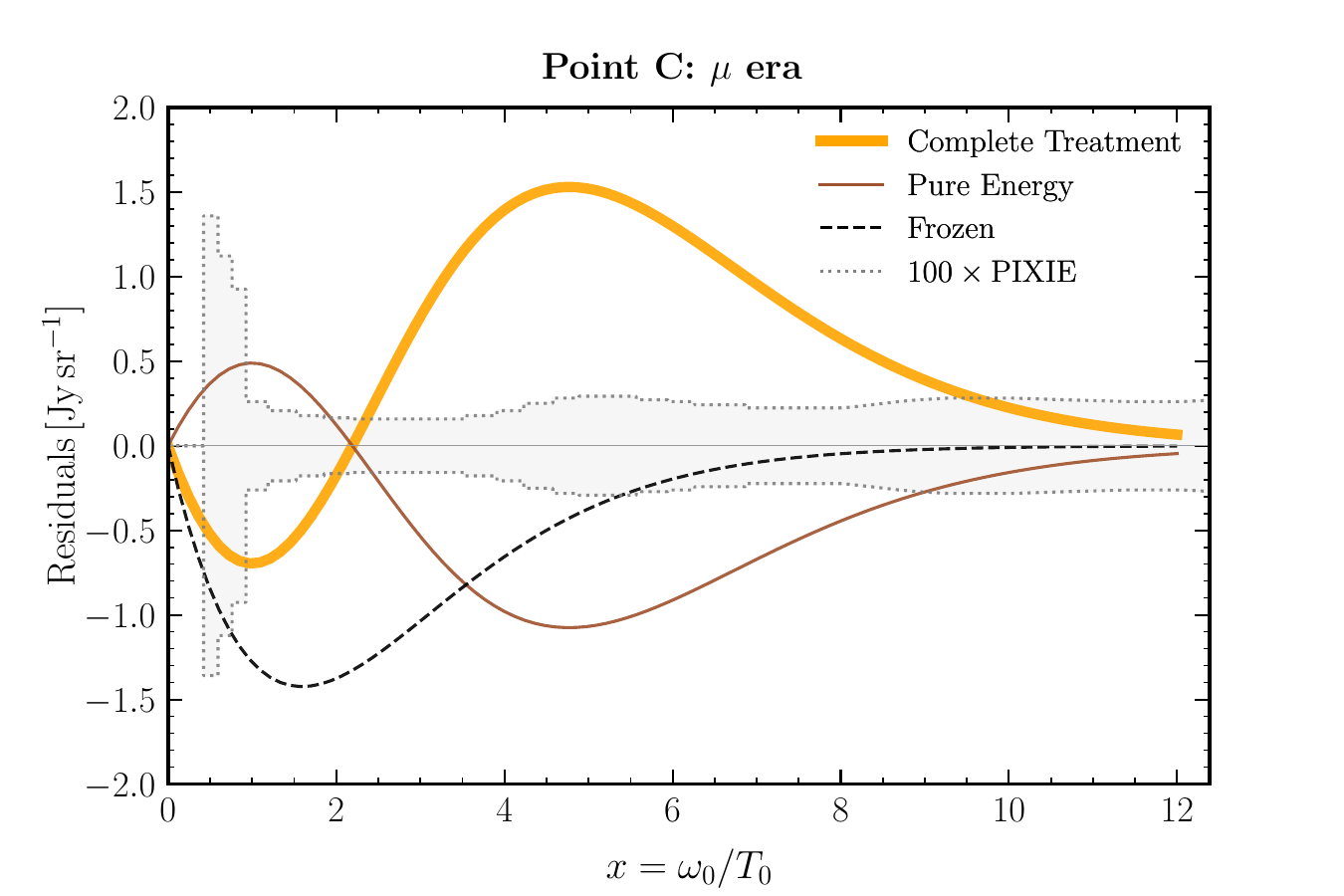}};
\end{tikzpicture}
\caption{(\textit{Top left}) Parameter points $A$, $B$ and $C$ as a function of $\map$ and $\epsilon$, and the CMB spectral distortion caused by $\gamma \rightarrow A'$ happening in points $A$ (\textit{top right}, $y$-era), $B$(\textit{bottom left}, $\mu$-$y$ transition era), and $C$ (\textit{bottom right}, $\mu$-era). We plot the predicted CMB spectral distortion using our method~(thick lines with the same colors as corresponding eras) labeled by the ``completed treatment'', the method assuming just free-streaming, labeled ``frozen''~(black dashed lines), and the method assuming the spectral distortion is given by a pure energy removal~(thin solid brown lines) for a direct comparison with \refcite{McDermott:2019lch}. This latter did not include a specific treatment for the $\mu$-$y$ transition era. Therefore, for point B we do not show the pure energy injection curve.  The computed distortions is to be compared with the projected sensitivity of PIXIE~\cite{Kogut:2024vbi} multiplied by a factor of 100 for clarity~(dotted lines)}.

\label{fig:distortion_plot}
\end{figure}

Importantly, the shape of the spectral distortion predicted by our more complete treatment differs significantly from the distortion expected from either \refcite{Mirizzi:2009iz} or \refcite{McDermott:2019lch}.
In \Fig{distortion_plot}, we show the spectral shape of a residual signal for a resonant conversion happening at three benchmark points: point ``A'' in the $y$-era (blue, top right), point ``B'' in the $\mu$-$y$ transition era (green, bottom left), and point ``C'' in the $\mu$-era (yellow, bottom right). 
We choose $\epsilon$ to be roughly three times smaller than the COBE-FIRAS constraint that we set in parameter space that is currently not ruled out.
In these three panels, the thick solid lines labeled by ``complete treatment'' are the shapes of the CMB spectral distortions acquired in our work. 
We compare this to the CMB spectral distortion assuming free-streaming photons after $\gamma \rightarrow A'$ conversion~(dashed black, labeled ``Frozen'', utilized in \refcite{Mirizzi:2009iz}) and the one assuming that $\gamma \rightarrow A'$ induces a pure energy removal~(solid brown, labeled ``Pure Energy,'' utilized in \refcite{McDermott:2019lch}). 
We also include the expected sensitivity from PIXIE,~\cite{Kogut:2024vbi} scaled up by a factor of 100 for clarity, to highlight the impact a next-generation CMB spectrum satellite would have on detecting $A'$.

At point ``A'',  resonant $\gamma \rightarrow A'$ conversion happens in the $y$-era. From the upper right panel of \Fig{distortion_plot}, we find that the spectral distortion shape from the complete treatment~(thick solid blue) is similar to the result under the assumption that photons free-stream after $\gamma \rightarrow A'$ conversion~(dashed black).  
It is instructive to see how taking the appropriate limit of $G_y$ in our calculations reproduces the distortion from simple free-streaming at low redshifts. 
Because $\Ps \simeq 1$ in the relevant frequency range, the free-streaming term in \Eq{Gy_small_y} dominates over the $\YYY$-term. Moreover, because $y_\gamma \ll 1$, the delta-function approximation that we used in \Eq{Gy_small_y} is legitimate (this is the correct limit obtained by taking the more general Green's function found in \Eq{Gy_Appx} and applying the narrow width approximation). Based on the discussions above, we can substitute \Eq{Gy_small_y} into \Eq{Delta_I_Greens} to obtain
\bea
\label{eq:Delta_I_AToAp_y}
y\text{ era:} \quad \Delta I_\gamma^{y}(x;\Tcmbtoday) \simeq - P_{\gamma \rightarrow A'}(x) \,\, \bgI(x;\Tcmbtoday) \,,
\eea
which is the same as the result assuming that photons only free-stream after the resonant $\gamma \rightarrow A'$ conversion. 
This result explains the smooth transition from our COBE-FIRAS constraint to the one shown in \refcite{Caputo:2020bdy} when $m_{A'} \lesssim 10^{-10}\eV$\@. One should also note that naively assuming that spectral distortions imparted on the CMB by $\gamma \to A'$ are equivalent to pure energy removal, as utilized in \refcite{McDermott:2019lch}, leads to the incorrect shape of the CMB distortion. 
This treatment is equivalent to setting $\Ps \simeq 0$ in our expression for $G_y$, such that $\Delta I_\gamma(x;\Tcmbtoday)|_\text{pure energy} \propto \YYY(x;\Tcmbtoday)$ instead. 
Therefore, the COBE-FIRAS constraint in \refcite{McDermott:2019lch} cannot smoothly transit to the low mass limits in \refcite{Caputo:2020bdy}, when $m_{A'} \lesssim \SI{e-10}{\eV}$.
This highlights the importance of the more accurate calculation that we perform in this paper.

The point ``C'' corresponds to the $\gamma \rightarrow A'$ resonant conversion in the $\mu$-era. From the lower right panel of \Fig{distortion_plot}, we find that the CMB spectral distortion~(thick solid orange) has a similar shape to the $\MMM$-function, but has the opposite sign as compared to the result assuming the spectral distortion comes from just pure energy removal~(solid brown) utilized in \refcite{McDermott:2019lch}. To quantitatively explain this, we can write the $\mu$-era distortion as
\bea
\label{eq:Delta_I_AToAp_mu}
\mu\text{ era:} \quad \Delta I^{\mu}_\gamma(x;\Tcmbtoday) = \mu_{\gamma \rightarrow A'} \, \MMM(x;\Tcmbtoday)\,,
\eea
where
\bea
\mu_{\gamma \rightarrow A'} \simeq - \alpha_\rho \frac{3}{\kappa_c} \vis(z_\res) \,\, \epsilon^2 \mathcal{F} \, \left( 1-  \Big\langle \Ps(x',z_\res) \, \frac{x_0}{x'}\Big\rangle\right)\,.
\eea
Here, $x_0 \simeq 3.6$, and
\bea
    \langle f(x') \rangle = \int dx' \frac{1}{\bgn} \frac{d \bgn}{dx'} f(x') \,,
\eea 
which corresponds to an average of $f(x')$ over the spectrum of photons. 
As we have discussed in \Sec{green_func}, the temperature shift term of \Eq{Gmu} can be removed because $T_0$ is a nuisance parameter varied during the data analysis. Here, we approximately have $P_{\gamma \rightarrow A'} \simeq \epsilon^2 \mathcal{F}/x'$ based on \Eq{trans_prob_RAD}. In the frequency range and redshift we are interested in, we find that $\Ps \simeq 1$. Therefore, we have
\bea
\label{eq:mu_AtoAp}
\mu_{\gamma \rightarrow A'} \propto - \left( 1 - \Big \langle  \frac{x_0}{x'} \Big\rangle \right) > 0.
\eea
This result is completely different from the spectral distortion calculated under the pure energy removal formalism assumption, as used in \refcite{McDermott:2019lch}. 
The pure energy removal process is recovered in the limit where no photons survive and everything cools the plasma, \textit{i.e.}\ $\Ps \simeq 0$, from which we obtain $\mu_{\gamma \rightarrow A'}|_\text{pure energy} < 0$, which has the opposite sign. 
This explains the apparent sign flip in the bottom right plot of \Fig{distortion_plot}.

\begin{figure}[t]
\centering
\includegraphics[width=0.8\columnwidth]{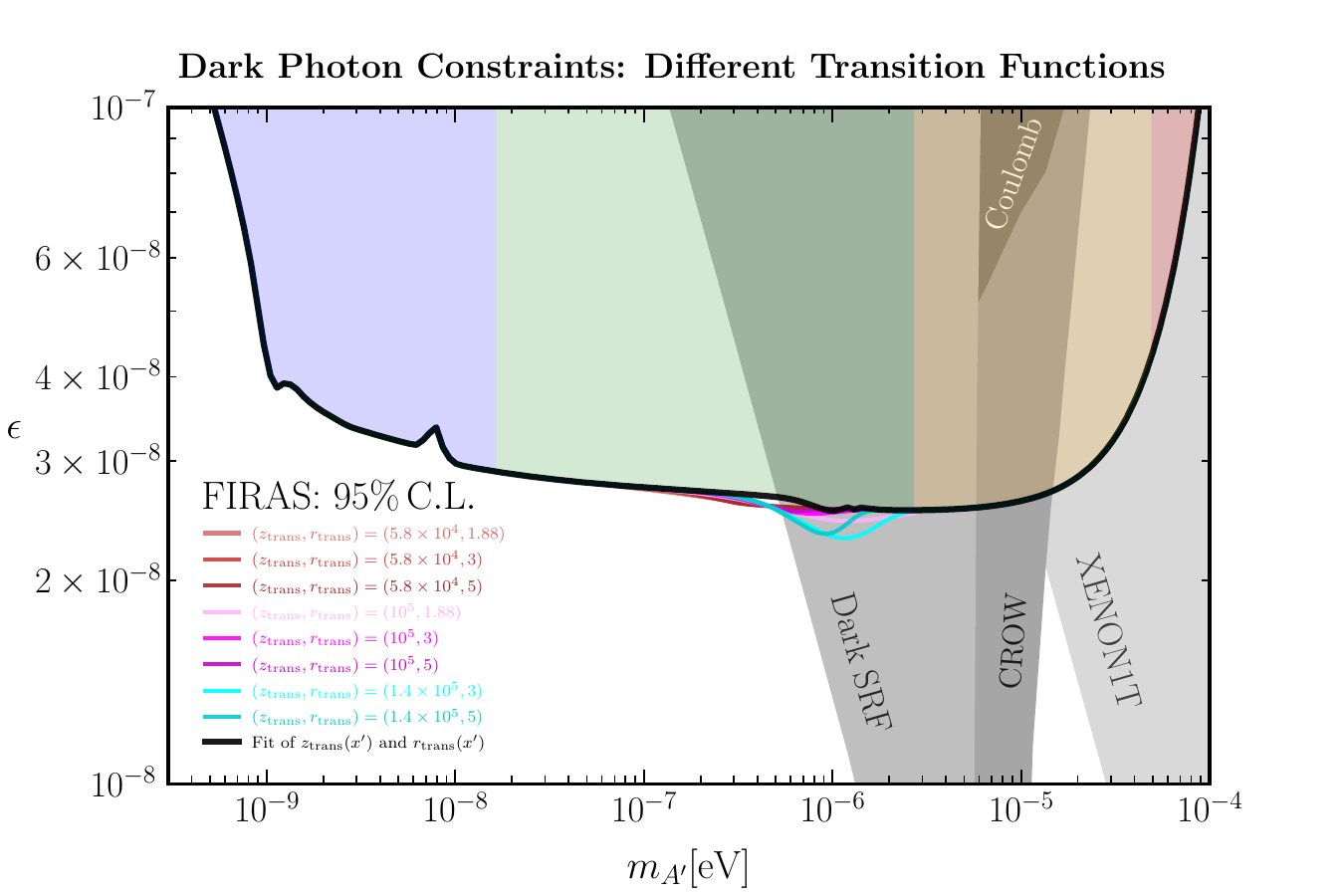}
\vspace{0.3cm}
\caption{The dark photon COBE-FIRAS constraint given different transition functions. As shown in \Eq{Tmu}, the transition function $\transfunc_\mu$ is parameterized by two parameters $z_\trans$ and $r_\trans$, which represent at which redshift and how fast the $\mu$-$y$ transition happens during the cosmological evolution. The black thick solid line is our fiducial constraint with $z_\trans(x')$ and $r_\trans(x')$ obtained by the fitting procedure described in the text. We can find that the dark photon COBE-FIRAS constraints with different $z_\trans$ and $r_\trans$ overlap in the dark photon mass range related to the whole free-streaming era, $y$-era, $\mu$-era, $T$ era, and most of the $\mu$-$y$ transition era. For the small portion of the $m_{A'}$ region where the COBE-FIRAS constraints with different parameterizations do not overlap, we draw this part of the constraint with the dashed line, which is the most conservative constraint among all the parameterizations of $\transfunc_\mu$.}
\label{fig:varyTmuplot}
\end{figure}

The bottom left plot of \Fig{distortion_plot} corresponds to point ``B'', with $\gamma \rightarrow A'$ happening in the $\mu$-$y$ transition era. The CMB spectral distortion~(thick blue line) is then obtained by plugging the Green's function, \Eq{Gtrans}, into~\Eq{I_dist_AToAp_calc} 
\bea
\label{eq:Delta_I_AToAp_trans}
& \mu-y \text{ transition era:} \quad \Delta I_\gamma^{\trans}(x;\Tcmbtoday) = -\int dx' \frac{1}{\bgn} \frac{d \bgn}{dx'} P_{\gamma \to A'}(x') G_\trans(x; x', z_\text{res};\Tcmbtoday)  \,.
\eea
Because the point ``B'' is at the side nearer to the $y$-era, the spectral distortion tends to be more like \Eq{Delta_I_AToAp_y}, which is close to the CMB spectral distortion under the free-streaming assumption~(dashed black). Our approximation for $\Delta I_\gamma^\trans$ in \Eq{Delta_I_AToAp_trans} depends on our choice of the parametrization of the transition function  $\transfunc_\mu$ inside $G_\trans$. 

\Fig{varyTmuplot} shows the different limits that we obtain under the different choices for the form of $\transfunc_\mu$ discussed in Sec.~\ref{sec:green_func}.
Our fiducial choice, with $z_\trans$ and $r_\trans$ varying over $x'$, is shown in black, while other choices of constant $z_\trans$ and $r_\trans$ are shown by the colored lines. 
We find that choosing different parametrizations of $\transfunc_\mu$ leads to similar limits in the range $\SI{2e-8}{\eV} \lesssim m_{A'} \lesssim \SI{2e-7}{\eV}$ in the $\mu$-$y$ transition region, where we therefore expect our limits to be robust. 
In this regime, the $\gamma \to A'$ transitions occur during a period when $G_\trans \simeq G_y$, in agreement with the results shown in \refcite{Chluba:2015hma}, and so the different parametrization choices are not important.
We therefore label the limit that we obtain as a solid green line in Fig.~\ref{fig:fullplot} and subsequent figures. 
In the region $\SI{3e-7}{\eV} \lesssim m_{A'}\lesssim \SI{3e-6}{\eV}$, we instead find differences in the limit, on the order of a few tens of percent, between different assumptions on $z_\trans$ and $r_\trans$. 
However, the fiducial approach that we adopt also gives the weakest constraints in this range, and so we adopt the fiducial result as our limit.
We denote our limit in Fig.~\ref{fig:fullplot} and subsequent figures with a dotted line, as an indication of the fact that our result may shift by a few tens of percent in $\epsilon$. 
A more comprehensive, numerical approach in this limited mass range is still highly desirable, particularly for predicting the exact spectral shapes of distortions during this epoch.

As a closing remark, we want to stress that characterizing the CMB spectral distortion accurately can help to distinguish between different particle physics models. Consider, for example, the signals induced by resonant $\gamma \rightarrow A'$ and $\gamma \rightarrow a$ transitions during the $\mu$-era, where $a$ denotes an axion-like particle. As discussed above, $\gamma \rightarrow A'$ induces a positive chemical potential shift. On the other hand, resonant $\gamma \rightarrow a$ transitions induce a negative chemical potential shift instead. This is because the transition probability from CMB photons to axions satisfies $P_{\gamma \rightarrow a} \propto x'$, when $\gamma \rightarrow a$ transitions happens in the RAD universe~\cite{Mirizzi:2009nq}. 
Doing a similar analysis as above, we have $\mu_{\gamma \rightarrow a} \propto - \left( \langle x'^2 \rangle - x_0 \langle x' \rangle \right) < 0$, which has a sign difference compared to the dark photon in \Eq{mu_AtoAp}. The spectral distortion is therefore highly model dependent, and, when accurately measured, can give information about the cause of any detected spectral distortion.

\section{Conclusions}
\label{sec:conclusion}

In this work, we have employed the Green's function method for photon injection and removal to perform an accurate determination of the COBE-FIRAS constraint on $\epsilon$ as a function of $\map$, due to spectral distortions arising from resonant $\gamma \rightarrow A'$ oscillations. Our updated limit transitions seamlessly to the free-streaming era constraint, when $m_{A'} \lesssim \SI{e-10}{\eV}$, and maintains consistency and smoothness during the $\mu$-$y$ transition era. Even without a full numerical treatment, our approximation for the Green's function during the $\mu-y$ transition era has an estimated $20\%$ theoretical uncertainty in the mass range of $3 \times 10^{-7} - 3 \times 10^{-6} \eV$ due to the different choices of transition functions. This improves on previous approaches in terms of the accuracy and consistency of the calculation. Moreover, our methodology accurately predicts the shape of the CMB spectral distortion across different eras, which differs significantly from the predictions in previous work, which made the assumption that the induced distortion is equivalent to a pure energy removal process from baryons~\cite{McDermott:2019lch}. For example, we find that for conversions occurring during the $\mu$-era, the $\gamma \rightarrow A'$ induces a positive chemical potential shift in the CMB spectrum, and not a negative shift; likewise, in the $y$-era, we found that $\gamma \rightarrow A'$ induces a distortion consistent with free streaming photons, differing from the $y$-distortion predicted by Ref.~\cite{McDermott:2019lch}. This is crucial for identifying definitive signatures of $\gamma \to A'$ conversions in upcoming experiments aiming to measure CMB spectral distortions. Our limits from COBE-FIRAS, and future projections for next-generation experiments such as PIXIE, form accurate benchmarks for experiments like DarkSRF, which are targeting dark photons in a similar mass range. 

The strong limits that we obtain for dark photons in this work demonstrate that CMB spectral distortions are an important probe of new physics in the early universe. While there are existing private code packages like CosmoTherm~\cite{Chluba:2011hw} that are capable of performing a full numerical treatment of spectral distortions, an open-source code would help to make spectral distortion calculations more accessible to the cosmology and high-energy theory communities.
In addition, the exquisite sensitivity of future CMB experiments like PIXIE means that any new physics signal needs to be disentangled from foreground distortions from reionization and the intracluster medium. 
A detailed analysis of how well we can recover a signal like those expected in $\gamma \to A'$ in the presence of these foregrounds, together with a more realistic treatment of the detector performance, may be of interest in providing a realistic assessment of the new-physics reach of PIXIE.

\vspace{0.8cm}
\noindent\textbf{Note added.} The results from this work were first presented by one of us in their PhD thesis~\cite{Gan:2024ele} and at a conference talk~\cite{TeVPA:2024}. While we were finalizing this submission, Ref.~\cite{Chluba:2024wui} appeared on the arXiv, which also studies DP-induced spectral distortions. Ref.~\cite{Chluba:2024wui} contains a full numerical treatment for spectral distortions, using the private code CosmoTherm~\cite{Chluba:2011hw}, by one of the authors. A fully numerical approach allows the authors to treat also the regime of large spectral distortions, where the Green's function method that we use breaks down. However, this is only important for $m_{A'} \gtrsim 10^{-4}\, \rm eV$ and kinetic mixing above $\sim 10^{-6}$. This region of the parameter space is already robustly excluded by other experiments, such as Xenon1T and solar emission constraints. For all parts of parameter space where our CMB spectral distortion limits are the most stringent constraint,  distortions remain small, and the comparison between our results and those of Ref.~\cite{Chluba:2011hw} shows excellent agreement between the Green's function method and the full numerical treatment. All of our code and data are made public, including the fits for the Green's functions in the $\mu-y$ transition era, which were not available before. These may also be applied in other contexts where photon injection or subtraction take different forms.

\tocless\section{Acknowledgments}
We want to thank Jens Chluba, David Dunsky, Patrick J. Fox, Junwu Huang, Gustavo Marques-Tavares, and Clayton Ristow for helpful discussions. HL was supported by the Kavli Institute for Cosmological Physics and the University of Chicago through an endowment from the Kavli Foundation and its founder Fred Kavli, and Fermilab operated by the Fermi Research Alliance, LLC under contract DE-AC02-07CH11359 with the U.S. Department of Energy, Office of Science, Office of High-Energy Physics. XG and GA are supported by the James Arthur Graduate Associate~(JAGA) Fellowship. The work of XG is supported in part by the Deutsche Forschungsgemeinschaft under Germany’s Excellence Strategy - EXC 2121 “Quantum Universe” - 390833306. JTR is supported by NSF grant PHY-2210498.  This work was performed in part at the Aspen Center for Physics, which is supported by NSF grant PHY-1607611. This research was supported in part by grant NSF PHY-2309135 to the Kavli Institute for Theoretical Physics~(KITP). The work presented in this paper was performed on computational resources managed and supported by Princeton Research Computing. This work was supported
in part through the NYU IT High Performance Computing resources, services, and staff expertise.

\vspace{15mm}

\appendix

\section{Profile Likelihood Test}
\label{appx:likelihood}

The analysis of the COBE-FIRAS data in our paper follows \refscite{Caputo:2020bdy,Caputo:2020rnx}. In this section, we give an introduction to this analysis, for completeness. To impose the constraint, or the future projections, on $\gamma \rightarrow A'$ oscillations using  data from CMB spectral measurements, we use the constructed Gaussian likelihood function\footnote{Since we are profiling over $T_0$, when calculating the CMB spectral distortion happening in the $\mu$-era, we only need to include the $\MMM$-term in \Eq{Gmu} when utilizing the Green's function method. This gives numerically more reliable results. }
\bea
\log \mathcal{L}(\da|m_{A'},\epsilon) = \max_{T_0} \left\{ - \frac{1}{2}  {\bm \delta} \vecI^T(m_{A'},\epsilon;T_0) \cdot \matcov^{-1} \cdot {\bm \delta} \vecI(m_{A'},\epsilon;T_0) \right\}.
\eea
$\matcov$ is the $N_\da \times N_\da$ covariance matrix. $N_\da$ is the number of the data points. ${\bm \delta} \vecI\,(m_{A'},\epsilon;T_0)$ is the $N_\da$ vector which is written as
\bea
{\bm \delta} \vecI(m_{A'},\epsilon;T_0) = \bgvecI(T_0) + {\bm \Delta} \vecI(m_{A'},\epsilon;T_0) - \bgvecI(T_{\da,0}) - \vecres,
\eea
where $T_{\da,0}=2.7255\text{K}$ is the measured CMB temperature, $\vecres$ is the $N_\da$ vector of residuals, and ${\bf \Delta} \vecI(m_{A'},\epsilon;T_0)$ is the $\gamma \rightarrow A'$-induced CMB spectral distortion for all frequency bins given the dark photon parameter $m_{A'}$ and $\epsilon$. 
Note that $\bgvecI(T_{\da,0}) + \vecres$ is the observed COBE-FIRAS spectrum, and $\bgvecI(T_0) + {\bm \Delta} \vecI(m_{A'},\epsilon;T_0)$ is the predicted CMB spectrum with distortions from dark photons with a blackbody temperature of $T_0$, so that ${\bm \delta} \vecI$ is the difference between the predicted and measured spectra.
In the COBE-FIRAS dataset~\cite{Fixsen:1996nj}, $N_\da = 43$ and the covariance matrix elements are
\bea
\mathcal{C}_{ij} = \sigma_i \sigma_j \mathcal{Q}_{|i-j|},
\eea
where $i,j = 1,\cdots,N_\da$. Here, $\sigma$ is the uncertainty and $\vecQ$ is the $N_\da \times 1$ vector which is
\bea
\vecQ = (\mathcal{Q}_0, \mathcal{Q}_1, \cdots, \mathcal{Q}_{N_\da-1}) = (1, 0.176, \cdots, 0.008). 
\eea
The test statistic is
\bea
\label{eq:TS_form}
\TS\,(m_{A'},\epsilon) = 2 \left[ \log \mathcal{L}(\da|m_{A'},\epsilon) - \max_{\epsilon} \log \mathcal{L}(\da|m_{A'},\epsilon) \right] \,,
\eea
which by Wilks' theorem~\cite{Wilks:1938dza,ParticleDataGroup:2022pth} follows a $\chi^2$-distribution with one degree of freedom. Therefore, to test $A'$ with $95\%$ confidence interval with a one-sided $\chi^2$ distribution, we choose $\TS = -2.71$ to impose the constraint on $\epsilon$ given $m_{A'}$. 

For the PIXIE projection, we use the sensitivity provided in Ref.~\cite{Kogut:2024vbi}, which assumes $N_\data = 313$, $\vecres = \bm{0}$, and a diagonal covariance matrix.

\section{Constants}
\label{appx:const}

In this section, we list the constants appearing in the calculation of the CMB spectral distortion and introduce their physical meanings. Firstly, we write down the following integrals
\begin{equation}
\left\{\begin{aligned}
\label{eq:Tk_Fk_def}
& \bbT_k = \int^\infty_0 dx \,\, x^k \cdot \bgf(x) \cdot \ttt(x) = (k+1)! \,\,\zeta(k+1) \,, \\
& \bbF_k = \int_0^\infty dx \,\, x^k \cdot \bgf(x) = k! \, \zeta(k+1) \,,
\end{aligned}
\right. 
\end{equation}
that are used in this paper. Given the fundamental ingredients shown in \Eq{Tk_Fk_def}, we can write all the constants appearing in the calculation of the CMB spectral distortion as
\bea
\label{eq:const}
\alpha_\rho = \frac{\bbF_2}{\bbF_3} \simeq 0.3702, \quad \alpha_\mu = \frac{\bbT_1}{\bbT_2}  \simeq 0.456, \quad x_0 = \frac{\bbT_3}{\bbT_2} = \frac{4}{3 \alpha_\rho} \simeq 3.6, \quad \kappa_c =  \frac{\bbT_1 \bbT_3  - \bbT_2^2 }{\bbF_2 \bbF_3 } \simeq 2.14185. 
\eea
Here, $\alpha_\rho$ quantifies the ratio between $\bgrho$ and $\bgn$, which is given by
\bea
\label{eq:alpha_rho_def}
\alpha_\rho = \frac{\bgn(T)\, \Tcmb}{\bgrho(T)}. 
\eea
$\alpha_\mu$ quantifies the relation between the nonzero chemical potential and the temperature shift for the pure energy injection~($\Ps=0$) during the $\mu$-era. In this case, the CMB photon number density does not change. From Eqs.~\ref{eq:mu_era_dist} and \ref{eq:mmm_def}, we have
\bea
\Ps = 0 : \quad \frac{t}{\mupara} = \frac{\int \dbar^3 \, \veck \bgf(x) \cdot \ttt(x)/x }{\int \dbar^3 \veck \, \bgf(x) \cdot \ttt(x)} = \alpha_\mu. 
\eea
$x_0$ is the dimensionless critical frequency for the photon injection or removal process. Let us take the monochromatic photon injection with the dimensionless frequency $x$, as an example. For the low-frequency photon satisfying $x < \Ps \cdot x_0$, the chemical potential from the photon injection flips sign compared to a pure energy injection. $\kappa_c$ comes from solving the equations for the number and energy density variations of the CMB photon caused by the photon injection. $3/\kappa_c \simeq 1.401$ is the numerical factor that frequently appears in former literature, such as \refscite{Chluba:2013vsa,Chluba:2013pya,Chluba:2015hma,Chluba:2019nxa,Hook:2023smg}, when describing the nonzero chemical potential developed during the $\mu$-era.

\section{Monochromatic Photon Injection in the $\mu$-Era}
\label{appx:mono_mu_era}

In this section, we derive the chemical potential and temperature shift in the $\mu$-era. As a simplified example, in this section, we only consider the monochromatic photon injection, with frequency $x_\inj$, at the redshift $z_\inj$, which gives
\bea
\Gamma_\inj(x,z) = \widetilde{\Gamma}_\inj \, \delta(x-x_\inj) \, \delta(z-z_\inj) \, . 
\eea
In \SubAppx{green_func_mu}, we present the Green's function formalism for photon injections, in the $\mu$-era, with continuous spectra and redshifts.

From \Eq{mu_era_dist}, we see that the variations in number and energy densities in the $\mu$-era are
\bea
\label{eq:Delta_n_rho_t_mu}
\left\{
\begin{aligned}
& \Ps(x_\inj,z_\inj) \, \Delta n_{\gamma,\inj} = g_\gamma \int \dbar^3 \veck \, \Delta f_\gamma = \bgn(T_\inj) \left( \frac{\bbT_2}{\bbF_2} t_\inj - \frac{\bbT_1}{\bbF_2} \mu_\inj + \cdots \right)\\
& \Delta \rho_{\gamma,\inj} = g_\gamma \int \dbar^3 \veck \,\, \wcmb \Delta f_\gamma = \bgrho(T_\inj) \left( \frac{\bbT_3}{\bbF_3} t_\inj - \frac{\bbT_2}{\bbF_3} \mu_\inj + \cdots \right)
\end{aligned}
\right.,
\eea
where $\mu_\inj$ and $t_\inj$ represent the chemical potential and temperature shifts, respectively, immediately after photon injection. $\Delta n_{\gamma,\inj}$ and $\Delta \rho_{\gamma,\inj}$ represent the number and energy densities of injected photons, which can be written as
\bea
\Delta n_{\gamma,\inj} = \frac{d \bgn}{d x_{\inj}} \frac{\widetilde{\Gamma}_\inj}{(1+z_\inj) H(z_\inj)} \quad \text{and} \quad \Delta \rho_{\gamma,\inj} = x_\inj T \, \Delta n_{\gamma,\inj}\, \, .
\eea
$\Ps$ is the photon survival probability, quantifying the ratio of remaining injected photons after the photon absorption by {\dcs} and {\br}. Solving \Eq{Delta_n_rho_t_mu} and utilizing \Eq{alpha_rho_def}, we have
\bea
\label{eq:mu_inj}
\mu_\inj = \alpha_\rho x_\inj \cdot \frac{3}{\kappa_c}  \left[ 1 - \Ps(x_\inj,z_\inj) \frac{x_0}{x_\inj} \right] \frac{\Delta n_{\gamma,\inj} }{\bgn} \, , 
\eea
and
\bea
t_\inj - \alpha_\mu \mu_\inj  = \alpha_\rho x_\inj \cdot \frac{\Ps}{4} \frac{x_0}{x_\inj} \frac{\Delta n_{\gamma,\inj}}{\bgn} \, \,  .
\eea
Because {\dcs} and {\br} tend to drive the chemical potential of CMB photons to zero,  the chemical potential today, $\mu_0$, is given by $\mu_\inj$ multiplied by the visibility function
\bea
\label{eq:visibility_func}
\vis(z) \simeq 0.983 \exp\left[ - \left(\frac{z}{z_\mu}\right)^{2.5} \right] \left[ 1 - 0.0381 \left( \frac{z}{z_\mu}\right)^{2.29} \right], \quad \text{where $z_\mu \simeq 2 \times 10^6$} \,  \, . 
\eea
The derivation of \Eq{visibility_func} can be found in Ref.~\cite{Chluba:2013kua}. Therefore, we have
\bea
\label{eq:mu_0_from_mu_inj}
\mu_0= \alpha_\rho x_\inj \cdot \frac{3}{\kappa_c} \, \vis(z_\inj) \, \left[ 1 - \Ps(x_\inj,z_\inj) \frac{x_0}{x_\inj} \right] \frac{\Delta n_{\gamma,\inj} }{\bgn} \,  \, . 
\eea
When $z_\inj \gg z_\mu$, $\vis \simeq 0$, and therefore $\mu_0  \simeq 0$. Alternatively, when $z_\inj \ll z_\mu$, $\vis \simeq 1$, implying that $\mu_0 \simeq \mu_\inj$. To illustrate the difference between pure photon injection~($\Ps \simeq 1$) and pure energy injection~($\Ps \simeq 0$), we compare the ratio of the chemical potentials of CMB photons, in both cases, and find
\bea
\frac{\mu_0|_{\Ps = 1}}{\mu_0|_{\Ps = 0}} \simeq 1 - \frac{x_0}{x_\inj} \, \,  . 
\eea
From this equation, we find that when $x_\inj < x_0$, the two chemical potentials have different signs. Therefore, for  photon injections in this frequency range, one clearly cannot naively use the pure energy injection formalism of \refcite{Chluba:2013vsa}. 

For generic photon injection/removal processes happening in the $\mu$-era, the $\mu$ distortion can be decomposed into a linear superposition of a series of  monochromatic photon injection/removal processes. In this case, by performing the replacements $x_\inj \rightarrow x'$, $z_\inj \rightarrow z'$, and $\widetilde{\Gamma}_\inj \rightarrow \Gamma_\inj(x',z')$, the induced chemical potential shift can be written as 
\bea
\mu = \int dz' \int dx' \, \alpha_\rho x' \left[1-P_s(x',z') \frac{x_0}{x'}\right] \frac{d \bgn/dx'}{\bgn} \frac{\Gamma_\inj(x',z')}{(1+z') H(z')} \, \,  .
\eea
The above calculation is actually the derivation of the factor in front of the first term in \Eq{Gmu}, which contains $\MMM$. The second term in \Eq{Gmu}, containing $\TTT$, is derived utilizing the normalization condition listed in \Eq{norm_condition_photon}, which is equivalent to imposing energy conservation for the photon injection/removal processes.

\section{Green's Functions}
\label{appx:green_func}

In this section, we introduce the calculation of the CMB spectral distortion, utilizing the Green's function method, and list the concrete forms of Green's functions used in our work. Here, the CMB spectral distortion is
\bea
\label{eq:I_dist_photon_inj_calc}
\Delta I_\gamma (x;\Tcmbtoday) = \int dx' \int dz'\, G(x;x',z';\Tcmbtoday) \, \Source(x',z') \, \, .
\eea
$\Source(x',z')$ is the source of the photon injection, and it is written as\footnote{To get the source term in \refcite{Chluba:2015hma}, we can do the replacement $\Source(x',z') \rightarrow 2\pi T_0 x' \cdot \Source(x',z')$. }
\bea
\label{eq:S_inj_general}
\Source(x',z') = \frac{1}{\bgn} \frac{d \bgn}{dx'} \frac{\Gammainj(z')}{(1+z') H(z')} \, \, .
\eea
Positive and negative $\Gamma_\inj$ represent the rate of the photon injection and removal, respectively.\footnote{For $\gamma \rightarrow A'$ resonant conversions, we can derive \Eq{S_ATpAp}, under the narrow width approximation, by substituting \Eq{trans_rate} into \Eq{S_inj_general}.}

\subsection{Functions}

In this subsection, we summarize the functions that appear in Green's functions and are distributed throughout the main text. For CMB photons with an exact blackbody phase space distribution, the intensity is
\bea
\bgI(\omega_{0};\Tcmbtoday) = \frac{\omega_0^3}{2\pi^2} \bgf(x;\Tcmbtoday) = \frac{\Tcmbtoday^3}{2\pi^2}\frac{x^3}{e^x-1} \, \, .
\eea
To describe the shape of the CMB spectral distortion, we have the dimensionless functions
\bea
\label{eq:T_M_Y_dimless}
\ttt(x) = \frac{x e^x}{e^x-1}, \quad \mmm(x) = \ttt(x) \left( \alpha_\mu - \frac{1}{x} \right), \quad \yyy(x) = \ttt(x) \left(x \frac{e^x+1}{e^x-1} -4\right) \, \, ,
\eea
where $\ttt(x)$ comes from the temperature shift, $\mmm(x)$  from the nonzero chemical potential, and $\yyy(x)$ from the SZ effect. To describe the absolute CMB spectral distortion in terms of intensity, we define the functions
\bea
\label{eq:T_M_Y_dim}
\begin{pmatrix}
\TTT(x;\Tcmbtoday)\\ 
\MMM(x;\Tcmbtoday)\\
\YYY(x;\Tcmbtoday)
\end{pmatrix} 
= \bgI(x;\Tcmbtoday) \times 
\begin{pmatrix} 
\ttt(x)\\
\mmm(x)\\
\yyy(x)\\
\end{pmatrix} \, \, ,
\eea
which have the same dimension as the CMB intensity and Green's functions. In this paper, we use Eqs.~\ref{eq:T_M_Y_dimless} and~\ref{eq:T_M_Y_dim} interchangeably.

\subsection{Green's Function Normalization}
\label{appx:green_func_norm}
We now derive the Green's function normalization shown in \Eq{norm_condition_photon}. 
First, the relation between the distortion to the intensity of the CMB, $\Delta I_\nu(x)$, and the total energy density of the distortion, $\Delta \rho_\gamma$, is
\bea
\label{eq:rho_distortion_1}
    \Delta \rho_\gamma = 2 T_0 \int dx \, \Delta I_\gamma(x) = 2T_0 \int dx \int dx' \int dz' \, G(x; x', z'; T_0) \,\Source(x', z')  \,,
\eea
where we have used the relation $x T_0 = 2 \pi \nu_0$. On the other hand, from  \Eq{S_inj_general}, we see that the injected energy density, in the frequency and redshift intervals $(x',x' + dx')$ and $(z', z' + dz')$, respectively, is given by $x' T_0 (1+z') \cdot \bgn(T_0) (1+z')^3 \cdot \Source(x', z') $. 
This energy density redshifts as $(1+z)^4$; we therefore see that 
\bea
\label{eq:rho_distortion_2}
    \Delta \rho_\gamma = \int dx' \int dz' \, x' \, T_0 \, \bgn(T_0) \, \Source(x', z')  \,.
\eea
Comparing Eqs.~\ref{eq:rho_distortion_1} and~\ref{eq:rho_distortion_2}, we find 
\bea
\label{eq:norm_condition_photon_Appx}
    2T_0 \int dx \, G(x; x', z'; T_0) = x' T_0 \overline{n}_\gamma(T_0) = x' \alpha_\rho \bgrho(T_0) \,,
\eea
where $\alpha_\rho$ is defined in \Eq{alpha_rho_def}. This equation is the same expression shown in \Eq{norm_condition_photon}, and corresponds to a normalization condition of the Green's function. 

\subsection{Green's Function in the $\mu$-Era}
\label{appx:green_func_mu}

For the CMB spectral distortion in the $\mu$-era, we have
\bea
\label{eq:Gmu_Appx}
G_\mu(x;x',z';\Tcmbtoday) = \alpha_\rho x' \cdot \frac{3}{\kappa_c} \, \vis(z')  \left[ 1 - \Ps (x',z') \frac{x_0}{x'} \right] \, \MMM(x;\Tcmbtoday) + \frac{\lambda(x',z')}{4} \, \TTT(x;\Tcmbtoday),
\eea
where 
\bea
\label{eq:lambda_func_Appx}
\lambda(x',z') = \alpha_\rho x' \cdot \left[ 1 - \left(1 - \Ps (x',z') \frac{x_0}{x'} \right) \, \vis(z') \right].
\eea
$\Ps(x',z')$ is the survival probability function of the injected photons. During the $\mu$-era, it can be approximately written as
\bea
\Ps(x',z') = e^{-\tauff(x',z')},
\eea
where $\tauff(x',z') \simeq x_c(z')/x'$.
$\tauff(x',z')$ is the optical depth describing the absorption of the injected photons caused by {\br} and {\dcs}, and $x_c(z')$ is the critical frequency for {\cs} to dominate over {\br} and {\dcs}. $\lambda(x',z')$ is determined by the normalization condition of \Eq{norm_condition_photon}.

In our work, because we utilize the COBE-FIRAS data to explore $\gamma \rightarrow A'$ oscillations, photon frequencies satisfying $x' \gg x_c(z')$ are the dominant contribution to the integration needed to calculate the CMB spectral distortion. Therefore, we approximately have $\Ps(x',z') \simeq 1$, which corresponds to a pure photon injection. For the injection of low-frequency photons satisfying $x'<x_0$, the nonzero chemical potential of the CMB photons has an extra minus sign compared with the chemical potential calculated under the framework of a pure energy injection where $\Ps = 0$. The opposite sign of $\mu_{\gamma \rightarrow A'}$ highlights the key difference between our approach, utilizing the full $\mu$-era Green's function~($P_s \simeq 1$, $\mu_{\gamma \rightarrow A'} >0$), and the previous approach, utilizing the $\mu$-era Green's function for  pure energy injections~($P_s \simeq 0$, $\mu_{\gamma \rightarrow A'} < 0$)~\cite{McDermott:2019lch}.

\subsection{Green's Function in the $y$-Era}
\label{appx:green_func_y}

We now introduce the Green's function in the $y$-Era, which is written as
\bea
\label{eq:Gy_Appx}
G_y(x;x',z';\Tcmbtoday) 
& = \alpha_\rho x' \cdot \left(1 - \Ps(x',z') \frac{e^{ \left(\alpha(x',z')+\beta(x',z') \right) \, y_\gamma(z') } }{1+x' y_\gamma(z')}\right) \frac{\YYY(x;\Tcmbtoday)}{4} \\
& \quad + \alpha_\rho x' \cdot \frac{\bgrho(\Tcmbtoday)}{2 \Tcmbtoday} \, \Ps(x',z') \, \deltalike(x;x',z') \, \, .
\eea
Here, $\alpha$ and $\beta$ are defined as
\bea
\alpha(x',z') = \frac{3-2f(x')}{\sqrt{1+x' y_\gamma(z')}}, \quad \beta(x',z') = \frac{1}{1 + x' y_\gamma(z') \left[ 1 - f(x') \right]} \, \, ,
\eea
where
\bea
f(x') = e^{-x'} \left( 1 + \frac{x'^2}{2} \right) \, \, . 
\eea
$F$ is defined as
\bea
\label{eq:deltalike_def_Appx}
\deltalike(x;x',z') = \frac{ e^{-\frac{1}{4 \beta(x',z') \, y_\gamma(z')} \left\{\log\left[x \left(1/x' + y_\gamma(z') \right)\right] - \alpha(x',z')\, y_\gamma(z') \right\}^2 } }{x'\sqrt{4 \pi \,\, \beta(x',z') \,\, y_\gamma(z')}} \, \, .
\eea
Here, $F(x;x',z')$ obeys
\bea
\label{eq:int_F_Appx}
\int_0^\infty dx \, F(x;x',z') = \frac{e^{ \left(\alpha(x',z')+\beta(x',z') \right) \, y_\gamma(z') } }{1+x' y_\gamma(z')} \, \, ,
\eea
which helps us to easily check that \Eq{Gy_Appx} satisfies the normalization condition of \Eq{norm_condition_photon}. 

In the late universe, such as during the free-streaming era, the $y$-era, or the late stage of the $\mu$-$y$ transition era, $y_\gamma \ll 1$ because {\cs} is inefficient. In such a situation, because $\alpha, \beta \sim \mathcal{O}(1)$, we can approximately write the second term in \Eq{Gy_Appx} as
\bea
y_\gamma \ll 1: \quad \deltalike(x;x',z') \simeq \delta(x-x') \, \, .
\eea
In this case, the Green's function in \Eq{Gy_Appx} can be approximately written as
\bea
\label{eq:Gy_small_y_Appx}
y_\gamma \ll 1: \quad G_y(x;x',z';\Tcmbtoday) \simeq \alpha_\rho x' \cdot \left( 1 - \Ps(x',z') \right) \, \frac{\YYY(x;\Tcmbtoday)}{4} + \alpha_\rho x' \cdot \frac{\bgrho(\Tcmbtoday)}{2 \Tcmbtoday} \, \Ps(x',z') \, \delta(x-x') \, \, .
\eea
In the earlier universe, $F(x;x',z')$ is broadened by {\cs}. By performing a numerical comparison, we find that as long as $x' y_\gamma, \alpha y_\gamma$, and $\beta y_\gamma \lesssim 1$, then $F(x;x',z')$ can be well approximated by the Gaussian function $\frac{e^{(\alpha+\beta)y_\gamma}}{1+x'y_\gamma} \, \frac{e^{-(x-x_\text{peak})^2/\sigma^2}}{\sqrt{\pi} \sigma}$ with standard deviation $\sigma$ and peak location $x_\text{peak}$. Here, the prefactor multiplying the Gaussian is acquired by imposing the same normalization as \Eq{int_F_Appx}. By matching the peak of the Gaussian and \Eq{deltalike_def_Appx}, we have
\bea
\label{eq:sigma_xpeak_Appx}
\sigma = \frac{e^{(\alpha+\beta)y_\gamma}}{1+x'y_\gamma} \sqrt{4 \beta y_\gamma}x', \quad x_\text{peak} = \frac{e^{\alpha y_\gamma}}{1+x' y_\gamma} \, x' \, \, . 
\eea
From the above equation, we can find that  larger $y_\gamma$~(more efficient {\cs}) or  larger $x'$ (larger photon frequency), corresponds to a wider $F(x;x',z')$ distribution. In addition, when $\alpha y_\gamma$ or $x' y_\gamma$ is larger than $\mathcal{O}(1)$, the peak of this distribution will have a significant deviation from $x=x'$. 

For the exploration of $\gamma \rightarrow A'$ using COBE-FIRAS data, we have $\tauff \ll 1$ because only the frequency range $x'\gtrsim 0.1$ inside the integration is relevant. This leads to $\Ps \simeq 1$, which reveals that the $\gamma \rightarrow A'$ process corresponds to pure photon removal. In this case, compared with the free-streaming term, the $y$ distortion term gives a subdominant contribution to the CMB spectral distortion. 

\subsection{Green's Function for Pure Energy Injection}
\label{appx:green_func_pure_energy}

In this section, we build the connection between the formalisms describing photon injection and pure energy injection which is utilized in \cite{McDermott:2019lch}.  By choosing the $\Ps \rightarrow 0$ limit, which is the case for injected low-frequency photons that are absorbed by the thermal bath, the photon injection Green's function developed in \refcite{Chluba:2015hma} becomes the pure energy injection Green's function developed in \refcite{Chluba:2013vsa}. Such an approach allows us to quantitatively demonstrate that the incorrectness in the previous work arises from adopting an opposite limit for $\gamma \rightarrow A'$, namely $P_s \simeq 0$, whereas the correct limit should be $P_s \simeq 1$.

Because $G(x,x',z';\Tcmbtoday)/\alpha_\rho x'$ is independent of $x'$ when $\Ps(x',z') = 0$, we can define the Green's function for the pure energy injection as
\bea
\label{eq:G_th_limit}
G^\Th(x;z';\Tcmbtoday) =  \frac{G(x;x',z';\Tcmbtoday)}{\alpha_\rho x'}\bigg|_{\Ps = 0} \, \, . 
\eea
Therefore, we can write \Eq{I_dist_photon_inj_calc} and \Eq{S_inj_general} as
\bea
\label{eq:I_dist_pure_energy}
\Delta I_\gamma(x;\Tcmbtoday) = \int dz' \,\, G^\Th(x;z';\Tcmbtoday) \frac{d(Q/\bgrho)}{dz'} \, \, ,
\eea
where the speed of the energy injection can be written as
\bea
\frac{d(Q/\bgrho)}{dz'} = \int dx' \frac{\wcmb'  d \bgn/dx'}{\bgrho} \, \frac{\Gammainj(x',z')}{(1+z') H(z')} \, \, . 
\eea
Applying \Eq{G_th_limit}, we can derive the Green's functions for the thermalized energy injection, which are written as
\bea
\label{eq:G_th_mu}
G^\Th_\mu(x;z';\Tcmbtoday) = \frac{3}{\kappa_c} \, \vis(z') \, \MMM(x;\Tcmbtoday) + \frac{1-\vis(z')}{4} \, \TTT(x;\Tcmbtoday) 
\eea
for the $\mu$ distortion, and
\bea
\label{eq:G_th_y}
G^\Th_y(x;z';\Tcmbtoday) = \frac{1}{4} \YYY(x;\Tcmbtoday) 
\eea
for the $y$ distortion. We can find that Eqs.~\ref{eq:I_dist_pure_energy},~\ref{eq:G_th_mu}, and~\ref{eq:G_th_y} are all consistent with the formulas shown in \refcite{Chluba:2013vsa}. We can also easily verify that both Eqs.~\ref{eq:G_th_mu} and~\ref{eq:G_th_y} satisfy the normalization condition of the pure energy injection, which is
\bea
\label{eq:norm_condition_energy_inj}
2 \Tcmbtoday \int dx \,\, G^\Th(x;z';\Tcmbtoday) = \bgrho(\Tcmbtoday) \, \, . 
\eea

\bibliography{References}

\begin{thebibliography}{114}%
\makeatletter
\providecommand \@ifxundefined [1]{%
 \@ifx{#1\undefined}
}%
\providecommand \@ifnum [1]{%
 \ifnum #1\expandafter \@firstoftwo
 \else \expandafter \@secondoftwo
 \fi
}%
\providecommand \@ifx [1]{%
 \ifx #1\expandafter \@firstoftwo
 \else \expandafter \@secondoftwo
 \fi
}%
\providecommand \natexlab [1]{#1}%
\providecommand \enquote  [1]{``#1''}%
\providecommand \bibnamefont  [1]{#1}%
\providecommand \bibfnamefont [1]{#1}%
\providecommand \citenamefont [1]{#1}%
\providecommand \href@noop [0]{\@secondoftwo}%
\providecommand \href [0]{\begingroup \@sanitize@url \@href}%
\providecommand \@href[1]{\@@startlink{#1}\@@href}%
\providecommand \@@href[1]{\endgroup#1\@@endlink}%
\providecommand \@sanitize@url [0]{\catcode `\\12\catcode `\$12\catcode
  `\&12\catcode `\#12\catcode `\^12\catcode `\_12\catcode `\%12\relax}%
\providecommand \@@startlink[1]{}%
\providecommand \@@endlink[0]{}%
\providecommand \url  [0]{\begingroup\@sanitize@url \@url }%
\providecommand \@url [1]{\endgroup\@href {#1}{\urlprefix }}%
\providecommand \urlprefix  [0]{URL }%
\providecommand \Eprint [0]{\href }%
\providecommand \doibase [0]{http://dx.doi.org/}%
\providecommand \selectlanguage [0]{\@gobble}%
\providecommand \bibinfo  [0]{\@secondoftwo}%
\providecommand \bibfield  [0]{\@secondoftwo}%
\providecommand \translation [1]{[#1]}%
\providecommand \BibitemOpen [0]{}%
\providecommand \bibitemStop [0]{}%
\providecommand \bibitemNoStop [0]{.\EOS\space}%
\providecommand \EOS [0]{\spacefactor3000\relax}%
\providecommand \BibitemShut  [1]{\csname bibitem#1\endcsname}%
\let\auto@bib@innerbib\@empty
\bibitem [{\citenamefont {Bartlett}\ \emph {et~al.}(1970)\citenamefont
  {Bartlett}, \citenamefont {Goldhagen},\ and\ \citenamefont
  {Phillips}}]{Bartlett:1970js}%
  \BibitemOpen
  \bibfield  {author} {\bibinfo {author} {\bibfnamefont {D.~F.}\ \bibnamefont
  {Bartlett}}, \bibinfo {author} {\bibfnamefont {P.~E.}\ \bibnamefont
  {Goldhagen}}, \ and\ \bibinfo {author} {\bibfnamefont {E.~A.}\ \bibnamefont
  {Phillips}},\ }\bibfield  {title} {\enquote {\bibinfo {title} {{Experimental
  Test of Coulomb's Law}},}\ }\href {\doibase 10.1103/PhysRevD.2.483}
  {\bibfield  {journal} {\bibinfo  {journal} {Phys. Rev. D}\ }\textbf {\bibinfo
  {volume} {2}},\ \bibinfo {pages} {483--487} (\bibinfo {year}
  {1970})}\BibitemShut {NoStop}%
\bibitem [{\citenamefont {Williams}\ \emph {et~al.}(1971)\citenamefont
  {Williams}, \citenamefont {Faller},\ and\ \citenamefont
  {Hill}}]{Williams:1971ms}%
  \BibitemOpen
  \bibfield  {author} {\bibinfo {author} {\bibfnamefont {E.~R.}\ \bibnamefont
  {Williams}}, \bibinfo {author} {\bibfnamefont {J.~E.}\ \bibnamefont
  {Faller}}, \ and\ \bibinfo {author} {\bibfnamefont {H.~A.}\ \bibnamefont
  {Hill}},\ }\bibfield  {title} {\enquote {\bibinfo {title} {{New experimental
  test of Coulomb's law: A Laboratory upper limit on the photon rest mass}},}\
  }\href {\doibase 10.1103/PhysRevLett.26.721} {\bibfield  {journal} {\bibinfo
  {journal} {Phys. Rev. Lett.}\ }\textbf {\bibinfo {volume} {26}},\ \bibinfo
  {pages} {721--724} (\bibinfo {year} {1971})}\BibitemShut {NoStop}%
\bibitem [{\citenamefont {Bartlett}\ and\ \citenamefont
  {Loegl}(1988)}]{Bartlett:1988yy}%
  \BibitemOpen
  \bibfield  {author} {\bibinfo {author} {\bibfnamefont {D.~F.}\ \bibnamefont
  {Bartlett}}\ and\ \bibinfo {author} {\bibfnamefont {S.}~\bibnamefont
  {Loegl}},\ }\bibfield  {title} {\enquote {\bibinfo {title} {{LIMITS ON AN
  ELECTROMAGNETIC FIFTH FORCE}},}\ }\href {\doibase
  10.1103/PhysRevLett.61.2285} {\bibfield  {journal} {\bibinfo  {journal}
  {Phys. Rev. Lett.}\ }\textbf {\bibinfo {volume} {61}},\ \bibinfo {pages}
  {2285--2287} (\bibinfo {year} {1988})}\BibitemShut {NoStop}%
\bibitem [{\citenamefont {Kroff}\ and\ \citenamefont
  {Malta}(2020)}]{Kroff:2020zhp}%
  \BibitemOpen
  \bibfield  {author} {\bibinfo {author} {\bibfnamefont {D.}~\bibnamefont
  {Kroff}}\ and\ \bibinfo {author} {\bibfnamefont {P.~C.}\ \bibnamefont
  {Malta}},\ }\bibfield  {title} {\enquote {\bibinfo {title} {{Constraining
  hidden photons via atomic force microscope measurements and the
  Plimpton-Lawton experiment}},}\ }\href {\doibase 10.1103/PhysRevD.102.095015}
  {\bibfield  {journal} {\bibinfo  {journal} {Phys. Rev. D}\ }\textbf {\bibinfo
  {volume} {102}},\ \bibinfo {pages} {095015} (\bibinfo {year} {2020})},\
  \Eprint {http://arxiv.org/abs/2008.02209} {arXiv:2008.02209 [hep-ph]}
  \BibitemShut {NoStop}%
\bibitem [{\citenamefont {Betz}\ \emph {et~al.}(2013)\citenamefont {Betz},
  \citenamefont {Caspers}, \citenamefont {Gasior}, \citenamefont {Thumm},\ and\
  \citenamefont {Rieger}}]{Betz:2013dza}%
  \BibitemOpen
  \bibfield  {author} {\bibinfo {author} {\bibfnamefont {M.}~\bibnamefont
  {Betz}}, \bibinfo {author} {\bibfnamefont {F.}~\bibnamefont {Caspers}},
  \bibinfo {author} {\bibfnamefont {M.}~\bibnamefont {Gasior}}, \bibinfo
  {author} {\bibfnamefont {M.}~\bibnamefont {Thumm}}, \ and\ \bibinfo {author}
  {\bibfnamefont {S.~W.}\ \bibnamefont {Rieger}},\ }\bibfield  {title}
  {\enquote {\bibinfo {title} {{First results of the CERN Resonant Weakly
  Interacting sub-eV Particle Search (CROWS)}},}\ }\href {\doibase
  10.1103/PhysRevD.88.075014} {\bibfield  {journal} {\bibinfo  {journal} {Phys.
  Rev. D}\ }\textbf {\bibinfo {volume} {88}},\ \bibinfo {pages} {075014}
  (\bibinfo {year} {2013})},\ \Eprint {http://arxiv.org/abs/1310.8098}
  {arXiv:1310.8098 [physics.ins-det]} \BibitemShut {NoStop}%
\bibitem [{\citenamefont {Berlin}\ \emph {et~al.}(2022)\citenamefont {Berlin}
  \emph {et~al.}}]{Berlin:2022hfx}%
  \BibitemOpen
  \bibfield  {author} {\bibinfo {author} {\bibfnamefont {Asher}\ \bibnamefont
  {Berlin}} \emph {et~al.},\ }\bibfield  {title} {\enquote {\bibinfo {title}
  {{Searches for New Particles, Dark Matter, and Gravitational Waves with SRF
  Cavities}},}\ }\href@noop {} {\  (\bibinfo {year} {2022})},\ \Eprint
  {http://arxiv.org/abs/2203.12714} {arXiv:2203.12714 [hep-ph]} \BibitemShut
  {NoStop}%
\bibitem [{\citenamefont {Ortiz}\ \emph {et~al.}(2022)\citenamefont {Ortiz}
  \emph {et~al.}}]{Ortiz:2020tgs}%
  \BibitemOpen
  \bibfield  {author} {\bibinfo {author} {\bibfnamefont {M.~Diaz}\ \bibnamefont
  {Ortiz}} \emph {et~al.},\ }\bibfield  {title} {\enquote {\bibinfo {title}
  {{Design of the ALPS II optical system}},}\ }\href {\doibase
  10.1016/j.dark.2022.100968} {\bibfield  {journal} {\bibinfo  {journal} {Phys.
  Dark Univ.}\ }\textbf {\bibinfo {volume} {35}},\ \bibinfo {pages} {100968}
  (\bibinfo {year} {2022})},\ \Eprint {http://arxiv.org/abs/2009.14294}
  {arXiv:2009.14294 [physics.optics]} \BibitemShut {NoStop}%
\bibitem [{\citenamefont {Miyazaki}\ \emph {et~al.}(2024)\citenamefont
  {Miyazaki}, \citenamefont {Lofnes}, \citenamefont {Caspers}, \citenamefont
  {Spagnolo}, \citenamefont {Jelonnek}, \citenamefont {Ruess}, \citenamefont
  {Steinmann},\ and\ \citenamefont {Thumm}}]{Miyazaki:2022kxl}%
  \BibitemOpen
  \bibfield  {author} {\bibinfo {author} {\bibfnamefont {Akira}\ \bibnamefont
  {Miyazaki}}, \bibinfo {author} {\bibfnamefont {Tor}\ \bibnamefont {Lofnes}},
  \bibinfo {author} {\bibfnamefont {Fritz}\ \bibnamefont {Caspers}}, \bibinfo
  {author} {\bibfnamefont {Paolo}\ \bibnamefont {Spagnolo}}, \bibinfo {author}
  {\bibfnamefont {John}\ \bibnamefont {Jelonnek}}, \bibinfo {author}
  {\bibfnamefont {Tobias}\ \bibnamefont {Ruess}}, \bibinfo {author}
  {\bibfnamefont {Johannes~L.}\ \bibnamefont {Steinmann}}, \ and\ \bibinfo
  {author} {\bibfnamefont {Manfred}\ \bibnamefont {Thumm}},\ }\bibfield
  {title} {\enquote {\bibinfo {title} {{Millimeter-Wave WISP Search with
  Coherent Light-Shining-Through-a-Wall Toward the STAX Project}},}\ }\href
  {\doibase 10.1002/andp.202200619} {\bibfield  {journal} {\bibinfo  {journal}
  {Annalen Phys.}\ }\textbf {\bibinfo {volume} {536}},\ \bibinfo {pages}
  {2200619} (\bibinfo {year} {2024})},\ \Eprint
  {http://arxiv.org/abs/2212.01139} {arXiv:2212.01139 [hep-ph]} \BibitemShut
  {NoStop}%
\bibitem [{\citenamefont {Berlin}\ \emph
  {et~al.}(2023{\natexlab{a}})\citenamefont {Berlin}, \citenamefont {Harnik},\
  and\ \citenamefont {Janish}}]{Berlin:2023mti}%
  \BibitemOpen
  \bibfield  {author} {\bibinfo {author} {\bibfnamefont {Asher}\ \bibnamefont
  {Berlin}}, \bibinfo {author} {\bibfnamefont {Roni}\ \bibnamefont {Harnik}}, \
  and\ \bibinfo {author} {\bibfnamefont {Ryan}\ \bibnamefont {Janish}},\
  }\bibfield  {title} {\enquote {\bibinfo {title} {{Light Shining Through a
  Thin Wall: Evanescent Hidden Photon Detection}},}\ }\href@noop {} {\
  (\bibinfo {year} {2023}{\natexlab{a}})},\ \Eprint
  {http://arxiv.org/abs/2303.00014} {arXiv:2303.00014 [hep-ph]} \BibitemShut
  {NoStop}%
\bibitem [{\citenamefont {Antel}\ \emph {et~al.}(2023)\citenamefont {Antel}
  \emph {et~al.}}]{Antel:2023hkf}%
  \BibitemOpen
  \bibfield  {author} {\bibinfo {author} {\bibfnamefont {C.}~\bibnamefont
  {Antel}} \emph {et~al.},\ }\bibfield  {title} {\enquote {\bibinfo {title}
  {{Feebly-interacting particles: FIPs 2022 Workshop Report}},}\ }\href
  {\doibase 10.1140/epjc/s10052-023-12168-5} {\bibfield  {journal} {\bibinfo
  {journal} {Eur. Phys. J. C}\ }\textbf {\bibinfo {volume} {83}},\ \bibinfo
  {pages} {1122} (\bibinfo {year} {2023})},\ \Eprint
  {http://arxiv.org/abs/2305.01715} {arXiv:2305.01715 [hep-ph]} \BibitemShut
  {NoStop}%
\bibitem [{\citenamefont {Redondo}(2008)}]{Redondo:2008aa}%
  \BibitemOpen
  \bibfield  {author} {\bibinfo {author} {\bibfnamefont {Javier}\ \bibnamefont
  {Redondo}},\ }\bibfield  {title} {\enquote {\bibinfo {title} {{Helioscope
  Bounds on Hidden Sector Photons}},}\ }\href {\doibase
  10.1088/1475-7516/2008/07/008} {\bibfield  {journal} {\bibinfo  {journal}
  {JCAP}\ }\textbf {\bibinfo {volume} {07}},\ \bibinfo {pages} {008} (\bibinfo
  {year} {2008})},\ \Eprint {http://arxiv.org/abs/0801.1527} {arXiv:0801.1527
  [hep-ph]} \BibitemShut {NoStop}%
\bibitem [{\citenamefont {Schwarz}\ \emph {et~al.}(2015)\citenamefont
  {Schwarz}, \citenamefont {Knabbe}, \citenamefont {Lindner}, \citenamefont
  {Redondo}, \citenamefont {Ringwald}, \citenamefont {Schneide}, \citenamefont
  {Susol},\ and\ \citenamefont {Wiedemann}}]{Schwarz:2015lqa}%
  \BibitemOpen
  \bibfield  {author} {\bibinfo {author} {\bibfnamefont {Matthias}\
  \bibnamefont {Schwarz}}, \bibinfo {author} {\bibfnamefont {Ernst-Axel}\
  \bibnamefont {Knabbe}}, \bibinfo {author} {\bibfnamefont {Axel}\ \bibnamefont
  {Lindner}}, \bibinfo {author} {\bibfnamefont {Javier}\ \bibnamefont
  {Redondo}}, \bibinfo {author} {\bibfnamefont {Andreas}\ \bibnamefont
  {Ringwald}}, \bibinfo {author} {\bibfnamefont {Magnus}\ \bibnamefont
  {Schneide}}, \bibinfo {author} {\bibfnamefont {Jaroslaw}\ \bibnamefont
  {Susol}}, \ and\ \bibinfo {author} {\bibfnamefont {G\"unter}\ \bibnamefont
  {Wiedemann}},\ }\bibfield  {title} {\enquote {\bibinfo {title} {{Results from
  the Solar Hidden Photon Search (SHIPS)}},}\ }\href {\doibase
  10.1088/1475-7516/2015/08/011} {\bibfield  {journal} {\bibinfo  {journal}
  {JCAP}\ }\textbf {\bibinfo {volume} {08}},\ \bibinfo {pages} {011} (\bibinfo
  {year} {2015})},\ \Eprint {http://arxiv.org/abs/1502.04490} {arXiv:1502.04490
  [hep-ph]} \BibitemShut {NoStop}%
\bibitem [{\citenamefont {Frerick}\ \emph {et~al.}(2023)\citenamefont
  {Frerick}, \citenamefont {Kahlhoefer},\ and\ \citenamefont
  {Schmidt-Hoberg}}]{Frerick:2022mjg}%
  \BibitemOpen
  \bibfield  {author} {\bibinfo {author} {\bibfnamefont {Jonas}\ \bibnamefont
  {Frerick}}, \bibinfo {author} {\bibfnamefont {Felix}\ \bibnamefont
  {Kahlhoefer}}, \ and\ \bibinfo {author} {\bibfnamefont {Kai}\ \bibnamefont
  {Schmidt-Hoberg}},\ }\bibfield  {title} {\enquote {\bibinfo {title} {{A' view
  of the sunrise: boosting helioscopes with angular information}},}\ }\href
  {\doibase 10.1088/1475-7516/2023/03/001} {\bibfield  {journal} {\bibinfo
  {journal} {JCAP}\ }\textbf {\bibinfo {volume} {03}},\ \bibinfo {pages} {001}
  (\bibinfo {year} {2023})},\ \Eprint {http://arxiv.org/abs/2211.00022}
  {arXiv:2211.00022 [hep-ph]} \BibitemShut {NoStop}%
\bibitem [{\citenamefont {O'Shea}\ \emph {et~al.}(2023)\citenamefont {O'Shea},
  \citenamefont {Giannotti}, \citenamefont {Irastorza}, \citenamefont
  {Plasencia}, \citenamefont {Redondo}, \citenamefont {Ruz},\ and\
  \citenamefont {Vogel}}]{OShea:2023gqn}%
  \BibitemOpen
  \bibfield  {author} {\bibinfo {author} {\bibfnamefont {T.}~\bibnamefont
  {O'Shea}}, \bibinfo {author} {\bibfnamefont {M.}~\bibnamefont {Giannotti}},
  \bibinfo {author} {\bibfnamefont {I.~G.}\ \bibnamefont {Irastorza}}, \bibinfo
  {author} {\bibfnamefont {L.~M.}\ \bibnamefont {Plasencia}}, \bibinfo {author}
  {\bibfnamefont {J.}~\bibnamefont {Redondo}}, \bibinfo {author} {\bibfnamefont
  {J.}~\bibnamefont {Ruz}}, \ and\ \bibinfo {author} {\bibfnamefont {J.~K.}\
  \bibnamefont {Vogel}},\ }\bibfield  {title} {\enquote {\bibinfo {title}
  {{Prospects on the Detection of Solar Dark Photons by the International Axion
  Observatory}},}\ }\href@noop {} {\  (\bibinfo {year} {2023})},\ \Eprint
  {http://arxiv.org/abs/2312.10150} {arXiv:2312.10150 [hep-ph]} \BibitemShut
  {NoStop}%
\bibitem [{\citenamefont {An}\ \emph {et~al.}(2015)\citenamefont {An},
  \citenamefont {Pospelov}, \citenamefont {Pradler},\ and\ \citenamefont
  {Ritz}}]{An:2014twa}%
  \BibitemOpen
  \bibfield  {author} {\bibinfo {author} {\bibfnamefont {Haipeng}\ \bibnamefont
  {An}}, \bibinfo {author} {\bibfnamefont {Maxim}\ \bibnamefont {Pospelov}},
  \bibinfo {author} {\bibfnamefont {Josef}\ \bibnamefont {Pradler}}, \ and\
  \bibinfo {author} {\bibfnamefont {Adam}\ \bibnamefont {Ritz}},\ }\bibfield
  {title} {\enquote {\bibinfo {title} {{Direct Detection Constraints on Dark
  Photon Dark Matter}},}\ }\href {\doibase 10.1016/j.physletb.2015.06.018}
  {\bibfield  {journal} {\bibinfo  {journal} {Phys. Lett. B}\ }\textbf
  {\bibinfo {volume} {747}},\ \bibinfo {pages} {331--338} (\bibinfo {year}
  {2015})},\ \Eprint {http://arxiv.org/abs/1412.8378} {arXiv:1412.8378
  [hep-ph]} \BibitemShut {NoStop}%
\bibitem [{\citenamefont {An}\ \emph {et~al.}(2020)\citenamefont {An},
  \citenamefont {Pospelov}, \citenamefont {Pradler},\ and\ \citenamefont
  {Ritz}}]{An:2020bxd}%
  \BibitemOpen
  \bibfield  {author} {\bibinfo {author} {\bibfnamefont {Haipeng}\ \bibnamefont
  {An}}, \bibinfo {author} {\bibfnamefont {Maxim}\ \bibnamefont {Pospelov}},
  \bibinfo {author} {\bibfnamefont {Josef}\ \bibnamefont {Pradler}}, \ and\
  \bibinfo {author} {\bibfnamefont {Adam}\ \bibnamefont {Ritz}},\ }\bibfield
  {title} {\enquote {\bibinfo {title} {{New limits on dark photons from solar
  emission and keV scale dark matter}},}\ }\href {\doibase
  10.1103/PhysRevD.102.115022} {\bibfield  {journal} {\bibinfo  {journal}
  {Phys. Rev. D}\ }\textbf {\bibinfo {volume} {102}},\ \bibinfo {pages}
  {115022} (\bibinfo {year} {2020})},\ \Eprint
  {http://arxiv.org/abs/2006.13929} {arXiv:2006.13929 [hep-ph]} \BibitemShut
  {NoStop}%
\bibitem [{\citenamefont {Lasenby}\ and\ \citenamefont
  {Van~Tilburg}(2021)}]{Lasenby:2020goo}%
  \BibitemOpen
  \bibfield  {author} {\bibinfo {author} {\bibfnamefont {Robert}\ \bibnamefont
  {Lasenby}}\ and\ \bibinfo {author} {\bibfnamefont {Ken}\ \bibnamefont
  {Van~Tilburg}},\ }\bibfield  {title} {\enquote {\bibinfo {title} {{Dark
  photons in the solar basin}},}\ }\href {\doibase 10.1103/PhysRevD.104.023020}
  {\bibfield  {journal} {\bibinfo  {journal} {Phys. Rev. D}\ }\textbf {\bibinfo
  {volume} {104}},\ \bibinfo {pages} {023020} (\bibinfo {year} {2021})},\
  \Eprint {http://arxiv.org/abs/2008.08594} {arXiv:2008.08594 [hep-ph]}
  \BibitemShut {NoStop}%
\bibitem [{\citenamefont {Aprile}\ \emph {et~al.}(2022)\citenamefont {Aprile}
  \emph {et~al.}}]{XENON:2021qze}%
  \BibitemOpen
  \bibfield  {author} {\bibinfo {author} {\bibfnamefont {E.}~\bibnamefont
  {Aprile}} \emph {et~al.} (\bibinfo {collaboration} {XENON}),\ }\bibfield
  {title} {\enquote {\bibinfo {title} {{Emission of single and few electrons in
  XENON1T and limits on light dark matter}},}\ }\href {\doibase
  10.1103/PhysRevD.106.022001} {\bibfield  {journal} {\bibinfo  {journal}
  {Phys. Rev. D}\ }\textbf {\bibinfo {volume} {106}},\ \bibinfo {pages}
  {022001} (\bibinfo {year} {2022})},\ \Eprint
  {http://arxiv.org/abs/2112.12116} {arXiv:2112.12116 [hep-ex]} \BibitemShut
  {NoStop}%
\bibitem [{\citenamefont {Pospelov}\ \emph {et~al.}(2008)\citenamefont
  {Pospelov}, \citenamefont {Ritz},\ and\ \citenamefont
  {Voloshin}}]{Pospelov:2008jk}%
  \BibitemOpen
  \bibfield  {author} {\bibinfo {author} {\bibfnamefont {Maxim}\ \bibnamefont
  {Pospelov}}, \bibinfo {author} {\bibfnamefont {Adam}\ \bibnamefont {Ritz}}, \
  and\ \bibinfo {author} {\bibfnamefont {Mikhail~B.}\ \bibnamefont
  {Voloshin}},\ }\bibfield  {title} {\enquote {\bibinfo {title} {{Bosonic
  super-WIMPs as keV-scale dark matter}},}\ }\href {\doibase
  10.1103/PhysRevD.78.115012} {\bibfield  {journal} {\bibinfo  {journal} {Phys.
  Rev. D}\ }\textbf {\bibinfo {volume} {78}},\ \bibinfo {pages} {115012}
  (\bibinfo {year} {2008})},\ \Eprint {http://arxiv.org/abs/0807.3279}
  {arXiv:0807.3279 [hep-ph]} \BibitemShut {NoStop}%
\bibitem [{\citenamefont {Redondo}\ and\ \citenamefont
  {Postma}(2009)}]{Redondo:2008ec}%
  \BibitemOpen
  \bibfield  {author} {\bibinfo {author} {\bibfnamefont {Javier}\ \bibnamefont
  {Redondo}}\ and\ \bibinfo {author} {\bibfnamefont {Marieke}\ \bibnamefont
  {Postma}},\ }\bibfield  {title} {\enquote {\bibinfo {title} {{Massive hidden
  photons as lukewarm dark matter}},}\ }\href {\doibase
  10.1088/1475-7516/2009/02/005} {\bibfield  {journal} {\bibinfo  {journal}
  {JCAP}\ }\textbf {\bibinfo {volume} {02}},\ \bibinfo {pages} {005} (\bibinfo
  {year} {2009})},\ \Eprint {http://arxiv.org/abs/0811.0326} {arXiv:0811.0326
  [hep-ph]} \BibitemShut {NoStop}%
\bibitem [{\citenamefont {Redondo}\ and\ \citenamefont
  {Raffelt}(2013)}]{Redondo:2013lna}%
  \BibitemOpen
  \bibfield  {author} {\bibinfo {author} {\bibfnamefont {Javier}\ \bibnamefont
  {Redondo}}\ and\ \bibinfo {author} {\bibfnamefont {Georg}\ \bibnamefont
  {Raffelt}},\ }\bibfield  {title} {\enquote {\bibinfo {title} {{Solar
  constraints on hidden photons re-visited}},}\ }\href {\doibase
  10.1088/1475-7516/2013/08/034} {\bibfield  {journal} {\bibinfo  {journal}
  {JCAP}\ }\textbf {\bibinfo {volume} {08}},\ \bibinfo {pages} {034} (\bibinfo
  {year} {2013})},\ \Eprint {http://arxiv.org/abs/1305.2920} {arXiv:1305.2920
  [hep-ph]} \BibitemShut {NoStop}%
\bibitem [{\citenamefont {An}\ \emph {et~al.}(2013)\citenamefont {An},
  \citenamefont {Pospelov},\ and\ \citenamefont {Pradler}}]{An:2013yfc}%
  \BibitemOpen
  \bibfield  {author} {\bibinfo {author} {\bibfnamefont {Haipeng}\ \bibnamefont
  {An}}, \bibinfo {author} {\bibfnamefont {Maxim}\ \bibnamefont {Pospelov}}, \
  and\ \bibinfo {author} {\bibfnamefont {Josef}\ \bibnamefont {Pradler}},\
  }\bibfield  {title} {\enquote {\bibinfo {title} {{New stellar constraints on
  dark photons}},}\ }\href {\doibase 10.1016/j.physletb.2013.07.008} {\bibfield
   {journal} {\bibinfo  {journal} {Phys. Lett. B}\ }\textbf {\bibinfo {volume}
  {725}},\ \bibinfo {pages} {190--195} (\bibinfo {year} {2013})},\ \Eprint
  {http://arxiv.org/abs/1302.3884} {arXiv:1302.3884 [hep-ph]} \BibitemShut
  {NoStop}%
\bibitem [{\citenamefont {Vinyoles}\ \emph {et~al.}(2015)\citenamefont
  {Vinyoles}, \citenamefont {Serenelli}, \citenamefont {Villante},
  \citenamefont {Basu}, \citenamefont {Redondo},\ and\ \citenamefont
  {Isern}}]{Vinyoles:2015mhi}%
  \BibitemOpen
  \bibfield  {author} {\bibinfo {author} {\bibfnamefont {Nuria}\ \bibnamefont
  {Vinyoles}}, \bibinfo {author} {\bibfnamefont {Aldo}\ \bibnamefont
  {Serenelli}}, \bibinfo {author} {\bibfnamefont {Francesco}\ \bibnamefont
  {Villante}}, \bibinfo {author} {\bibfnamefont {Sarbani}\ \bibnamefont
  {Basu}}, \bibinfo {author} {\bibfnamefont {Javier}\ \bibnamefont {Redondo}},
  \ and\ \bibinfo {author} {\bibfnamefont {Jordi}\ \bibnamefont {Isern}},\
  }\bibfield  {title} {\enquote {\bibinfo {title} {{New Axion and Hidden Photon
  Constraints from a Solar Data Global Fit}},}\ }in\ \href {\doibase
  10.3204/DESY-PROC-2015-02/vinyoles_nuria} {\emph {\bibinfo {booktitle} {{11th
  Patras Workshop on Axions, WIMPs and WISPs}}}}\ (\bibinfo {year} {2015})\
  pp.\ \bibinfo {pages} {92--95}\BibitemShut {NoStop}%
\bibitem [{\citenamefont {Redondo}(2015)}]{Redondo:2015iea}%
  \BibitemOpen
  \bibfield  {author} {\bibinfo {author} {\bibfnamefont {Javier}\ \bibnamefont
  {Redondo}},\ }\bibfield  {title} {\enquote {\bibinfo {title} {{Atlas of solar
  hidden photon emission}},}\ }\href {\doibase 10.1088/1475-7516/2015/07/024}
  {\bibfield  {journal} {\bibinfo  {journal} {JCAP}\ }\textbf {\bibinfo
  {volume} {07}},\ \bibinfo {pages} {024} (\bibinfo {year} {2015})},\ \Eprint
  {http://arxiv.org/abs/1501.07292} {arXiv:1501.07292 [hep-ph]} \BibitemShut
  {NoStop}%
\bibitem [{\citenamefont {Hardy}\ and\ \citenamefont
  {Lasenby}(2017)}]{Hardy:2016kme}%
  \BibitemOpen
  \bibfield  {author} {\bibinfo {author} {\bibfnamefont {Edward}\ \bibnamefont
  {Hardy}}\ and\ \bibinfo {author} {\bibfnamefont {Robert}\ \bibnamefont
  {Lasenby}},\ }\bibfield  {title} {\enquote {\bibinfo {title} {{Stellar
  cooling bounds on new light particles: plasma mixing effects}},}\ }\href
  {\doibase 10.1007/JHEP02(2017)033} {\bibfield  {journal} {\bibinfo  {journal}
  {JHEP}\ }\textbf {\bibinfo {volume} {02}},\ \bibinfo {pages} {033} (\bibinfo
  {year} {2017})},\ \Eprint {http://arxiv.org/abs/1611.05852} {arXiv:1611.05852
  [hep-ph]} \BibitemShut {NoStop}%
\bibitem [{\citenamefont {Li}\ and\ \citenamefont {Xu}(2023)}]{Li:2023vpv}%
  \BibitemOpen
  \bibfield  {author} {\bibinfo {author} {\bibfnamefont {Shao-Ping}\
  \bibnamefont {Li}}\ and\ \bibinfo {author} {\bibfnamefont {Xun-Jie}\
  \bibnamefont {Xu}},\ }\bibfield  {title} {\enquote {\bibinfo {title}
  {{Production rates of dark photons and Z' in the Sun and stellar cooling
  bounds}},}\ }\href {\doibase 10.1088/1475-7516/2023/09/009} {\bibfield
  {journal} {\bibinfo  {journal} {JCAP}\ }\textbf {\bibinfo {volume} {09}},\
  \bibinfo {pages} {009} (\bibinfo {year} {2023})},\ \Eprint
  {http://arxiv.org/abs/2304.12907} {arXiv:2304.12907 [hep-ph]} \BibitemShut
  {NoStop}%
\bibitem [{\citenamefont {Cardoso}\ \emph {et~al.}(2018)\citenamefont
  {Cardoso}, \citenamefont {Dias}, \citenamefont {Hartnett}, \citenamefont
  {Middleton}, \citenamefont {Pani},\ and\ \citenamefont
  {Santos}}]{Cardoso:2018tly}%
  \BibitemOpen
  \bibfield  {author} {\bibinfo {author} {\bibfnamefont {Vitor}\ \bibnamefont
  {Cardoso}}, \bibinfo {author} {\bibfnamefont {\'Oscar J.~C.}\ \bibnamefont
  {Dias}}, \bibinfo {author} {\bibfnamefont {Gavin~S.}\ \bibnamefont
  {Hartnett}}, \bibinfo {author} {\bibfnamefont {Matthew}\ \bibnamefont
  {Middleton}}, \bibinfo {author} {\bibfnamefont {Paolo}\ \bibnamefont {Pani}},
  \ and\ \bibinfo {author} {\bibfnamefont {Jorge~E.}\ \bibnamefont {Santos}},\
  }\bibfield  {title} {\enquote {\bibinfo {title} {{Constraining the mass of
  dark photons and axion-like particles through black-hole superradiance}},}\
  }\href {\doibase 10.1088/1475-7516/2018/03/043} {\bibfield  {journal}
  {\bibinfo  {journal} {JCAP}\ }\textbf {\bibinfo {volume} {03}},\ \bibinfo
  {pages} {043} (\bibinfo {year} {2018})},\ \Eprint
  {http://arxiv.org/abs/1801.01420} {arXiv:1801.01420 [gr-qc]} \BibitemShut
  {NoStop}%
\bibitem [{\citenamefont {Davoudiasl}\ and\ \citenamefont
  {Denton}(2019)}]{Davoudiasl:2019nlo}%
  \BibitemOpen
  \bibfield  {author} {\bibinfo {author} {\bibfnamefont {Hooman}\ \bibnamefont
  {Davoudiasl}}\ and\ \bibinfo {author} {\bibfnamefont {Peter~B}\ \bibnamefont
  {Denton}},\ }\bibfield  {title} {\enquote {\bibinfo {title} {{Ultralight
  Boson Dark Matter and Event Horizon Telescope Observations of M87*}},}\
  }\href {\doibase 10.1103/PhysRevLett.123.021102} {\bibfield  {journal}
  {\bibinfo  {journal} {Phys. Rev. Lett.}\ }\textbf {\bibinfo {volume} {123}},\
  \bibinfo {pages} {021102} (\bibinfo {year} {2019})},\ \Eprint
  {http://arxiv.org/abs/1904.09242} {arXiv:1904.09242 [astro-ph.CO]}
  \BibitemShut {NoStop}%
\bibitem [{\citenamefont {\"Unal}\ \emph {et~al.}(2021)\citenamefont {\"Unal},
  \citenamefont {Pacucci},\ and\ \citenamefont {Loeb}}]{Unal:2020jiy}%
  \BibitemOpen
  \bibfield  {author} {\bibinfo {author} {\bibfnamefont {Caner}\ \bibnamefont
  {\"Unal}}, \bibinfo {author} {\bibfnamefont {Fabio}\ \bibnamefont {Pacucci}},
  \ and\ \bibinfo {author} {\bibfnamefont {Abraham}\ \bibnamefont {Loeb}},\
  }\bibfield  {title} {\enquote {\bibinfo {title} {{Properties of ultralight
  bosons from heavy quasar spins via superradiance}},}\ }\href {\doibase
  10.1088/1475-7516/2021/05/007} {\bibfield  {journal} {\bibinfo  {journal}
  {JCAP}\ }\textbf {\bibinfo {volume} {05}},\ \bibinfo {pages} {007} (\bibinfo
  {year} {2021})},\ \Eprint {http://arxiv.org/abs/2012.12790} {arXiv:2012.12790
  [hep-ph]} \BibitemShut {NoStop}%
\bibitem [{\citenamefont {Siemonsen}\ \emph {et~al.}(2023)\citenamefont
  {Siemonsen}, \citenamefont {Mondino}, \citenamefont {Egana-Ugrinovic},
  \citenamefont {Huang}, \citenamefont {Baryakhtar},\ and\ \citenamefont
  {East}}]{Siemonsen:2022ivj}%
  \BibitemOpen
  \bibfield  {author} {\bibinfo {author} {\bibfnamefont {Nils}\ \bibnamefont
  {Siemonsen}}, \bibinfo {author} {\bibfnamefont {Cristina}\ \bibnamefont
  {Mondino}}, \bibinfo {author} {\bibfnamefont {Daniel}\ \bibnamefont
  {Egana-Ugrinovic}}, \bibinfo {author} {\bibfnamefont {Junwu}\ \bibnamefont
  {Huang}}, \bibinfo {author} {\bibfnamefont {Masha}\ \bibnamefont
  {Baryakhtar}}, \ and\ \bibinfo {author} {\bibfnamefont {William~E.}\
  \bibnamefont {East}},\ }\bibfield  {title} {\enquote {\bibinfo {title} {{Dark
  photon superradiance: Electrodynamics and multimessenger signals}},}\ }\href
  {\doibase 10.1103/PhysRevD.107.075025} {\bibfield  {journal} {\bibinfo
  {journal} {Phys. Rev. D}\ }\textbf {\bibinfo {volume} {107}},\ \bibinfo
  {pages} {075025} (\bibinfo {year} {2023})},\ \Eprint
  {http://arxiv.org/abs/2212.09772} {arXiv:2212.09772 [astro-ph.HE]}
  \BibitemShut {NoStop}%
\bibitem [{\citenamefont {Davis}\ \emph {et~al.}(1975)\citenamefont {Davis},
  \citenamefont {Goldhaber},\ and\ \citenamefont {Nieto}}]{Davis:1975mn}%
  \BibitemOpen
  \bibfield  {author} {\bibinfo {author} {\bibfnamefont {Leverett}\
  \bibnamefont {Davis}, \bibfnamefont {Jr.}}, \bibinfo {author} {\bibfnamefont
  {Alfred~S.}\ \bibnamefont {Goldhaber}}, \ and\ \bibinfo {author}
  {\bibfnamefont {Michael~Martin}\ \bibnamefont {Nieto}},\ }\bibfield  {title}
  {\enquote {\bibinfo {title} {{Limit on the photon mass deduced from
  Pioneer-10 observations of Jupiter's magnetic field}},}\ }\href {\doibase
  10.1103/PhysRevLett.35.1402} {\bibfield  {journal} {\bibinfo  {journal}
  {Phys. Rev. Lett.}\ }\textbf {\bibinfo {volume} {35}},\ \bibinfo {pages}
  {1402--1405} (\bibinfo {year} {1975})}\BibitemShut {NoStop}%
\bibitem [{\citenamefont {Marocco}(2021)}]{Marocco:2021dku}%
  \BibitemOpen
  \bibfield  {author} {\bibinfo {author} {\bibfnamefont {Giacomo}\ \bibnamefont
  {Marocco}},\ }\bibfield  {title} {\enquote {\bibinfo {title} {{Dark photon
  limits from magnetic fields and astrophysical plasmas}},}\ }\href@noop {} {\
  (\bibinfo {year} {2021})},\ \Eprint {http://arxiv.org/abs/2110.02875}
  {arXiv:2110.02875 [hep-ph]} \BibitemShut {NoStop}%
\bibitem [{\citenamefont {Yan}\ \emph {et~al.}(2023)\citenamefont {Yan},
  \citenamefont {Li},\ and\ \citenamefont {Fan}}]{Yan:2023kdg}%
  \BibitemOpen
  \bibfield  {author} {\bibinfo {author} {\bibfnamefont {Shi}\ \bibnamefont
  {Yan}}, \bibinfo {author} {\bibfnamefont {Lingfeng}\ \bibnamefont {Li}}, \
  and\ \bibinfo {author} {\bibfnamefont {JiJi}\ \bibnamefont {Fan}},\
  }\bibfield  {title} {\enquote {\bibinfo {title} {{Constraints on photon mass
  and dark photon from the Jovian magnetic field}},}\ }\href@noop {} {\
  (\bibinfo {year} {2023})},\ \Eprint {http://arxiv.org/abs/2312.06746}
  {arXiv:2312.06746 [hep-ph]} \BibitemShut {NoStop}%
\bibitem [{\citenamefont {Goldhaber}\ and\ \citenamefont
  {Nieto}(1971)}]{Goldhaber:1971mr}%
  \BibitemOpen
  \bibfield  {author} {\bibinfo {author} {\bibfnamefont {Alfred~S.}\
  \bibnamefont {Goldhaber}}\ and\ \bibinfo {author} {\bibfnamefont
  {Michael~Martin}\ \bibnamefont {Nieto}},\ }\bibfield  {title} {\enquote
  {\bibinfo {title} {{Terrestrial and extra-terrestrial limits on the photon
  mass}},}\ }\href {\doibase 10.1103/RevModPhys.43.277} {\bibfield  {journal}
  {\bibinfo  {journal} {Rev. Mod. Phys.}\ }\textbf {\bibinfo {volume} {43}},\
  \bibinfo {pages} {277--296} (\bibinfo {year} {1971})}\BibitemShut {NoStop}%
\bibitem [{\citenamefont {Fischbach}\ \emph {et~al.}(1994)\citenamefont
  {Fischbach}, \citenamefont {Kloor}, \citenamefont {Langel}, \citenamefont
  {Liu},\ and\ \citenamefont {Peredo}}]{Fischbach:1994ir}%
  \BibitemOpen
  \bibfield  {author} {\bibinfo {author} {\bibfnamefont {E.}~\bibnamefont
  {Fischbach}}, \bibinfo {author} {\bibfnamefont {H.}~\bibnamefont {Kloor}},
  \bibinfo {author} {\bibfnamefont {R.~A.}\ \bibnamefont {Langel}}, \bibinfo
  {author} {\bibfnamefont {A.~T.~Y.}\ \bibnamefont {Liu}}, \ and\ \bibinfo
  {author} {\bibfnamefont {M.}~\bibnamefont {Peredo}},\ }\bibfield  {title}
  {\enquote {\bibinfo {title} {{New geomagnetic limits on the photon mass and
  on long range forces coexisting with electromagnetism}},}\ }\href {\doibase
  10.1103/PhysRevLett.73.514} {\bibfield  {journal} {\bibinfo  {journal} {Phys.
  Rev. Lett.}\ }\textbf {\bibinfo {volume} {73}},\ \bibinfo {pages} {514--517}
  (\bibinfo {year} {1994})}\BibitemShut {NoStop}%
\bibitem [{\citenamefont {Kloor}\ \emph {et~al.}(1994)\citenamefont {Kloor},
  \citenamefont {Fischbach}, \citenamefont {Talmadge},\ and\ \citenamefont
  {Greene}}]{Kloor:1994xm}%
  \BibitemOpen
  \bibfield  {author} {\bibinfo {author} {\bibfnamefont {H.}~\bibnamefont
  {Kloor}}, \bibinfo {author} {\bibfnamefont {E.}~\bibnamefont {Fischbach}},
  \bibinfo {author} {\bibfnamefont {C.}~\bibnamefont {Talmadge}}, \ and\
  \bibinfo {author} {\bibfnamefont {G.~L.}\ \bibnamefont {Greene}},\ }\bibfield
   {title} {\enquote {\bibinfo {title} {{Limits on new forces coexisting with
  electromagnetism}},}\ }\href {\doibase 10.1103/PhysRevD.49.2098} {\bibfield
  {journal} {\bibinfo  {journal} {Phys. Rev. D}\ }\textbf {\bibinfo {volume}
  {49}},\ \bibinfo {pages} {2098--2113} (\bibinfo {year} {1994})}\BibitemShut
  {NoStop}%
\bibitem [{\citenamefont {Caputo}\ \emph {et~al.}(2021)\citenamefont {Caputo},
  \citenamefont {Millar}, \citenamefont {O'Hare},\ and\ \citenamefont
  {Vitagliano}}]{Caputo:2021eaa}%
  \BibitemOpen
  \bibfield  {author} {\bibinfo {author} {\bibfnamefont {Andrea}\ \bibnamefont
  {Caputo}}, \bibinfo {author} {\bibfnamefont {Alexander~J.}\ \bibnamefont
  {Millar}}, \bibinfo {author} {\bibfnamefont {Ciaran A.~J.}\ \bibnamefont
  {O'Hare}}, \ and\ \bibinfo {author} {\bibfnamefont {Edoardo}\ \bibnamefont
  {Vitagliano}},\ }\bibfield  {title} {\enquote {\bibinfo {title} {{Dark photon
  limits: A handbook}},}\ }\href {\doibase 10.1103/PhysRevD.104.095029}
  {\bibfield  {journal} {\bibinfo  {journal} {Phys. Rev. D}\ }\textbf {\bibinfo
  {volume} {104}},\ \bibinfo {pages} {095029} (\bibinfo {year} {2021})},\
  \Eprint {http://arxiv.org/abs/2105.04565} {arXiv:2105.04565 [hep-ph]}
  \BibitemShut {NoStop}%
\bibitem [{\citenamefont {Holdom}(1986)}]{Holdom:1985ag}%
  \BibitemOpen
  \bibfield  {author} {\bibinfo {author} {\bibfnamefont {Bob}\ \bibnamefont
  {Holdom}},\ }\bibfield  {title} {\enquote {\bibinfo {title} {{Two U(1)'s and
  Epsilon Charge Shifts}},}\ }\href {\doibase 10.1016/0370-2693(86)91377-8}
  {\bibfield  {journal} {\bibinfo  {journal} {Phys. Lett. B}\ }\textbf
  {\bibinfo {volume} {166}},\ \bibinfo {pages} {196--198} (\bibinfo {year}
  {1986})}\BibitemShut {NoStop}%
\bibitem [{\citenamefont {Mirizzi}\ \emph
  {et~al.}(2009{\natexlab{a}})\citenamefont {Mirizzi}, \citenamefont
  {Redondo},\ and\ \citenamefont {Sigl}}]{Mirizzi:2009iz}%
  \BibitemOpen
  \bibfield  {author} {\bibinfo {author} {\bibfnamefont {Alessandro}\
  \bibnamefont {Mirizzi}}, \bibinfo {author} {\bibfnamefont {Javier}\
  \bibnamefont {Redondo}}, \ and\ \bibinfo {author} {\bibfnamefont {Gunter}\
  \bibnamefont {Sigl}},\ }\bibfield  {title} {\enquote {\bibinfo {title}
  {{Microwave Background Constraints on Mixing of Photons with Hidden
  Photons}},}\ }\href {\doibase 10.1088/1475-7516/2009/03/026} {\bibfield
  {journal} {\bibinfo  {journal} {JCAP}\ }\textbf {\bibinfo {volume} {03}},\
  \bibinfo {pages} {026} (\bibinfo {year} {2009}{\natexlab{a}})},\ \Eprint
  {http://arxiv.org/abs/0901.0014} {arXiv:0901.0014 [hep-ph]} \BibitemShut
  {NoStop}%
\bibitem [{\citenamefont {Mirizzi}\ \emph
  {et~al.}(2009{\natexlab{b}})\citenamefont {Mirizzi}, \citenamefont
  {Redondo},\ and\ \citenamefont {Sigl}}]{Mirizzi:2009nq}%
  \BibitemOpen
  \bibfield  {author} {\bibinfo {author} {\bibfnamefont {Alessandro}\
  \bibnamefont {Mirizzi}}, \bibinfo {author} {\bibfnamefont {Javier}\
  \bibnamefont {Redondo}}, \ and\ \bibinfo {author} {\bibfnamefont {Gunter}\
  \bibnamefont {Sigl}},\ }\bibfield  {title} {\enquote {\bibinfo {title}
  {{Constraining resonant photon-axion conversions in the Early Universe}},}\
  }\href {\doibase 10.1088/1475-7516/2009/08/001} {\bibfield  {journal}
  {\bibinfo  {journal} {JCAP}\ }\textbf {\bibinfo {volume} {08}},\ \bibinfo
  {pages} {001} (\bibinfo {year} {2009}{\natexlab{b}})},\ \Eprint
  {http://arxiv.org/abs/0905.4865} {arXiv:0905.4865 [hep-ph]} \BibitemShut
  {NoStop}%
\bibitem [{\citenamefont {Kunze}\ and\ \citenamefont
  {V\'azquez-Mozo}(2015)}]{Kunze:2015noa}%
  \BibitemOpen
  \bibfield  {author} {\bibinfo {author} {\bibfnamefont {Kerstin~E.}\
  \bibnamefont {Kunze}}\ and\ \bibinfo {author} {\bibfnamefont {Miguel~\'A.}\
  \bibnamefont {V\'azquez-Mozo}},\ }\bibfield  {title} {\enquote {\bibinfo
  {title} {{Constraints on hidden photons from current and future observations
  of CMB spectral distortions}},}\ }\href {\doibase
  10.1088/1475-7516/2015/12/028} {\bibfield  {journal} {\bibinfo  {journal}
  {JCAP}\ }\textbf {\bibinfo {volume} {12}},\ \bibinfo {pages} {028} (\bibinfo
  {year} {2015})},\ \Eprint {http://arxiv.org/abs/1507.02614} {arXiv:1507.02614
  [astro-ph.CO]} \BibitemShut {NoStop}%
\bibitem [{\citenamefont {McDermott}\ and\ \citenamefont
  {Witte}(2020)}]{McDermott:2019lch}%
  \BibitemOpen
  \bibfield  {author} {\bibinfo {author} {\bibfnamefont {Samuel~D.}\
  \bibnamefont {McDermott}}\ and\ \bibinfo {author} {\bibfnamefont {Samuel~J.}\
  \bibnamefont {Witte}},\ }\bibfield  {title} {\enquote {\bibinfo {title}
  {{Cosmological evolution of light dark photon dark matter}},}\ }\href
  {\doibase 10.1103/PhysRevD.101.063030} {\bibfield  {journal} {\bibinfo
  {journal} {Phys. Rev. D}\ }\textbf {\bibinfo {volume} {101}},\ \bibinfo
  {pages} {063030} (\bibinfo {year} {2020})},\ \Eprint
  {http://arxiv.org/abs/1911.05086} {arXiv:1911.05086 [hep-ph]} \BibitemShut
  {NoStop}%
\bibitem [{\citenamefont {Caputo}\ \emph
  {et~al.}(2020{\natexlab{a}})\citenamefont {Caputo}, \citenamefont {Liu},
  \citenamefont {Mishra-Sharma},\ and\ \citenamefont
  {Ruderman}}]{Caputo:2020bdy}%
  \BibitemOpen
  \bibfield  {author} {\bibinfo {author} {\bibfnamefont {Andrea}\ \bibnamefont
  {Caputo}}, \bibinfo {author} {\bibfnamefont {Hongwan}\ \bibnamefont {Liu}},
  \bibinfo {author} {\bibfnamefont {Siddharth}\ \bibnamefont {Mishra-Sharma}},
  \ and\ \bibinfo {author} {\bibfnamefont {Joshua~T.}\ \bibnamefont
  {Ruderman}},\ }\bibfield  {title} {\enquote {\bibinfo {title} {{Dark Photon
  Oscillations in Our Inhomogeneous Universe}},}\ }\href {\doibase
  10.1103/PhysRevLett.125.221303} {\bibfield  {journal} {\bibinfo  {journal}
  {Phys. Rev. Lett.}\ }\textbf {\bibinfo {volume} {125}},\ \bibinfo {pages}
  {221303} (\bibinfo {year} {2020}{\natexlab{a}})},\ \Eprint
  {http://arxiv.org/abs/2002.05165} {arXiv:2002.05165 [astro-ph.CO]}
  \BibitemShut {NoStop}%
\bibitem [{\citenamefont {Caputo}\ \emph
  {et~al.}(2020{\natexlab{b}})\citenamefont {Caputo}, \citenamefont {Liu},
  \citenamefont {Mishra-Sharma},\ and\ \citenamefont
  {Ruderman}}]{Caputo:2020rnx}%
  \BibitemOpen
  \bibfield  {author} {\bibinfo {author} {\bibfnamefont {Andrea}\ \bibnamefont
  {Caputo}}, \bibinfo {author} {\bibfnamefont {Hongwan}\ \bibnamefont {Liu}},
  \bibinfo {author} {\bibfnamefont {Siddharth}\ \bibnamefont {Mishra-Sharma}},
  \ and\ \bibinfo {author} {\bibfnamefont {Joshua~T.}\ \bibnamefont
  {Ruderman}},\ }\bibfield  {title} {\enquote {\bibinfo {title} {{Modeling Dark
  Photon Oscillations in Our Inhomogeneous Universe}},}\ }\href {\doibase
  10.1103/PhysRevD.102.103533} {\bibfield  {journal} {\bibinfo  {journal}
  {Phys. Rev. D}\ }\textbf {\bibinfo {volume} {102}},\ \bibinfo {pages}
  {103533} (\bibinfo {year} {2020}{\natexlab{b}})},\ \Eprint
  {http://arxiv.org/abs/2004.06733} {arXiv:2004.06733 [astro-ph.CO]}
  \BibitemShut {NoStop}%
\bibitem [{\citenamefont {Garcia}\ \emph {et~al.}(2020)\citenamefont {Garcia},
  \citenamefont {Bondarenko}, \citenamefont {Ploeckinger}, \citenamefont
  {Pradler},\ and\ \citenamefont {Sokolenko}}]{Garcia:2020qrp}%
  \BibitemOpen
  \bibfield  {author} {\bibinfo {author} {\bibfnamefont {Andres~Aramburo}\
  \bibnamefont {Garcia}}, \bibinfo {author} {\bibfnamefont {Kyrylo}\
  \bibnamefont {Bondarenko}}, \bibinfo {author} {\bibfnamefont {Sylvia}\
  \bibnamefont {Ploeckinger}}, \bibinfo {author} {\bibfnamefont {Josef}\
  \bibnamefont {Pradler}}, \ and\ \bibinfo {author} {\bibfnamefont {Anastasia}\
  \bibnamefont {Sokolenko}},\ }\bibfield  {title} {\enquote {\bibinfo {title}
  {{Effective photon mass and (dark) photon conversion in the inhomogeneous
  Universe}},}\ }\href {\doibase 10.1088/1475-7516/2020/10/011} {\bibfield
  {journal} {\bibinfo  {journal} {JCAP}\ }\textbf {\bibinfo {volume} {10}},\
  \bibinfo {pages} {011} (\bibinfo {year} {2020})},\ \Eprint
  {http://arxiv.org/abs/2003.10465} {arXiv:2003.10465 [astro-ph.CO]}
  \BibitemShut {NoStop}%
\bibitem [{\citenamefont {P\^{i}rvu}\ \emph {et~al.}(2024)\citenamefont
  {P\^{i}rvu}, \citenamefont {Huang},\ and\ \citenamefont
  {Johnson}}]{Pirvu:2023lch}%
  \BibitemOpen
  \bibfield  {author} {\bibinfo {author} {\bibfnamefont {Dalila}\ \bibnamefont
  {P\^{i}rvu}}, \bibinfo {author} {\bibfnamefont {Junwu}\ \bibnamefont
  {Huang}}, \ and\ \bibinfo {author} {\bibfnamefont {Matthew~C.}\ \bibnamefont
  {Johnson}},\ }\bibfield  {title} {\enquote {\bibinfo {title} {{Patchy
  screening of the CMB from dark photons}},}\ }\href {\doibase
  10.1088/1475-7516/2024/01/019} {\bibfield  {journal} {\bibinfo  {journal}
  {JCAP}\ }\textbf {\bibinfo {volume} {01}},\ \bibinfo {pages} {019} (\bibinfo
  {year} {2024})},\ \Eprint {http://arxiv.org/abs/2307.15124} {arXiv:2307.15124
  [hep-ph]} \BibitemShut {NoStop}%
\bibitem [{\citenamefont {Aramburo-Garcia}\ \emph {et~al.}(2024)\citenamefont
  {Aramburo-Garcia}, \citenamefont {Bondarenko}, \citenamefont {Boyarsky},
  \citenamefont {Kashko}, \citenamefont {Pradler}, \citenamefont {Sokolenko},
  \citenamefont {Kugel}, \citenamefont {Schaller},\ and\ \citenamefont
  {Schaye}}]{Aramburo-Garcia:2024cbz}%
  \BibitemOpen
  \bibfield  {author} {\bibinfo {author} {\bibfnamefont {Andres}\ \bibnamefont
  {Aramburo-Garcia}}, \bibinfo {author} {\bibfnamefont {Kyrylo}\ \bibnamefont
  {Bondarenko}}, \bibinfo {author} {\bibfnamefont {Alexey}\ \bibnamefont
  {Boyarsky}}, \bibinfo {author} {\bibfnamefont {Pavlo}\ \bibnamefont
  {Kashko}}, \bibinfo {author} {\bibfnamefont {Josef}\ \bibnamefont {Pradler}},
  \bibinfo {author} {\bibfnamefont {Anastasia}\ \bibnamefont {Sokolenko}},
  \bibinfo {author} {\bibfnamefont {Roi}\ \bibnamefont {Kugel}}, \bibinfo
  {author} {\bibfnamefont {Matthieu}\ \bibnamefont {Schaller}}, \ and\ \bibinfo
  {author} {\bibfnamefont {Joop}\ \bibnamefont {Schaye}},\ }\bibfield  {title}
  {\enquote {\bibinfo {title} {{Dark photon constraints from CMB temperature
  anisotropies}},}\ }\href@noop {} {\  (\bibinfo {year} {2024})},\ \Eprint
  {http://arxiv.org/abs/2405.05104} {arXiv:2405.05104 [astro-ph.CO]}
  \BibitemShut {NoStop}%
\bibitem [{\citenamefont {McCarthy}\ \emph {et~al.}(2024)\citenamefont
  {McCarthy}, \citenamefont {Pirvu}, \citenamefont {Hill}, \citenamefont
  {Huang}, \citenamefont {Johnson},\ and\ \citenamefont
  {Rogers}}]{McCarthy:2024ozh}%
  \BibitemOpen
  \bibfield  {author} {\bibinfo {author} {\bibfnamefont {Fiona}\ \bibnamefont
  {McCarthy}}, \bibinfo {author} {\bibfnamefont {Dalila}\ \bibnamefont
  {Pirvu}}, \bibinfo {author} {\bibfnamefont {J.~Colin}\ \bibnamefont {Hill}},
  \bibinfo {author} {\bibfnamefont {Junwu}\ \bibnamefont {Huang}}, \bibinfo
  {author} {\bibfnamefont {Matthew~C.}\ \bibnamefont {Johnson}}, \ and\
  \bibinfo {author} {\bibfnamefont {Keir~K.}\ \bibnamefont {Rogers}},\
  }\bibfield  {title} {\enquote {\bibinfo {title} {{Dark photon limits from
  patchy dark screening of the cosmic microwave background}},}\ }\href@noop {}
  {\  (\bibinfo {year} {2024})},\ \Eprint {http://arxiv.org/abs/2406.02546}
  {arXiv:2406.02546 [hep-ph]} \BibitemShut {NoStop}%
\bibitem [{\citenamefont {Fixsen}\ \emph {et~al.}(1996)\citenamefont {Fixsen},
  \citenamefont {Cheng}, \citenamefont {Gales}, \citenamefont {Mather},
  \citenamefont {Shafer},\ and\ \citenamefont {Wright}}]{Fixsen:1996nj}%
  \BibitemOpen
  \bibfield  {author} {\bibinfo {author} {\bibfnamefont {D.~J.}\ \bibnamefont
  {Fixsen}}, \bibinfo {author} {\bibfnamefont {E.~S.}\ \bibnamefont {Cheng}},
  \bibinfo {author} {\bibfnamefont {J.~M.}\ \bibnamefont {Gales}}, \bibinfo
  {author} {\bibfnamefont {John~C.}\ \bibnamefont {Mather}}, \bibinfo {author}
  {\bibfnamefont {R.~A.}\ \bibnamefont {Shafer}}, \ and\ \bibinfo {author}
  {\bibfnamefont {E.~L.}\ \bibnamefont {Wright}},\ }\bibfield  {title}
  {\enquote {\bibinfo {title} {{The Cosmic Microwave Background spectrum from
  the full COBE FIRAS data set}},}\ }\href {\doibase 10.1086/178173} {\bibfield
   {journal} {\bibinfo  {journal} {Astrophys. J.}\ }\textbf {\bibinfo {volume}
  {473}},\ \bibinfo {pages} {576} (\bibinfo {year} {1996})},\ \Eprint
  {http://arxiv.org/abs/astro-ph/9605054} {arXiv:astro-ph/9605054} \BibitemShut
  {NoStop}%
\bibitem [{\citenamefont {Kogut}\ \emph {et~al.}(2011)\citenamefont {Kogut}
  \emph {et~al.}}]{Kogut:2011xw}%
  \BibitemOpen
  \bibfield  {author} {\bibinfo {author} {\bibfnamefont {A.}~\bibnamefont
  {Kogut}} \emph {et~al.},\ }\bibfield  {title} {\enquote {\bibinfo {title}
  {{The Primordial Inflation Explorer (PIXIE): A Nulling Polarimeter for Cosmic
  Microwave Background Observations}},}\ }\href {\doibase
  10.1088/1475-7516/2011/07/025} {\bibfield  {journal} {\bibinfo  {journal}
  {JCAP}\ }\textbf {\bibinfo {volume} {07}},\ \bibinfo {pages} {025} (\bibinfo
  {year} {2011})},\ \Eprint {http://arxiv.org/abs/1105.2044} {arXiv:1105.2044
  [astro-ph.CO]} \BibitemShut {NoStop}%
\bibitem [{\citenamefont {Kogut}\ \emph {et~al.}(2024)\citenamefont {Kogut}
  \emph {et~al.}}]{Kogut:2024vbi}%
  \BibitemOpen
  \bibfield  {author} {\bibinfo {author} {\bibfnamefont {Alan}\ \bibnamefont
  {Kogut}} \emph {et~al.},\ }\bibfield  {title} {\enquote {\bibinfo {title}
  {{The Primordial Inflation Explorer (PIXIE): Mission Design and Science
  Goals}},}\ }\href@noop {} {\  (\bibinfo {year} {2024})},\ \Eprint
  {http://arxiv.org/abs/2405.20403} {arXiv:2405.20403 [astro-ph.CO]}
  \BibitemShut {NoStop}%
\bibitem [{\citenamefont {Andr\'e}\ \emph {et~al.}(2014)\citenamefont {Andr\'e}
  \emph {et~al.}}]{PRISM:2013fvg}%
  \BibitemOpen
  \bibfield  {author} {\bibinfo {author} {\bibfnamefont {Philippe}\
  \bibnamefont {Andr\'e}} \emph {et~al.} (\bibinfo {collaboration} {PRISM}),\
  }\bibfield  {title} {\enquote {\bibinfo {title} {{PRISM (Polarized Radiation
  Imaging and Spectroscopy Mission): An Extended White Paper}},}\ }\href
  {\doibase 10.1088/1475-7516/2014/02/006} {\bibfield  {journal} {\bibinfo
  {journal} {JCAP}\ }\textbf {\bibinfo {volume} {02}},\ \bibinfo {pages} {006}
  (\bibinfo {year} {2014})},\ \Eprint {http://arxiv.org/abs/1310.1554}
  {arXiv:1310.1554 [astro-ph.CO]} \BibitemShut {NoStop}%
\bibitem [{\citenamefont {Chluba}\ \emph {et~al.}(2021)\citenamefont {Chluba}
  \emph {et~al.}}]{Chluba:2019nxa}%
  \BibitemOpen
  \bibfield  {author} {\bibinfo {author} {\bibfnamefont {J.}~\bibnamefont
  {Chluba}} \emph {et~al.},\ }\bibfield  {title} {\enquote {\bibinfo {title}
  {{New horizons in cosmology with spectral distortions of the cosmic microwave
  background}},}\ }\href {\doibase 10.1007/s10686-021-09729-5} {\bibfield
  {journal} {\bibinfo  {journal} {Exper. Astron.}\ }\textbf {\bibinfo {volume}
  {51}},\ \bibinfo {pages} {1515--1554} (\bibinfo {year} {2021})},\ \Eprint
  {http://arxiv.org/abs/1909.01593} {arXiv:1909.01593 [astro-ph.CO]}
  \BibitemShut {NoStop}%
\bibitem [{\citenamefont {Sabyr}\ \emph {et~al.}(2024)\citenamefont {Sabyr},
  \citenamefont {Sierra}, \citenamefont {Hill},\ and\ \citenamefont
  {McMahon}}]{Sabyr:2024lgg}%
  \BibitemOpen
  \bibfield  {author} {\bibinfo {author} {\bibfnamefont {Alina}\ \bibnamefont
  {Sabyr}}, \bibinfo {author} {\bibfnamefont {Carlos}\ \bibnamefont {Sierra}},
  \bibinfo {author} {\bibfnamefont {J.~Colin}\ \bibnamefont {Hill}}, \ and\
  \bibinfo {author} {\bibfnamefont {Jeffrey~J.}\ \bibnamefont {McMahon}},\
  }\bibfield  {title} {\enquote {\bibinfo {title} {{SPECTER: An Instrument
  Concept for CMB Spectral Distortion Measurements with Enhanced
  Sensitivity}},}\ }\href@noop {} {\  (\bibinfo {year} {2024})},\ \Eprint
  {http://arxiv.org/abs/2409.12188} {arXiv:2409.12188 [astro-ph.CO]}
  \BibitemShut {NoStop}%
\bibitem [{\citenamefont {Mirizzi}\ \emph {et~al.}(2005)\citenamefont
  {Mirizzi}, \citenamefont {Raffelt},\ and\ \citenamefont
  {Serpico}}]{Mirizzi:2005ng}%
  \BibitemOpen
  \bibfield  {author} {\bibinfo {author} {\bibfnamefont {Alessandro}\
  \bibnamefont {Mirizzi}}, \bibinfo {author} {\bibfnamefont {Georg~G.}\
  \bibnamefont {Raffelt}}, \ and\ \bibinfo {author} {\bibfnamefont
  {Pasquale~D.}\ \bibnamefont {Serpico}},\ }\bibfield  {title} {\enquote
  {\bibinfo {title} {{Photon-axion conversion as a mechanism for supernova
  dimming: Limits from CMB spectral distortion}},}\ }\href {\doibase
  10.1103/PhysRevD.72.023501} {\bibfield  {journal} {\bibinfo  {journal} {Phys.
  Rev. D}\ }\textbf {\bibinfo {volume} {72}},\ \bibinfo {pages} {023501}
  (\bibinfo {year} {2005})},\ \Eprint {http://arxiv.org/abs/astro-ph/0506078}
  {arXiv:astro-ph/0506078} \BibitemShut {NoStop}%
\bibitem [{\citenamefont {Tashiro}\ \emph {et~al.}(2013)\citenamefont
  {Tashiro}, \citenamefont {Silk},\ and\ \citenamefont
  {Marsh}}]{Tashiro:2013yea}%
  \BibitemOpen
  \bibfield  {author} {\bibinfo {author} {\bibfnamefont {Hiroyuki}\
  \bibnamefont {Tashiro}}, \bibinfo {author} {\bibfnamefont {Joseph}\
  \bibnamefont {Silk}}, \ and\ \bibinfo {author} {\bibfnamefont {David J.~E.}\
  \bibnamefont {Marsh}},\ }\bibfield  {title} {\enquote {\bibinfo {title}
  {{Constraints on primordial magnetic fields from CMB distortions in the
  axiverse}},}\ }\href {\doibase 10.1103/PhysRevD.88.125024} {\bibfield
  {journal} {\bibinfo  {journal} {Phys. Rev. D}\ }\textbf {\bibinfo {volume}
  {88}},\ \bibinfo {pages} {125024} (\bibinfo {year} {2013})},\ \Eprint
  {http://arxiv.org/abs/1308.0314} {arXiv:1308.0314 [astro-ph.CO]} \BibitemShut
  {NoStop}%
\bibitem [{\citenamefont {Ejlli}\ and\ \citenamefont
  {Dolgov}(2014)}]{Ejlli:2013uda}%
  \BibitemOpen
  \bibfield  {author} {\bibinfo {author} {\bibfnamefont {Damian}\ \bibnamefont
  {Ejlli}}\ and\ \bibinfo {author} {\bibfnamefont {Alexander~D.}\ \bibnamefont
  {Dolgov}},\ }\bibfield  {title} {\enquote {\bibinfo {title} {{CMB constraints
  on mass and coupling constant of light pseudoscalar particles}},}\ }\href
  {\doibase 10.1103/PhysRevD.90.063514} {\bibfield  {journal} {\bibinfo
  {journal} {Phys. Rev. D}\ }\textbf {\bibinfo {volume} {90}},\ \bibinfo
  {pages} {063514} (\bibinfo {year} {2014})},\ \Eprint
  {http://arxiv.org/abs/1312.3558} {arXiv:1312.3558 [hep-ph]} \BibitemShut
  {NoStop}%
\bibitem [{\citenamefont {Mukherjee}\ \emph {et~al.}(2018)\citenamefont
  {Mukherjee}, \citenamefont {Khatri},\ and\ \citenamefont
  {Wandelt}}]{Mukherjee:2018oeb}%
  \BibitemOpen
  \bibfield  {author} {\bibinfo {author} {\bibfnamefont {Suvodip}\ \bibnamefont
  {Mukherjee}}, \bibinfo {author} {\bibfnamefont {Rishi}\ \bibnamefont
  {Khatri}}, \ and\ \bibinfo {author} {\bibfnamefont {Benjamin~D.}\
  \bibnamefont {Wandelt}},\ }\bibfield  {title} {\enquote {\bibinfo {title}
  {{Polarized anisotropic spectral distortions of the CMB: Galactic and
  extragalactic constraints on photon-axion conversion}},}\ }\href {\doibase
  10.1088/1475-7516/2018/04/045} {\bibfield  {journal} {\bibinfo  {journal}
  {JCAP}\ }\textbf {\bibinfo {volume} {04}},\ \bibinfo {pages} {045} (\bibinfo
  {year} {2018})},\ \Eprint {http://arxiv.org/abs/1801.09701} {arXiv:1801.09701
  [astro-ph.CO]} \BibitemShut {NoStop}%
\bibitem [{\citenamefont {Ali-Ha\"\i{}moud}(2021)}]{Ali-Haimoud:2021lka}%
  \BibitemOpen
  \bibfield  {author} {\bibinfo {author} {\bibfnamefont {Yacine}\ \bibnamefont
  {Ali-Ha\"\i{}moud}},\ }\bibfield  {title} {\enquote {\bibinfo {title}
  {{Testing dark matter interactions with CMB spectral distortions}},}\ }\href
  {\doibase 10.1103/PhysRevD.103.043541} {\bibfield  {journal} {\bibinfo
  {journal} {Phys. Rev. D}\ }\textbf {\bibinfo {volume} {103}},\ \bibinfo
  {pages} {043541} (\bibinfo {year} {2021})},\ \Eprint
  {http://arxiv.org/abs/2101.04070} {arXiv:2101.04070 [astro-ph.CO]}
  \BibitemShut {NoStop}%
\bibitem [{\citenamefont {Chluba}\ \emph {et~al.}(2019)\citenamefont {Chluba}
  \emph {et~al.}}]{Chluba:2019kpb}%
  \BibitemOpen
  \bibfield  {author} {\bibinfo {author} {\bibfnamefont {J.}~\bibnamefont
  {Chluba}} \emph {et~al.},\ }\bibfield  {title} {\enquote {\bibinfo {title}
  {{Spectral Distortions of the CMB as a Probe of Inflation, Recombination,
  Structure Formation and Particle Physics}: {Astro2020 Science White
  Paper}},}\ }\href@noop {} {\bibfield  {journal} {\bibinfo  {journal} {Bull.
  Am. Astron. Soc.}\ }\textbf {\bibinfo {volume} {51}},\ \bibinfo {pages} {184}
  (\bibinfo {year} {2019})},\ \Eprint {http://arxiv.org/abs/1903.04218}
  {arXiv:1903.04218 [astro-ph.CO]} \BibitemShut {NoStop}%
\bibitem [{\citenamefont {Zener}(1932)}]{Zener:1932ws}%
  \BibitemOpen
  \bibfield  {author} {\bibinfo {author} {\bibfnamefont {Clarence}\
  \bibnamefont {Zener}},\ }\bibfield  {title} {\enquote {\bibinfo {title}
  {{Nonadiabatic crossing of energy levels}},}\ }\href {\doibase
  10.1098/rspa.1932.0165} {\bibfield  {journal} {\bibinfo  {journal} {Proc.
  Roy. Soc. Lond. A}\ }\textbf {\bibinfo {volume} {137}},\ \bibinfo {pages}
  {696--702} (\bibinfo {year} {1932})}\BibitemShut {NoStop}%
\bibitem [{\citenamefont {Landau}(1932)}]{landau1932theorie}%
  \BibitemOpen
  \bibfield  {author} {\bibinfo {author} {\bibfnamefont {Lev~Davidovich}\
  \bibnamefont {Landau}},\ }\bibfield  {title} {\enquote {\bibinfo {title}
  {{Zur Theorie der Energieubertragung. II}},}\ }\href@noop {} {\bibfield
  {journal} {\bibinfo  {journal} {Z. Sowjetunion}\ }\textbf {\bibinfo {volume}
  {2}},\ \bibinfo {pages} {46--51} (\bibinfo {year} {1932})}\BibitemShut
  {NoStop}%
\bibitem [{\citenamefont {Bondarenko}\ \emph {et~al.}(2020)\citenamefont
  {Bondarenko}, \citenamefont {Pradler},\ and\ \citenamefont
  {Sokolenko}}]{Bondarenko:2020moh}%
  \BibitemOpen
  \bibfield  {author} {\bibinfo {author} {\bibfnamefont {Kyrylo}\ \bibnamefont
  {Bondarenko}}, \bibinfo {author} {\bibfnamefont {Josef}\ \bibnamefont
  {Pradler}}, \ and\ \bibinfo {author} {\bibfnamefont {Anastasia}\ \bibnamefont
  {Sokolenko}},\ }\bibfield  {title} {\enquote {\bibinfo {title} {{Constraining
  dark photons and their connection to 21 cm cosmology with CMB data}},}\
  }\href {\doibase 10.1016/j.physletb.2020.135420} {\bibfield  {journal}
  {\bibinfo  {journal} {Phys. Lett. B}\ }\textbf {\bibinfo {volume} {805}},\
  \bibinfo {pages} {135420} (\bibinfo {year} {2020})},\ \Eprint
  {http://arxiv.org/abs/2002.08942} {arXiv:2002.08942 [astro-ph.CO]}
  \BibitemShut {NoStop}%
\bibitem [{\citenamefont {Arias}\ \emph {et~al.}(2012)\citenamefont {Arias},
  \citenamefont {Cadamuro}, \citenamefont {Goodsell}, \citenamefont {Jaeckel},
  \citenamefont {Redondo},\ and\ \citenamefont {Ringwald}}]{Arias:2012az}%
  \BibitemOpen
  \bibfield  {author} {\bibinfo {author} {\bibfnamefont {Paola}\ \bibnamefont
  {Arias}}, \bibinfo {author} {\bibfnamefont {Davide}\ \bibnamefont
  {Cadamuro}}, \bibinfo {author} {\bibfnamefont {Mark}\ \bibnamefont
  {Goodsell}}, \bibinfo {author} {\bibfnamefont {Joerg}\ \bibnamefont
  {Jaeckel}}, \bibinfo {author} {\bibfnamefont {Javier}\ \bibnamefont
  {Redondo}}, \ and\ \bibinfo {author} {\bibfnamefont {Andreas}\ \bibnamefont
  {Ringwald}},\ }\bibfield  {title} {\enquote {\bibinfo {title} {{WISPy Cold
  Dark Matter}},}\ }\href {\doibase 10.1088/1475-7516/2012/06/013} {\bibfield
  {journal} {\bibinfo  {journal} {JCAP}\ }\textbf {\bibinfo {volume} {06}},\
  \bibinfo {pages} {013} (\bibinfo {year} {2012})},\ \Eprint
  {http://arxiv.org/abs/1201.5902} {arXiv:1201.5902 [hep-ph]} \BibitemShut
  {NoStop}%
\bibitem [{\citenamefont {Dubovsky}\ and\ \citenamefont
  {Hern\'andez-Chifflet}(2015)}]{Dubovsky:2015cca}%
  \BibitemOpen
  \bibfield  {author} {\bibinfo {author} {\bibfnamefont {Sergei}\ \bibnamefont
  {Dubovsky}}\ and\ \bibinfo {author} {\bibfnamefont {Guzm\'an}\ \bibnamefont
  {Hern\'andez-Chifflet}},\ }\bibfield  {title} {\enquote {\bibinfo {title}
  {{Heating up the Galaxy with Hidden Photons}},}\ }\href {\doibase
  10.1088/1475-7516/2015/12/054} {\bibfield  {journal} {\bibinfo  {journal}
  {JCAP}\ }\textbf {\bibinfo {volume} {12}},\ \bibinfo {pages} {054} (\bibinfo
  {year} {2015})},\ \Eprint {http://arxiv.org/abs/1509.00039} {arXiv:1509.00039
  [hep-ph]} \BibitemShut {NoStop}%
\bibitem [{\citenamefont {Wadekar}\ and\ \citenamefont
  {Farrar}(2021)}]{Wadekar:2019mpc}%
  \BibitemOpen
  \bibfield  {author} {\bibinfo {author} {\bibfnamefont {Digvijay}\
  \bibnamefont {Wadekar}}\ and\ \bibinfo {author} {\bibfnamefont {Glennys~R.}\
  \bibnamefont {Farrar}},\ }\bibfield  {title} {\enquote {\bibinfo {title}
  {{Gas-rich dwarf galaxies as a new probe of dark matter interactions with
  ordinary matter}},}\ }\href {\doibase 10.1103/PhysRevD.103.123028} {\bibfield
   {journal} {\bibinfo  {journal} {Phys. Rev. D}\ }\textbf {\bibinfo {volume}
  {103}},\ \bibinfo {pages} {123028} (\bibinfo {year} {2021})},\ \Eprint
  {http://arxiv.org/abs/1903.12190} {arXiv:1903.12190 [hep-ph]} \BibitemShut
  {NoStop}%
\bibitem [{\citenamefont {Witte}\ \emph {et~al.}(2020)\citenamefont {Witte},
  \citenamefont {Rosauro-Alcaraz}, \citenamefont {McDermott},\ and\
  \citenamefont {Poulin}}]{Witte:2020rvb}%
  \BibitemOpen
  \bibfield  {author} {\bibinfo {author} {\bibfnamefont {Samuel~J.}\
  \bibnamefont {Witte}}, \bibinfo {author} {\bibfnamefont {Salvador}\
  \bibnamefont {Rosauro-Alcaraz}}, \bibinfo {author} {\bibfnamefont
  {Samuel~D.}\ \bibnamefont {McDermott}}, \ and\ \bibinfo {author}
  {\bibfnamefont {Vivian}\ \bibnamefont {Poulin}},\ }\bibfield  {title}
  {\enquote {\bibinfo {title} {{Dark photon dark matter in the presence of
  inhomogeneous structure}},}\ }\href {\doibase 10.1007/JHEP06(2020)132}
  {\bibfield  {journal} {\bibinfo  {journal} {JHEP}\ }\textbf {\bibinfo
  {volume} {06}},\ \bibinfo {pages} {132} (\bibinfo {year} {2020})},\ \Eprint
  {http://arxiv.org/abs/2003.13698} {arXiv:2003.13698 [astro-ph.CO]}
  \BibitemShut {NoStop}%
\bibitem [{\citenamefont {An}\ \emph {et~al.}(2021)\citenamefont {An},
  \citenamefont {Huang}, \citenamefont {Liu},\ and\ \citenamefont
  {Xue}}]{An:2020jmf}%
  \BibitemOpen
  \bibfield  {author} {\bibinfo {author} {\bibfnamefont {Haipeng}\ \bibnamefont
  {An}}, \bibinfo {author} {\bibfnamefont {Fa~Peng}\ \bibnamefont {Huang}},
  \bibinfo {author} {\bibfnamefont {Jia}\ \bibnamefont {Liu}}, \ and\ \bibinfo
  {author} {\bibfnamefont {Wei}\ \bibnamefont {Xue}},\ }\bibfield  {title}
  {\enquote {\bibinfo {title} {{Radio-frequency Dark Photon Dark Matter across
  the Sun}},}\ }\href {\doibase 10.1103/PhysRevLett.126.181102} {\bibfield
  {journal} {\bibinfo  {journal} {Phys. Rev. Lett.}\ }\textbf {\bibinfo
  {volume} {126}},\ \bibinfo {pages} {181102} (\bibinfo {year} {2021})},\
  \Eprint {http://arxiv.org/abs/2010.15836} {arXiv:2010.15836 [hep-ph]}
  \BibitemShut {NoStop}%
\bibitem [{\citenamefont {An}\ \emph {et~al.}(2023)\citenamefont {An},
  \citenamefont {Ge}, \citenamefont {Guo}, \citenamefont {Huang}, \citenamefont
  {Liu},\ and\ \citenamefont {Lu}}]{An:2022hhb}%
  \BibitemOpen
  \bibfield  {author} {\bibinfo {author} {\bibfnamefont {Haipeng}\ \bibnamefont
  {An}}, \bibinfo {author} {\bibfnamefont {Shuailiang}\ \bibnamefont {Ge}},
  \bibinfo {author} {\bibfnamefont {Wen-Qing}\ \bibnamefont {Guo}}, \bibinfo
  {author} {\bibfnamefont {Xiaoyuan}\ \bibnamefont {Huang}}, \bibinfo {author}
  {\bibfnamefont {Jia}\ \bibnamefont {Liu}}, \ and\ \bibinfo {author}
  {\bibfnamefont {Zhiyao}\ \bibnamefont {Lu}},\ }\bibfield  {title} {\enquote
  {\bibinfo {title} {{Direct Detection of Dark Photon Dark Matter Using Radio
  Telescopes}},}\ }\href {\doibase 10.1103/PhysRevLett.130.181001} {\bibfield
  {journal} {\bibinfo  {journal} {Phys. Rev. Lett.}\ }\textbf {\bibinfo
  {volume} {130}},\ \bibinfo {pages} {181001} (\bibinfo {year} {2023})},\
  \Eprint {http://arxiv.org/abs/2207.05767} {arXiv:2207.05767 [hep-ph]}
  \BibitemShut {NoStop}%
\bibitem [{\citenamefont {An}\ \emph {et~al.}(2024{\natexlab{a}})\citenamefont
  {An}, \citenamefont {Ge}, \citenamefont {Liu},\ and\ \citenamefont
  {Liu}}]{An:2024wmc}%
  \BibitemOpen
  \bibfield  {author} {\bibinfo {author} {\bibfnamefont {Haipeng}\ \bibnamefont
  {An}}, \bibinfo {author} {\bibfnamefont {Shuailiang}\ \bibnamefont {Ge}},
  \bibinfo {author} {\bibfnamefont {Jia}\ \bibnamefont {Liu}}, \ and\ \bibinfo
  {author} {\bibfnamefont {Mingzhe}\ \bibnamefont {Liu}},\ }\bibfield  {title}
  {\enquote {\bibinfo {title} {{In-situ Measurements of Dark Photon Dark Matter
  using Parker Solar Probe: Going beyond the Radio Window}},}\ }\href@noop {}
  {\  (\bibinfo {year} {2024}{\natexlab{a}})},\ \Eprint
  {http://arxiv.org/abs/2405.12285} {arXiv:2405.12285 [hep-ph]} \BibitemShut
  {NoStop}%
\bibitem [{\citenamefont {An}\ \emph {et~al.}(2024{\natexlab{b}})\citenamefont
  {An}, \citenamefont {Ge}, \citenamefont {Liu},\ and\ \citenamefont
  {Lu}}]{An:2024kls}%
  \BibitemOpen
  \bibfield  {author} {\bibinfo {author} {\bibfnamefont {Haipeng}\ \bibnamefont
  {An}}, \bibinfo {author} {\bibfnamefont {Shuailiang}\ \bibnamefont {Ge}},
  \bibinfo {author} {\bibfnamefont {Jia}\ \bibnamefont {Liu}}, \ and\ \bibinfo
  {author} {\bibfnamefont {Zhiyao}\ \bibnamefont {Lu}},\ }\bibfield  {title}
  {\enquote {\bibinfo {title} {{Direct Detection of Dark Photon Dark Matter
  with the James Webb Space Telescope}},}\ }\href@noop {} {\  (\bibinfo {year}
  {2024}{\natexlab{b}})},\ \Eprint {http://arxiv.org/abs/2402.17140}
  {arXiv:2402.17140 [hep-ph]} \BibitemShut {NoStop}%
\bibitem [{\citenamefont {Chluba}(2015)}]{Chluba:2015hma}%
  \BibitemOpen
  \bibfield  {author} {\bibinfo {author} {\bibfnamefont {Jens}\ \bibnamefont
  {Chluba}},\ }\bibfield  {title} {\enquote {\bibinfo {title} {{Green's
  function of the cosmological thermalization problem \textendash{} II. Effect
  of photon injection and constraints}},}\ }\href {\doibase
  10.1093/mnras/stv2243} {\bibfield  {journal} {\bibinfo  {journal} {Mon. Not.
  Roy. Astron. Soc.}\ }\textbf {\bibinfo {volume} {454}},\ \bibinfo {pages}
  {4182--4196} (\bibinfo {year} {2015})},\ \Eprint
  {http://arxiv.org/abs/1506.06582} {arXiv:1506.06582 [astro-ph.CO]}
  \BibitemShut {NoStop}%
\bibitem [{\citenamefont {Romanenko}\ \emph {et~al.}(2023)\citenamefont
  {Romanenko} \emph {et~al.}}]{Romanenko:2023irv}%
  \BibitemOpen
  \bibfield  {author} {\bibinfo {author} {\bibfnamefont {A.}~\bibnamefont
  {Romanenko}} \emph {et~al.},\ }\bibfield  {title} {\enquote {\bibinfo {title}
  {{Search for Dark Photons with Superconducting Radio Frequency Cavities}},}\
  }\href {\doibase 10.1103/PhysRevLett.130.261801} {\bibfield  {journal}
  {\bibinfo  {journal} {Phys. Rev. Lett.}\ }\textbf {\bibinfo {volume} {130}},\
  \bibinfo {pages} {261801} (\bibinfo {year} {2023})},\ \Eprint
  {http://arxiv.org/abs/2301.11512} {arXiv:2301.11512 [hep-ex]} \BibitemShut
  {NoStop}%
\bibitem [{\citenamefont {Gan}(2024{\natexlab{a}})}]{Gan:2024ele}%
  \BibitemOpen
  \bibfield  {author} {\bibinfo {author} {\bibfnamefont {Xucheng}\ \bibnamefont
  {Gan}},\ }\bibfield  {title} {\enquote {\bibinfo {title} {{The Hidden
  Universe Odyssey: From Theoretical Foundations to Cosmological
  Detections}},}\ }\href {https://www.proquest.com/docview/3072082208}
  {\bibfield  {journal} {\bibinfo  {journal} {PhD Thesis}\ } (\bibinfo {year}
  {2024}{\natexlab{a}})}\BibitemShut {NoStop}%
\bibitem [{\citenamefont {Gan}(2024{\natexlab{b}})}]{TeVPA:2024}%
  \BibitemOpen
  \bibfield  {author} {\bibinfo {author} {\bibfnamefont {Xucheng}\ \bibnamefont
  {Gan}},\ }\bibfield  {title} {\enquote {\bibinfo {title} {{CMB} spectral
  distortions from dark photon oscillation},}\ }in\ \href
  {https://indico.uchicago.edu/event/427/contributions/1398/} {\emph {\bibinfo
  {booktitle} {Proceedings of the 2024 TeV Particle Astrophysics Conference
  (TeVPA 2024)}}}\ (\bibinfo {organization} {University of Chicago},\ \bibinfo
  {year} {2024})\ \bibinfo {note} {conference presentation}\BibitemShut
  {NoStop}%
\bibitem [{\citenamefont {Weinberg}(2013)}]{Weinberg:1996kr}%
  \BibitemOpen
  \bibfield  {author} {\bibinfo {author} {\bibfnamefont {Steven}\ \bibnamefont
  {Weinberg}},\ }\href {\doibase 10.1017/CBO9781139644174} {\emph {\bibinfo
  {title} {{The quantum theory of fields. Vol. 2: Modern applications}}}}\
  (\bibinfo  {publisher} {Cambridge University Press},\ \bibinfo {year}
  {2013})\BibitemShut {NoStop}%
\bibitem [{\citenamefont {Burgess}\ \emph {et~al.}(2008)\citenamefont
  {Burgess}, \citenamefont {Conlon}, \citenamefont {Hung}, \citenamefont {Kom},
  \citenamefont {Maharana},\ and\ \citenamefont {Quevedo}}]{Burgess:2008ri}%
  \BibitemOpen
  \bibfield  {author} {\bibinfo {author} {\bibfnamefont {C.~P.}\ \bibnamefont
  {Burgess}}, \bibinfo {author} {\bibfnamefont {J.~P.}\ \bibnamefont {Conlon}},
  \bibinfo {author} {\bibfnamefont {L-Y.}\ \bibnamefont {Hung}}, \bibinfo
  {author} {\bibfnamefont {C.~H.}\ \bibnamefont {Kom}}, \bibinfo {author}
  {\bibfnamefont {Anshuman}\ \bibnamefont {Maharana}}, \ and\ \bibinfo {author}
  {\bibfnamefont {F.}~\bibnamefont {Quevedo}},\ }\bibfield  {title} {\enquote
  {\bibinfo {title} {{Continuous Global Symmetries and Hyperweak Interactions
  in String Compactifications}},}\ }\href {\doibase
  10.1088/1126-6708/2008/07/073} {\bibfield  {journal} {\bibinfo  {journal}
  {JHEP}\ }\textbf {\bibinfo {volume} {07}},\ \bibinfo {pages} {073} (\bibinfo
  {year} {2008})},\ \Eprint {http://arxiv.org/abs/0805.4037} {arXiv:0805.4037
  [hep-th]} \BibitemShut {NoStop}%
\bibitem [{\citenamefont {Gherghetta}\ \emph {et~al.}(2019)\citenamefont
  {Gherghetta}, \citenamefont {Kersten}, \citenamefont {Olive},\ and\
  \citenamefont {Pospelov}}]{Gherghetta:2019coi}%
  \BibitemOpen
  \bibfield  {author} {\bibinfo {author} {\bibfnamefont {Tony}\ \bibnamefont
  {Gherghetta}}, \bibinfo {author} {\bibfnamefont {J\"orn}\ \bibnamefont
  {Kersten}}, \bibinfo {author} {\bibfnamefont {Keith}\ \bibnamefont {Olive}},
  \ and\ \bibinfo {author} {\bibfnamefont {Maxim}\ \bibnamefont {Pospelov}},\
  }\bibfield  {title} {\enquote {\bibinfo {title} {{Evaluating the price of
  tiny kinetic mixing}},}\ }\href {\doibase 10.1103/PhysRevD.100.095001}
  {\bibfield  {journal} {\bibinfo  {journal} {Phys. Rev. D}\ }\textbf {\bibinfo
  {volume} {100}},\ \bibinfo {pages} {095001} (\bibinfo {year} {2019})},\
  \Eprint {http://arxiv.org/abs/1909.00696} {arXiv:1909.00696 [hep-ph]}
  \BibitemShut {NoStop}%
\bibitem [{\citenamefont {Gan}\ and\ \citenamefont {Liu}(2023)}]{Gan:2023wnp}%
  \BibitemOpen
  \bibfield  {author} {\bibinfo {author} {\bibfnamefont {Xucheng}\ \bibnamefont
  {Gan}}\ and\ \bibinfo {author} {\bibfnamefont {Di}~\bibnamefont {Liu}},\
  }\bibfield  {title} {\enquote {\bibinfo {title} {{Cosmologically varying
  kinetic mixing}},}\ }\href {\doibase 10.1007/JHEP11(2023)031} {\bibfield
  {journal} {\bibinfo  {journal} {JHEP}\ }\textbf {\bibinfo {volume} {11}},\
  \bibinfo {pages} {031} (\bibinfo {year} {2023})},\ \Eprint
  {http://arxiv.org/abs/2302.03056} {arXiv:2302.03056 [hep-ph]} \BibitemShut
  {NoStop}%
\bibitem [{\citenamefont {Gan}\ and\ \citenamefont {Tsai}(2023)}]{Gan:2023jbs}%
  \BibitemOpen
  \bibfield  {author} {\bibinfo {author} {\bibfnamefont {Xucheng}\ \bibnamefont
  {Gan}}\ and\ \bibinfo {author} {\bibfnamefont {Yu-Dai}\ \bibnamefont
  {Tsai}},\ }\bibfield  {title} {\enquote {\bibinfo {title} {{Cosmic
  Millicharge Background and Reheating Probes}},}\ }\href@noop {} {\  (\bibinfo
  {year} {2023})},\ \Eprint {http://arxiv.org/abs/2308.07951} {arXiv:2308.07951
  [hep-ph]} \BibitemShut {NoStop}%
\bibitem [{\citenamefont {Iles}\ \emph {et~al.}(2024)\citenamefont {Iles},
  \citenamefont {Heeba},\ and\ \citenamefont {Schutz}}]{Iles:2024zka}%
  \BibitemOpen
  \bibfield  {author} {\bibinfo {author} {\bibfnamefont {Ella}\ \bibnamefont
  {Iles}}, \bibinfo {author} {\bibfnamefont {Saniya}\ \bibnamefont {Heeba}}, \
  and\ \bibinfo {author} {\bibfnamefont {Katelin}\ \bibnamefont {Schutz}},\
  }\bibfield  {title} {\enquote {\bibinfo {title} {{Direct Detection of the
  Millicharged Background}},}\ }\href@noop {} {\  (\bibinfo {year} {2024})},\
  \Eprint {http://arxiv.org/abs/2407.21096} {arXiv:2407.21096 [hep-ph]}
  \BibitemShut {NoStop}%
\bibitem [{\citenamefont {Berlin}\ \emph
  {et~al.}(2023{\natexlab{b}})\citenamefont {Berlin}, \citenamefont {Dror},
  \citenamefont {Gan},\ and\ \citenamefont {Ruderman}}]{Berlin:2022hmt}%
  \BibitemOpen
  \bibfield  {author} {\bibinfo {author} {\bibfnamefont {Asher}\ \bibnamefont
  {Berlin}}, \bibinfo {author} {\bibfnamefont {Jeff~A.}\ \bibnamefont {Dror}},
  \bibinfo {author} {\bibfnamefont {Xucheng}\ \bibnamefont {Gan}}, \ and\
  \bibinfo {author} {\bibfnamefont {Joshua~T.}\ \bibnamefont {Ruderman}},\
  }\bibfield  {title} {\enquote {\bibinfo {title} {{Millicharged relics reveal
  massless dark photons}},}\ }\href {\doibase 10.1007/JHEP05(2023)046}
  {\bibfield  {journal} {\bibinfo  {journal} {JHEP}\ }\textbf {\bibinfo
  {volume} {05}},\ \bibinfo {pages} {046} (\bibinfo {year}
  {2023}{\natexlab{b}})},\ \Eprint {http://arxiv.org/abs/2211.05139}
  {arXiv:2211.05139 [hep-ph]} \BibitemShut {NoStop}%
\bibitem [{\citenamefont {Berlin}\ \emph
  {et~al.}(2023{\natexlab{c}})\citenamefont {Berlin}, \citenamefont
  {Tito~D'Agnolo}, \citenamefont {Ellis},\ and\ \citenamefont
  {Radkovski}}]{Berlin:2023gvx}%
  \BibitemOpen
  \bibfield  {author} {\bibinfo {author} {\bibfnamefont {Asher}\ \bibnamefont
  {Berlin}}, \bibinfo {author} {\bibfnamefont {Raffaele}\ \bibnamefont
  {Tito~D'Agnolo}}, \bibinfo {author} {\bibfnamefont {Sebastian A.~R.}\
  \bibnamefont {Ellis}}, \ and\ \bibinfo {author} {\bibfnamefont {Jury~I.}\
  \bibnamefont {Radkovski}},\ }\bibfield  {title} {\enquote {\bibinfo {title}
  {{Signals of millicharged dark matter in light-shining-through-wall
  experiments}},}\ }\href {\doibase 10.1007/JHEP08(2023)017} {\bibfield
  {journal} {\bibinfo  {journal} {JHEP}\ }\textbf {\bibinfo {volume} {08}},\
  \bibinfo {pages} {017} (\bibinfo {year} {2023}{\natexlab{c}})},\ \Eprint
  {http://arxiv.org/abs/2305.05684} {arXiv:2305.05684 [hep-ph]} \BibitemShut
  {NoStop}%
\bibitem [{\citenamefont {Chang}\ \emph {et~al.}(2022)\citenamefont {Chang},
  \citenamefont {Kaplan}, \citenamefont {Rajendran}, \citenamefont {Ramani},\
  and\ \citenamefont {Tanin}}]{Chang:2022gcs}%
  \BibitemOpen
  \bibfield  {author} {\bibinfo {author} {\bibfnamefont {Jae~Hyeok}\
  \bibnamefont {Chang}}, \bibinfo {author} {\bibfnamefont {David~E.}\
  \bibnamefont {Kaplan}}, \bibinfo {author} {\bibfnamefont {Surjeet}\
  \bibnamefont {Rajendran}}, \bibinfo {author} {\bibfnamefont {Harikrishnan}\
  \bibnamefont {Ramani}}, \ and\ \bibinfo {author} {\bibfnamefont {Erwin~H.}\
  \bibnamefont {Tanin}},\ }\bibfield  {title} {\enquote {\bibinfo {title}
  {{Dark Solar Wind}},}\ }\href {\doibase 10.1103/PhysRevLett.129.211101}
  {\bibfield  {journal} {\bibinfo  {journal} {Phys. Rev. Lett.}\ }\textbf
  {\bibinfo {volume} {129}},\ \bibinfo {pages} {211101} (\bibinfo {year}
  {2022})},\ \Eprint {http://arxiv.org/abs/2205.11527} {arXiv:2205.11527
  [hep-ph]} \BibitemShut {NoStop}%
\bibitem [{\citenamefont {Fiorillo}\ and\ \citenamefont
  {Vitagliano}(2024)}]{Fiorillo:2024upk}%
  \BibitemOpen
  \bibfield  {author} {\bibinfo {author} {\bibfnamefont {Damiano F.~G.}\
  \bibnamefont {Fiorillo}}\ and\ \bibinfo {author} {\bibfnamefont {Edoardo}\
  \bibnamefont {Vitagliano}},\ }\bibfield  {title} {\enquote {\bibinfo {title}
  {{Self-interacting dark sectors in supernovae are fluid}},}\ }\href@noop {}
  {\  (\bibinfo {year} {2024})},\ \Eprint {http://arxiv.org/abs/2404.07714}
  {arXiv:2404.07714 [hep-ph]} \BibitemShut {NoStop}%
\bibitem [{\citenamefont {Born}\ and\ \citenamefont
  {Wolf}(1999)}]{Born:1999ory}%
  \BibitemOpen
  \bibfield  {author} {\bibinfo {author} {\bibfnamefont {Max}\ \bibnamefont
  {Born}}\ and\ \bibinfo {author} {\bibfnamefont {Emil}\ \bibnamefont {Wolf}},\
  }\href {\doibase 10.1017/CBO9781139644181} {\emph {\bibinfo {title}
  {{Principles of optics}}}}\ (\bibinfo  {publisher} {Cambridge Univ. Pr.},\
  \bibinfo {year} {1999})\BibitemShut {NoStop}%
\bibitem [{\citenamefont {Ali-Haimoud}\ and\ \citenamefont
  {Hirata}(2011)}]{Ali-Haimoud:2010hou}%
  \BibitemOpen
  \bibfield  {author} {\bibinfo {author} {\bibfnamefont {Yacine}\ \bibnamefont
  {Ali-Haimoud}}\ and\ \bibinfo {author} {\bibfnamefont {Christopher~M.}\
  \bibnamefont {Hirata}},\ }\bibfield  {title} {\enquote {\bibinfo {title}
  {{HyRec: A fast and highly accurate primordial hydrogen and helium
  recombination code}},}\ }\href {\doibase 10.1103/PhysRevD.83.043513}
  {\bibfield  {journal} {\bibinfo  {journal} {Phys. Rev. D}\ }\textbf {\bibinfo
  {volume} {83}},\ \bibinfo {pages} {043513} (\bibinfo {year} {2011})},\
  \Eprint {http://arxiv.org/abs/1011.3758} {arXiv:1011.3758 [astro-ph.CO]}
  \BibitemShut {NoStop}%
\bibitem [{\citenamefont {Lee}\ and\ \citenamefont
  {Ali-Ha\"\i{}moud}(2020)}]{Lee:2020obi}%
  \BibitemOpen
  \bibfield  {author} {\bibinfo {author} {\bibfnamefont {Nanoom}\ \bibnamefont
  {Lee}}\ and\ \bibinfo {author} {\bibfnamefont {Yacine}\ \bibnamefont
  {Ali-Ha\"\i{}moud}},\ }\bibfield  {title} {\enquote {\bibinfo {title}
  {{HYREC-2: a highly accurate sub-millisecond recombination code}},}\ }\href
  {\doibase 10.1103/PhysRevD.102.083517} {\bibfield  {journal} {\bibinfo
  {journal} {Phys. Rev. D}\ }\textbf {\bibinfo {volume} {102}},\ \bibinfo
  {pages} {083517} (\bibinfo {year} {2020})},\ \Eprint
  {http://arxiv.org/abs/2007.14114} {arXiv:2007.14114 [astro-ph.CO]}
  \BibitemShut {NoStop}%
\bibitem [{\citenamefont {Aghanim}\ \emph {et~al.}(2020)\citenamefont {Aghanim}
  \emph {et~al.}}]{Planck:2018vyg}%
  \BibitemOpen
  \bibfield  {author} {\bibinfo {author} {\bibfnamefont {N.}~\bibnamefont
  {Aghanim}} \emph {et~al.} (\bibinfo {collaboration} {Planck}),\ }\bibfield
  {title} {\enquote {\bibinfo {title} {{Planck 2018 results. VI. Cosmological
  parameters}},}\ }\href {\doibase 10.1051/0004-6361/201833910} {\bibfield
  {journal} {\bibinfo  {journal} {Astron. Astrophys.}\ }\textbf {\bibinfo
  {volume} {641}},\ \bibinfo {pages} {A6} (\bibinfo {year} {2020})},\ \bibinfo
  {note} {[Erratum: Astron.Astrophys. 652, C4 (2021)]},\ \Eprint
  {http://arxiv.org/abs/1807.06209} {arXiv:1807.06209 [astro-ph.CO]}
  \BibitemShut {NoStop}%
\bibitem [{\citenamefont {Parke}(1986)}]{Parke:1986jy}%
  \BibitemOpen
  \bibfield  {author} {\bibinfo {author} {\bibfnamefont {Stephen~J.}\
  \bibnamefont {Parke}},\ }\bibfield  {title} {\enquote {\bibinfo {title}
  {{Nonadiabatic Level Crossing in Resonant Neutrino Oscillations}},}\ }\href
  {\doibase 10.1103/PhysRevLett.57.1275} {\bibfield  {journal} {\bibinfo
  {journal} {Phys. Rev. Lett.}\ }\textbf {\bibinfo {volume} {57}},\ \bibinfo
  {pages} {1275--1278} (\bibinfo {year} {1986})}\BibitemShut {NoStop}%
\bibitem [{\citenamefont {Kuo}\ and\ \citenamefont
  {Pantaleone}(1989)}]{Kuo:1989qe}%
  \BibitemOpen
  \bibfield  {author} {\bibinfo {author} {\bibfnamefont {Tzee-Ke}\ \bibnamefont
  {Kuo}}\ and\ \bibinfo {author} {\bibfnamefont {James~T.}\ \bibnamefont
  {Pantaleone}},\ }\bibfield  {title} {\enquote {\bibinfo {title} {{Neutrino
  Oscillations in Matter}},}\ }\href {\doibase 10.1103/RevModPhys.61.937}
  {\bibfield  {journal} {\bibinfo  {journal} {Rev. Mod. Phys.}\ }\textbf
  {\bibinfo {volume} {61}},\ \bibinfo {pages} {937} (\bibinfo {year}
  {1989})}\BibitemShut {NoStop}%
\bibitem [{\citenamefont {Brahma}\ \emph {et~al.}(2023)\citenamefont {Brahma},
  \citenamefont {Berlin},\ and\ \citenamefont {Schutz}}]{Brahma:2023zcw}%
  \BibitemOpen
  \bibfield  {author} {\bibinfo {author} {\bibfnamefont {Nirmalya}\
  \bibnamefont {Brahma}}, \bibinfo {author} {\bibfnamefont {Asher}\
  \bibnamefont {Berlin}}, \ and\ \bibinfo {author} {\bibfnamefont {Katelin}\
  \bibnamefont {Schutz}},\ }\bibfield  {title} {\enquote {\bibinfo {title}
  {{Photon-dark photon conversion with multiple level crossings}},}\ }\href
  {\doibase 10.1103/PhysRevD.108.095045} {\bibfield  {journal} {\bibinfo
  {journal} {Phys. Rev. D}\ }\textbf {\bibinfo {volume} {108}},\ \bibinfo
  {pages} {095045} (\bibinfo {year} {2023})},\ \Eprint
  {http://arxiv.org/abs/2308.08586} {arXiv:2308.08586 [hep-ph]} \BibitemShut
  {NoStop}%
\bibitem [{\citenamefont {{Illarionov}}\ and\ \citenamefont
  {{Siuniaev}}(1975)}]{Illarionov1975}%
  \BibitemOpen
  \bibfield  {author} {\bibinfo {author} {\bibfnamefont {A.~F.}\ \bibnamefont
  {{Illarionov}}}\ and\ \bibinfo {author} {\bibfnamefont {R.~A.}\ \bibnamefont
  {{Siuniaev}}},\ }\bibfield  {title} {\enquote {\bibinfo {title}
  {{Comptonization, the background-radiation spectrum, and the thermal history
  of the universe}},}\ }\href@noop {} {\bibfield  {journal} {\bibinfo
  {journal} {sovast}\ }\textbf {\bibinfo {volume} {18}},\ \bibinfo {pages}
  {691--699} (\bibinfo {year} {1975})}\BibitemShut {NoStop}%
\bibitem [{\citenamefont {{Chan}}\ and\ \citenamefont
  {{Jones}}(1975)}]{ChanJones1975}%
  \BibitemOpen
  \bibfield  {author} {\bibinfo {author} {\bibfnamefont {K.~L.}\ \bibnamefont
  {{Chan}}}\ and\ \bibinfo {author} {\bibfnamefont {B.~J.~T.}\ \bibnamefont
  {{Jones}}},\ }\bibfield  {title} {\enquote {\bibinfo {title} {{The evolution
  of the cosmic radiation spectrum under the influence of turbulent heating. I.
  Theory.}}}\ }\href {\doibase 10.1086/153811} {\bibfield  {journal} {\bibinfo
  {journal} {\apj}\ }\textbf {\bibinfo {volume} {200}},\ \bibinfo {pages}
  {454--470} (\bibinfo {year} {1975})}\BibitemShut {NoStop}%
\bibitem [{\citenamefont {{Danese}}\ and\ \citenamefont {{de
  Zotti}}(1980)}]{DaneseDeZotti80}%
  \BibitemOpen
  \bibfield  {author} {\bibinfo {author} {\bibfnamefont {L.}~\bibnamefont
  {{Danese}}}\ and\ \bibinfo {author} {\bibfnamefont {G.}~\bibnamefont {{de
  Zotti}}},\ }\bibfield  {title} {\enquote {\bibinfo {title} {{On distortions
  in the Rayleigh-Jeans region of the cosmic background radiation spectrum}},}\
  }\href@noop {} {\bibfield  {journal} {\bibinfo  {journal} {aap}\ }\textbf
  {\bibinfo {volume} {84}},\ \bibinfo {pages} {364} (\bibinfo {year}
  {1980})}\BibitemShut {NoStop}%
\bibitem [{\citenamefont {{Danese}}\ and\ \citenamefont {{de
  Zotti}}(1982)}]{DaneseDeZotti82}%
  \BibitemOpen
  \bibfield  {author} {\bibinfo {author} {\bibfnamefont {L.}~\bibnamefont
  {{Danese}}}\ and\ \bibinfo {author} {\bibfnamefont {G.}~\bibnamefont {{de
  Zotti}}},\ }\bibfield  {title} {\enquote {\bibinfo {title} {{Double Compton
  process and the spectrum of the microwave background}},}\ }\href@noop {}
  {\bibfield  {journal} {\bibinfo  {journal} {aap}\ }\textbf {\bibinfo {volume}
  {107}},\ \bibinfo {pages} {39--42} (\bibinfo {year} {1982})}\BibitemShut
  {NoStop}%
\bibitem [{\citenamefont {{Sunyaev}}\ and\ \citenamefont
  {{Zeldovich}}(1970)}]{Sunyaev1970}%
  \BibitemOpen
  \bibfield  {author} {\bibinfo {author} {\bibfnamefont {R.~A.}\ \bibnamefont
  {{Sunyaev}}}\ and\ \bibinfo {author} {\bibfnamefont {Ya.~B.}\ \bibnamefont
  {{Zeldovich}}},\ }\bibfield  {title} {\enquote {\bibinfo {title} {{The
  interaction of matter and radiation in the hot model of the Universe, II}},}\
  }\href {\doibase 10.1007/BF00653472} {\bibfield  {journal} {\bibinfo
  {journal} {apss}\ }\textbf {\bibinfo {volume} {7}},\ \bibinfo {pages}
  {20--30} (\bibinfo {year} {1970})}\BibitemShut {NoStop}%
\bibitem [{\citenamefont {{Zeldovich}}\ and\ \citenamefont
  {{Sunyaev}}(1969)}]{Zeldovich69}%
  \BibitemOpen
  \bibfield  {author} {\bibinfo {author} {\bibfnamefont {Ya.~B.}\ \bibnamefont
  {{Zeldovich}}}\ and\ \bibinfo {author} {\bibfnamefont {R.~A.}\ \bibnamefont
  {{Sunyaev}}},\ }\bibfield  {title} {\enquote {\bibinfo {title} {{The
  Interaction of Matter and Radiation in a Hot-Model Universe}},}\ }\href
  {\doibase 10.1007/BF00661821} {\bibfield  {journal} {\bibinfo  {journal}
  {apss}\ }\textbf {\bibinfo {volume} {4}},\ \bibinfo {pages} {301--316}
  (\bibinfo {year} {1969})}\BibitemShut {NoStop}%
\bibitem [{\citenamefont {{Daly}}(1991)}]{Daly91}%
  \BibitemOpen
  \bibfield  {author} {\bibinfo {author} {\bibfnamefont {R.~A.}\ \bibnamefont
  {{Daly}}},\ }\bibfield  {title} {\enquote {\bibinfo {title} {{Spectral
  Distortions of the Microwave Background Radiation Resulting from the Damping
  of Pressure Waves}},}\ }\href {\doibase 10.1086/169866} {\bibfield  {journal}
  {\bibinfo  {journal} {\apj}\ }\textbf {\bibinfo {volume} {371}},\ \bibinfo
  {pages} {14} (\bibinfo {year} {1991})}\BibitemShut {NoStop}%
\bibitem [{\citenamefont {Hu}\ and\ \citenamefont {Silk}(1993)}]{Hu:1992dc}%
  \BibitemOpen
  \bibfield  {author} {\bibinfo {author} {\bibfnamefont {Wayne}\ \bibnamefont
  {Hu}}\ and\ \bibinfo {author} {\bibfnamefont {Joseph}\ \bibnamefont {Silk}},\
  }\bibfield  {title} {\enquote {\bibinfo {title} {{Thermalization and spectral
  distortions of the cosmic background radiation}},}\ }\href {\doibase
  10.1103/PhysRevD.48.485} {\bibfield  {journal} {\bibinfo  {journal} {Phys.
  Rev. D}\ }\textbf {\bibinfo {volume} {48}},\ \bibinfo {pages} {485--502}
  (\bibinfo {year} {1993})}\BibitemShut {NoStop}%
\bibitem [{\citenamefont {Hu}(1995)}]{Hu:1995em}%
  \BibitemOpen
  \bibfield  {author} {\bibinfo {author} {\bibfnamefont {Wayne~T.}\
  \bibnamefont {Hu}},\ }\emph {\bibinfo {title} {{Wandering in the Background:
  A CMB Explorer}}},\ \href@noop {} {\bibinfo {type} {Other thesis}} (\bibinfo
  {year} {1995}),\ \Eprint {http://arxiv.org/abs/astro-ph/9508126}
  {arXiv:astro-ph/9508126} \BibitemShut {NoStop}%
\bibitem [{\citenamefont {Chluba}\ and\ \citenamefont
  {Sunyaev}(2012)}]{Chluba:2011hw}%
  \BibitemOpen
  \bibfield  {author} {\bibinfo {author} {\bibfnamefont {J.}~\bibnamefont
  {Chluba}}\ and\ \bibinfo {author} {\bibfnamefont {R.~A.}\ \bibnamefont
  {Sunyaev}},\ }\bibfield  {title} {\enquote {\bibinfo {title} {{The evolution
  of CMB spectral distortions in the early Universe}},}\ }\href {\doibase
  10.1111/j.1365-2966.2011.19786.x} {\bibfield  {journal} {\bibinfo  {journal}
  {Mon. Not. Roy. Astron. Soc.}\ }\textbf {\bibinfo {volume} {419}},\ \bibinfo
  {pages} {1294--1314} (\bibinfo {year} {2012})},\ \Eprint
  {http://arxiv.org/abs/1109.6552} {arXiv:1109.6552 [astro-ph.CO]} \BibitemShut
  {NoStop}%
\bibitem [{\citenamefont {Chluba}\ and\ \citenamefont
  {Jeong}(2014)}]{Chluba:2013pya}%
  \BibitemOpen
  \bibfield  {author} {\bibinfo {author} {\bibfnamefont {Jens}\ \bibnamefont
  {Chluba}}\ and\ \bibinfo {author} {\bibfnamefont {Donghui}\ \bibnamefont
  {Jeong}},\ }\bibfield  {title} {\enquote {\bibinfo {title} {{Teasing bits of
  information out of the CMB energy spectrum}},}\ }\href {\doibase
  10.1093/mnras/stt2327} {\bibfield  {journal} {\bibinfo  {journal} {Mon. Not.
  Roy. Astron. Soc.}\ }\textbf {\bibinfo {volume} {438}},\ \bibinfo {pages}
  {2065--2082} (\bibinfo {year} {2014})},\ \Eprint
  {http://arxiv.org/abs/1306.5751} {arXiv:1306.5751 [astro-ph.CO]} \BibitemShut
  {NoStop}%
\bibitem [{\citenamefont {Chluba}(2013)}]{Chluba:2013vsa}%
  \BibitemOpen
  \bibfield  {author} {\bibinfo {author} {\bibfnamefont {Jens}\ \bibnamefont
  {Chluba}},\ }\bibfield  {title} {\enquote {\bibinfo {title} {{Green's
  function of the cosmological thermalization problem}},}\ }\href {\doibase
  10.1093/mnras/stt1025} {\bibfield  {journal} {\bibinfo  {journal} {Mon. Not.
  Roy. Astron. Soc.}\ }\textbf {\bibinfo {volume} {434}},\ \bibinfo {pages}
  {352} (\bibinfo {year} {2013})},\ \Eprint {http://arxiv.org/abs/1304.6120}
  {arXiv:1304.6120 [astro-ph.CO]} \BibitemShut {NoStop}%
\bibitem [{\citenamefont {Acharya}\ and\ \citenamefont
  {Khatri}(2019)}]{Acharya:2018iwh}%
  \BibitemOpen
  \bibfield  {author} {\bibinfo {author} {\bibfnamefont {Sandeep~Kumar}\
  \bibnamefont {Acharya}}\ and\ \bibinfo {author} {\bibfnamefont {Rishi}\
  \bibnamefont {Khatri}},\ }\bibfield  {title} {\enquote {\bibinfo {title}
  {{Rich structure of non-thermal relativistic CMB spectral distortions from
  high energy particle cascades at redshifts $z\lesssim 2\times 10^5$}},}\
  }\href {\doibase 10.1103/PhysRevD.99.043520} {\bibfield  {journal} {\bibinfo
  {journal} {Phys. Rev. D}\ }\textbf {\bibinfo {volume} {99}},\ \bibinfo
  {pages} {043520} (\bibinfo {year} {2019})},\ \Eprint
  {http://arxiv.org/abs/1808.02897} {arXiv:1808.02897 [astro-ph.CO]}
  \BibitemShut {NoStop}%
\bibitem [{\citenamefont {Acharya}\ and\ \citenamefont
  {Chluba}(2022)}]{Acharya:2021zhq}%
  \BibitemOpen
  \bibfield  {author} {\bibinfo {author} {\bibfnamefont {Sandeep~Kumar}\
  \bibnamefont {Acharya}}\ and\ \bibinfo {author} {\bibfnamefont {Jens}\
  \bibnamefont {Chluba}},\ }\bibfield  {title} {\enquote {\bibinfo {title}
  {{CMB spectral distortions from continuous large energy release}},}\ }\href
  {\doibase 10.1093/mnras/stac2137} {\bibfield  {journal} {\bibinfo  {journal}
  {Mon. Not. Roy. Astron. Soc.}\ }\textbf {\bibinfo {volume} {515}},\ \bibinfo
  {pages} {5775--5789} (\bibinfo {year} {2022})},\ \Eprint
  {http://arxiv.org/abs/2112.06699} {arXiv:2112.06699 [astro-ph.CO]}
  \BibitemShut {NoStop}%
\bibitem [{\citenamefont {Chluba}\ \emph {et~al.}(2020)\citenamefont {Chluba},
  \citenamefont {Ravenni},\ and\ \citenamefont {Acharya}}]{Chluba:2020oip}%
  \BibitemOpen
  \bibfield  {author} {\bibinfo {author} {\bibfnamefont {Jens}\ \bibnamefont
  {Chluba}}, \bibinfo {author} {\bibfnamefont {Andrea}\ \bibnamefont
  {Ravenni}}, \ and\ \bibinfo {author} {\bibfnamefont {Sandeep~Kumar}\
  \bibnamefont {Acharya}},\ }\bibfield  {title} {\enquote {\bibinfo {title}
  {{Thermalization of large energy release in the early Universe}},}\ }\href
  {\doibase 10.1093/mnras/staa2131} {\bibfield  {journal} {\bibinfo  {journal}
  {Mon. Not. Roy. Astron. Soc.}\ }\textbf {\bibinfo {volume} {498}},\ \bibinfo
  {pages} {959--980} (\bibinfo {year} {2020})},\ \Eprint
  {http://arxiv.org/abs/2005.11325} {arXiv:2005.11325 [astro-ph.CO]}
  \BibitemShut {NoStop}%
\bibitem [{\citenamefont {Lesgourgues}(2011)}]{Lesgourgues:2011re}%
  \BibitemOpen
  \bibfield  {author} {\bibinfo {author} {\bibfnamefont {Julien}\ \bibnamefont
  {Lesgourgues}},\ }\bibfield  {title} {\enquote {\bibinfo {title} {{The Cosmic
  Linear Anisotropy Solving System (CLASS) I: Overview}},}\ }\href@noop {} {\
  (\bibinfo {year} {2011})},\ \Eprint {http://arxiv.org/abs/1104.2932}
  {arXiv:1104.2932 [astro-ph.IM]} \BibitemShut {NoStop}%
\bibitem [{\citenamefont {Blas}\ \emph {et~al.}(2011)\citenamefont {Blas},
  \citenamefont {Lesgourgues},\ and\ \citenamefont {Tram}}]{Blas:2011rf}%
  \BibitemOpen
  \bibfield  {author} {\bibinfo {author} {\bibfnamefont {Diego}\ \bibnamefont
  {Blas}}, \bibinfo {author} {\bibfnamefont {Julien}\ \bibnamefont
  {Lesgourgues}}, \ and\ \bibinfo {author} {\bibfnamefont {Thomas}\
  \bibnamefont {Tram}},\ }\bibfield  {title} {\enquote {\bibinfo {title} {{The
  Cosmic Linear Anisotropy Solving System (CLASS) II: Approximation
  schemes}},}\ }\href {\doibase 10.1088/1475-7516/2011/07/034} {\bibfield
  {journal} {\bibinfo  {journal} {JCAP}\ }\textbf {\bibinfo {volume} {07}},\
  \bibinfo {pages} {034} (\bibinfo {year} {2011})},\ \Eprint
  {http://arxiv.org/abs/1104.2933} {arXiv:1104.2933 [astro-ph.CO]} \BibitemShut
  {NoStop}%
\bibitem [{\citenamefont {Chluba}\ \emph {et~al.}(2024)\citenamefont {Chluba},
  \citenamefont {Cyr},\ and\ \citenamefont {Johnson}}]{Chluba:2024wui}%
  \BibitemOpen
  \bibfield  {author} {\bibinfo {author} {\bibfnamefont {Jens}\ \bibnamefont
  {Chluba}}, \bibinfo {author} {\bibfnamefont {Bryce}\ \bibnamefont {Cyr}}, \
  and\ \bibinfo {author} {\bibfnamefont {Matthew~C.}\ \bibnamefont {Johnson}},\
  }\bibfield  {title} {\enquote {\bibinfo {title} {{Revisiting Dark Photon
  Constraints from CMB Spectral Distortions}},}\ }\href@noop {} {\  (\bibinfo
  {year} {2024})},\ \Eprint {http://arxiv.org/abs/2409.12115} {arXiv:2409.12115
  [astro-ph.CO]} \BibitemShut {NoStop}%
\bibitem [{\citenamefont {Wilks}(1938)}]{Wilks:1938dza}%
  \BibitemOpen
  \bibfield  {author} {\bibinfo {author} {\bibfnamefont {S.~S.}\ \bibnamefont
  {Wilks}},\ }\bibfield  {title} {\enquote {\bibinfo {title} {{The Large-Sample
  Distribution of the Likelihood Ratio for Testing Composite Hypotheses}},}\
  }\href {\doibase 10.1214/aoms/1177732360} {\bibfield  {journal} {\bibinfo
  {journal} {Annals Math. Statist.}\ }\textbf {\bibinfo {volume} {9}},\
  \bibinfo {pages} {60--62} (\bibinfo {year} {1938})}\BibitemShut {NoStop}%
\bibitem [{\citenamefont {Workman}\ \emph {et~al.}(2022)\citenamefont {Workman}
  \emph {et~al.}}]{ParticleDataGroup:2022pth}%
  \BibitemOpen
  \bibfield  {author} {\bibinfo {author} {\bibfnamefont {R.~L.}\ \bibnamefont
  {Workman}} \emph {et~al.} (\bibinfo {collaboration} {Particle Data Group}),\
  }\bibfield  {title} {\enquote {\bibinfo {title} {{Review of Particle
  Physics}},}\ }\href {\doibase 10.1093/ptep/ptac097} {\bibfield  {journal}
  {\bibinfo  {journal} {PTEP}\ }\textbf {\bibinfo {volume} {2022}},\ \bibinfo
  {pages} {083C01} (\bibinfo {year} {2022})}\BibitemShut {NoStop}%
\bibitem [{\citenamefont {Hook}\ \emph {et~al.}(2024)\citenamefont {Hook},
  \citenamefont {Marques-Tavares},\ and\ \citenamefont
  {Ristow}}]{Hook:2023smg}%
  \BibitemOpen
  \bibfield  {author} {\bibinfo {author} {\bibfnamefont {Anson}\ \bibnamefont
  {Hook}}, \bibinfo {author} {\bibfnamefont {Gustavo}\ \bibnamefont
  {Marques-Tavares}}, \ and\ \bibinfo {author} {\bibfnamefont {Clayton}\
  \bibnamefont {Ristow}},\ }\bibfield  {title} {\enquote {\bibinfo {title}
  {{CMB spectral distortions from an axion-dark photon-photon interaction}},}\
  }\href {\doibase 10.1007/JHEP05(2024)086} {\bibfield  {journal} {\bibinfo
  {journal} {JHEP}\ }\textbf {\bibinfo {volume} {05}},\ \bibinfo {pages} {086}
  (\bibinfo {year} {2024})},\ \Eprint {http://arxiv.org/abs/2306.13135}
  {arXiv:2306.13135 [hep-ph]} \BibitemShut {NoStop}%
\bibitem [{\citenamefont {Chluba}(2014)}]{Chluba:2013kua}%
  \BibitemOpen
  \bibfield  {author} {\bibinfo {author} {\bibfnamefont {Jens}\ \bibnamefont
  {Chluba}},\ }\bibfield  {title} {\enquote {\bibinfo {title} {{Refined
  approximations for the distortion visibility function and
  \ensuremath{\mu}-type spectral distortions}},}\ }\href {\doibase
  10.1093/mnras/stu414} {\bibfield  {journal} {\bibinfo  {journal} {Mon. Not.
  Roy. Astron. Soc.}\ }\textbf {\bibinfo {volume} {440}},\ \bibinfo {pages}
  {2544--2563} (\bibinfo {year} {2014})},\ \Eprint
  {http://arxiv.org/abs/1312.6030} {arXiv:1312.6030 [astro-ph.CO]} \BibitemShut
  {NoStop}%
\end{thebibliography}%

\end{document}